\documentclass[fleqn,usenatbib]{mnras}

\usepackage{amssymb}

\usepackage{newtxtext,newtxmath}

\usepackage[T1]{fontenc}
\usepackage{comment}
\usepackage{booktabs}
\usepackage{lscape}
\usepackage{rotating}
\usepackage{longtable} 
\usepackage{pdflscape} 
\usepackage{caption}
\usepackage{threeparttable}         
\usepackage{threeparttablex} 

\DeclareRobustCommand{\VAN}[3]{#2}
\let\VANthebibliography\thebibliography
\def\thebibliography{\DeclareRobustCommand{\VAN}[3]{##3}\VANthebibliography}

\usepackage{graphicx}	
\usepackage{amsmath}	

\usepackage{etoolbox}

\defcitealias{Garcia2019}{G19}
\defcitealias{Britavskiy2019}{BR19}
\defcitealias{Camacho2016}{C16}
\defcitealias{Kaufer2004}{K04}

\newcommand{\ha}{$\rm H\alpha$}
\newcommand{\hb}{$\rm H\beta$}
\newcommand{\hg}{$\rm H\gamma$}
\newcommand{\hd}{$\rm H\delta$}
\newcommand{\HeI}{He~{\sc i}}
\newcommand{\HeII}{He~{\sc ii}}
\newcommand{\HeIandII}{He~{\sc i+ii}}
\newcommand{\SiII}{Si~{\sc ii}}
\newcommand{\SiIII}{Si~{\sc iii}}
\newcommand{\SiIV}{Si~{\sc iv}}

\newcommand{\MgII}{Mg~{\sc ii}}
\newcommand{\FeI}{Fe~{\sc i}}
\newcommand{\FeII}{Fe~{\sc ii}}
\newcommand{\CrII}{Cr~{\sc ii}}
\newcommand{\NIII}{N~{\sc iii}}

\newcommand{\NV}{N~{\sc v}}
\newcommand{\CIII}{C~{\sc iii}}
\newcommand{\CIV}{C~{\sc iv}}
\newcommand{\OII}{O~{\sc ii}}

\newcommand{\EBV}{$\mathrm{E}(B-V)_\mathrm{fg}$}
\newcommand{\vrad}{\mbox{$v_{\mathrm{rad}}$}}
\newcommand{\Teff}{\mbox{$T_{\mathrm{eff}}$}}
\newcommand{\vsini}{$ v \sin i$}

\newcommand{\kms}{\mbox{km~s$^{-1}$}}
\newcommand{\Msun}{\mbox{M$_{\odot}$}}

\newcommand{\Zsun}{\mbox{Z$_{\odot}$}}
\newcommand{\Osun}{\mbox{O$_{\odot}$}}
\newcommand{\Fesun}{\mbox{Fe$_{\odot}$}}
\newcommand{\h}{$^{\rm h}$}
\newcommand{\m}{$^{\rm m}$}
\newcommand{\s}{$^{\rm s}$}

\newcommand{\TotalStars}{174}
\newcommand{\ForegroundStars}{5}

\newcommand{\LateStarsSexA}{4}

\newcommand{\AFStars}{6}
\newcommand{\OBStars}{159}

\newcommand{\Ostars}{71}

\newcommand{\LowOBStars}{41}

\newcommand{\EarlyBI}{22}

\newcommand{\BAI}{38}
\newcommand{\RedI}{4}
\newcommand{\SBStars}{33}
\newcommand{\CHEStars}{18}
\newcommand{\SBCHEStars}{4}
\newcommand{\BoxStars}{79}
\newcommand{\OBoxStars}{50}
\newcommand{\BBoxStars}{26}
\newcommand{\OBAboxStars}{3}

\newcommand{\MiniOBoxStars}{21}

\newcommand{\OoutBoxStars}{17}

\newcommand{\MiniBoxStarsWithoutOBA}{25}

\newcommand{\PrecisionBox}{66}

\newcommand{\PrecisionMiniBox}{84}

\newcommand{\DefineRemark}[2]{%
  \expandafter\newcommand\csname rmk-#1\endcsname{#2}%
}
\newcommand{\Remark}[1]{\csname rmk-#1\endcsname}
\DefineRemark{LSS1.OB122}{s116}
\DefineRemark{LSS1.OB221}{s091}
\DefineRemark{LSS1.OB323}{s077}
\DefineRemark{LSS1.OB324}{s111}
\DefineRemark{LSS1.OB326}{s014}
\DefineRemark{LSS1.OB422}{s074}
\DefineRemark{LSS1.OB521}{s022}
\DefineRemark{LSS1.OB523}{s038}
\DefineRemark{LSS1.OB525}{s119}
\DefineRemark{LSS1.OB625}{s047}
\DefineRemark{MOS1.s01}{s041}
\DefineRemark{MOS1.s02}{s003}
\DefineRemark{MOS1.s03}{s029}
\DefineRemark{MOS1.s04}{s004}
\DefineRemark{MOS1.s06}{s028}
\DefineRemark{MOS1.s07}{s042}
\DefineRemark{MOS1.s08}{s071}
\DefineRemark{MOS1.s10}{s092}
\DefineRemark{MOS1.s11}{s061}
\DefineRemark{MOS1.s12}{s099}
\DefineRemark{MOS1.s13}{s063}
\DefineRemark{MOS1.s14}{s089}
\DefineRemark{MOS2.s01}{s059}
\DefineRemark{MOS2.s03}{s094}
\DefineRemark{MOS2.s05}{s141}
\DefineRemark{MOS2.s06}{s067}
\DefineRemark{MOS2.s07}{s013}
\DefineRemark{MOS2.s08}{s100}
\DefineRemark{MOS2.s09}{s078}
\DefineRemark{MOS2.s12}{s016}
\DefineRemark{MOS2.s13}{s018}
\DefineRemark{MOS2.s15}{s007}
\DefineRemark{MOS2.s17}{s051}
\DefineRemark{MOS2.s18}{s037}
\DefineRemark{MOS2.s19}{s021}
\DefineRemark{MOS2.s22}{s052}
\DefineRemark{MOS2.s24}{s069}
\DefineRemark{MOS2.s25}{s009}
\DefineRemark{MOS2.s26}{s137}
\DefineRemark{MOS2.s27}{s012}
\DefineRemark{MOS2.s28}{s055}
\DefineRemark{MOS2.s29}{s032}
\DefineRemark{MOS2.s30}{s050}
\DefineRemark{MOS2.s31}{s082}
\DefineRemark{MOS2.s32}{s066}
\DefineRemark{MOS2.s33}{s158}
\DefineRemark{MOS2.s34}{s031}
\DefineRemark{MOS2.s35}{s056}
\DefineRemark{MOS2.s36}{s072}
\DefineRemark{MOS2.s37}{s133}
\DefineRemark{MOS2.s38}{s131}
\DefineRemark{MOS2.s39}{s064}
\DefineRemark{MOS2.s40}{s030}
\DefineRemark{LSS2.OB112}{s173}
\DefineRemark{LSS2.OB121}{s128}
\DefineRemark{LSS2.OB122}{s017}
\DefineRemark{LSS2.OB123}{s130}
\DefineRemark{LSS2.OB124}{s156}
\DefineRemark{LSS2.OB125}{s144}
\DefineRemark{LSS2.OB126}{s140}
\DefineRemark{LSS2.OB127}{s049}
\DefineRemark{LSS2.OB1213}{s129}
\DefineRemark{LSS2.OB326}{s103}
\DefineRemark{LSS2.OB327}{s062}
\DefineRemark{LSS2.OB329}{s132}
\DefineRemark{LSS2.OB3211}{s114}
\DefineRemark{LSS3.OB01104}{s006}
\DefineRemark{LSS3.OB01201}{s165}
\DefineRemark{LSS3.OB01202}{s142}
\DefineRemark{LSS3.OB01203}{s080}
\DefineRemark{LSS3.OB01204}{s121}
\DefineRemark{LSS3.OB01205}{s101}
\DefineRemark{LSS3.OB01207}{s087}
\DefineRemark{LSS3.OB02203}{s115}
\DefineRemark{LSS3.OB02205}{s155}
\DefineRemark{LSS3.OB02208}{s117}
\DefineRemark{LSS3.OB02209}{s096}
\DefineRemark{LSS3.OB03102}{s143}
\DefineRemark{LSS3.OB03104}{s019}
\DefineRemark{LSS3.OB03105}{s150}
\DefineRemark{LSS3.OB03106}{s163}
\DefineRemark{LSS3.OB03107}{s153}
\DefineRemark{LSS3.OB03108}{s170}
\DefineRemark{LSS3.OB03109}{s148}
\DefineRemark{LSS3.OB03202}{s005}
\DefineRemark{LSS3.OB04201}{s040}
\DefineRemark{LSS3.OB04202}{s167}
\DefineRemark{LSS3.OB05101}{s174}
\DefineRemark{LSS3.OB05207}{s160}
\DefineRemark{LSS3.OB05209}{s034}
\DefineRemark{LSS3.OB06101}{s154}
\DefineRemark{LSS3.OB06102}{s134}
\DefineRemark{LSS3.OB06201}{s166}
\DefineRemark{LSS3.OB06202}{s076}
\DefineRemark{LSS3.OB06206}{s162}
\DefineRemark{LSS3.OB06208}{s109}
\DefineRemark{LSS3.OB11201}{s001}
\DefineRemark{LSS3.OB11202}{s023}
\DefineRemark{LSS3.OB11203}{s172}
\DefineRemark{LSS3.OB11205}{s113}
\DefineRemark{LSS3.OB11206}{s122}
\DefineRemark{LSS3.OB11207}{s123}
\DefineRemark{LSS3.OB13202}{s048}
\DefineRemark{LSS3.OB13203}{s090}
\DefineRemark{LSS3.OB13205}{s015}
\DefineRemark{LSS3.OB13207}{s138}
\DefineRemark{LSS3.OB13209}{s039}
\DefineRemark{LSS3.OB15102}{s088}
\DefineRemark{LSS3.OB15204}{s046}
\DefineRemark{LSS3.OB15206}{s084}
\DefineRemark{LSS3.OB15207}{s075}
\DefineRemark{LSS3.OB15208}{s145}
\DefineRemark{LSS3.OB15209}{s152}
\DefineRemark{LSS3.OB15210}{s147}
\DefineRemark{LSS3.OB17212}{s026}
\DefineRemark{LSS3.OB31201}{s139}
\DefineRemark{LSS3.OB31202}{s053}
\DefineRemark{LSS3.OB31210}{s058}
\DefineRemark{LSS3.OB31211}{s169}
\DefineRemark{LSS3.OB44202}{s157}
\DefineRemark{LSS3.OB44209}{s073}
\DefineRemark{LSS3.OB44211}{s027}
\DefineRemark{LSS3.OB44213}{s010}
\DefineRemark{LSS3.OB45203}{s035}
\DefineRemark{LSS3.OB45205}{s151}
\DefineRemark{LSS3.OB46203}{s044}
\DefineRemark{LSS3.OB46204}{s136}
\DefineRemark{LSS3.OB46205}{s164}
\DefineRemark{LSS3.OB46206}{s081}
\DefineRemark{LSS3.OB46207}{s126}
\DefineRemark{LSS3.OB47103}{s060}
\DefineRemark{LSS3.OB47203}{s079}
\DefineRemark{LSS3.OB01102}{s106}
\DefineRemark{LSS3.OB04205}{s105}
\DefineRemark{LSS3.OB02207}{s057}
\DefineRemark{LSS3.OB03201}{s083}
\DefineRemark{LSS3.OB17206}{s086}
\DefineRemark{LSS3.OB15203}{s036}
\DefineRemark{LSS3.OB13204}{s108}
\DefineRemark{LSS3.OB15205}{s054}
\DefineRemark{LSS3.OB17201}{s097}
\DefineRemark{LSS3.OB17203}{s068}
\DefineRemark{LSS3.OB18203}{s093}
\DefineRemark{LSS3.OB31203}{s070}
\DefineRemark{LSS3.OB44207}{s043}
\DefineRemark{LSS3.OB44208}{s085}
\DefineRemark{LSS3.OB49211}{s124}
\DefineRemark{LSS3.OB02206}{s112}
\DefineRemark{LSS3.OB17204}{s168}
\DefineRemark{LSS3.OB47204}{s159}
\DefineRemark{LSS3.OB17205}{s033}
\DefineRemark{LSS3.OB17210}{s149}
\DefineRemark{LSS3.OB31204}{s024}
\DefineRemark{LSS3.OB49201}{s025}
\DefineRemark{LSS3.OB49203}{s161}
\DefineRemark{LSS3.OB49205}{s065}
\DefineRemark{LSS3.OB49207}{s118}
\DefineRemark{LSS3.OB49209}{s095}
\DefineRemark{LSS3.OB49212}{s146}
\DefineRemark{LSS3.OB01101}{s120}
\DefineRemark{LSS3.OB02202}{s002}
\DefineRemark{LSS3.OB13208}{s127}
\DefineRemark{LSS3.OB46201}{s171}
\DefineRemark{LSS3.OB15201}{s045}
\DefineRemark{LSS3.OB31209}{s102}
\DefineRemark{LSS3.OB04101}{s125}
\DefineRemark{LSS3.OB04203}{s098}
\DefineRemark{LSS3.OB04206}{s107}
\DefineRemark{LSS3.OB05202}{s110}
\DefineRemark{LSS3.OB05204}{s104}
\DefineRemark{LSS3.OB11204}{s011}
\DefineRemark{LSS3.OB17202}{s020}
\DefineRemark{LSS3.OB49206}{s008}
\DefineRemark{LSS3.OB44206}{s135}

\title{A new reference catalogue for the very metal-poor Universe: +150 OB stars in Sextans~A}

\author[]{
M. Lorenzo,$^{1,2}$\thanks{E-mail: mlorenzo@cab.inta-csic.es}
M. Garcia,$^{1}$
F. Najarro$^{1}$
A. Herrero,$^{3, 4}$
M. Cerviño,$^{1}$
N. Castro,$^{5}$
\\
$^{1}$Centro de Astrobiología (CAB), CSIC-INTA, Carretera de Ajalvir km 4, 28850 Torrejón de Ardoz, Madrid, Spain\\
$^{2}$Departamento de Física Teórica, Universidad Autónoma de Madrid (UAM), Campus de Cantoblanco, E-28049 Madrid, Spain\\
$^{3}$Instituto de Astrofísica de Canarias, La Laguna, Tenerife, Spain\\
$^{4}$Departamento de Astrofísica, Universidad de La Laguna, Tenerife, Spain\\
$^{5}$Leibniz-Institut für Astrophysik Potsdam (AIP), An der Sternwarte 16, 14482, Potsdam, Germany
}

\date{Accepted XXX. Received YYY; in original form ZZZ}

\pubyear{2022}

\begin{document}
\label{firstpage}
\pagerange{\pageref{firstpage}--\pageref{lastpage}}
\maketitle

\begin{abstract}
Local Group (LG) very metal-poor massive stars are the best proxy for the First Stars of the Universe and fundamental to modelling the evolution of early galaxies. These stars may follow new evolutionary pathways restricted to very low metallicities, such as chemically homogeneous evolution (CHE). However, given the great distance leap needed to reach very metal-poor galaxies of the LG and vicinity, no comprehensive spectroscopic studies have been carried out at metallicities lower than the Small Magellanic Cloud (SMC, \mbox{Z = 1/5 \Zsun}) until now. After five observing campaigns at the 10.4-m Gran Telescopio Canarias, we have assembled a low-resolution (\mbox{R $\sim$ 1000}) spectroscopic collection of more than 150 OB stars in the 1/10 \Zsun~galaxy Sextans~A, 
increasing by an order of magnitude the number of massive stars known in this galaxy.
The catalogue includes \BAI~BA-type supergiants, \RedI~red supergiants, and the first candidate 1/10 \Zsun~binary systems, CHE sources and systems hosting stripped stars.
The sample massive stars mainly overlap the higher concentrations of neutral gas of Sextans~A. However, we find some sources in low H~{\sc i} column-density regions.
The colour-magnitude diagram of the galaxy presents large dispersion, which suggests uneven, internal extinction in Sextans~A.
This is the largest catalogue of OB-type stars ever produced at sub-SMC metallicities. This sample constitutes a fundamental first step to unveiling the evolutionary pathways and fates of very metal-poor massive stars, analyzing the dependence of radiation-driven winds with metallicity, and studying binary systems in an environment analogue to the early Universe.





\end{abstract}





\section{Introduction}

Massive stars ($M$ > 8~\Msun) are powerful drivers of galaxy evolution. Their stellar winds, supernova explosions and mighty production of ionizing radiation stir and impact the surrounding interstellar medium to such an extent that they dominate the integrated light of star-forming regions. Massive star properties and spectra are thus essential to interpret evidence from galaxies at all redshifts, while they are also pivotal ingredients to model the chemodynamical evolution of galaxies with time.

Yet, very little is known about the evolution of very metal-poor massive stars ($Z$ $\leq$ 1/10 \Zsun), key to interpreting the processes and events of the early epochs of the Universe.
The increasing number of open questions in the field is demanding a large census of resolved massive stars in environments with very low metallicity. 
Such a sample can be used, after quantitative spectroscopic analysis, to study galaxies at the peak of the cosmic star formation \citep[$z$~$\sim$~2,][]{MadauDickinson2014} or earlier.
Among these questions are the evolutionary pathways of massive stars at metallicities lower than the Small Magellanic Cloud  \citep{Groh2019, Szecsi2022}, the incidence of chemically homogeneous evolution \citep[CHE, ][]{Szecsi2015}, CHE stars as possible ionizing sources of the extreme \mbox{\HeII~4686} nebular lines detected in some very metal-poor galaxies \citep{Kehrig2015,Sobral2019,Wofford2021}, and the relative role of radiation driven winds and binary mass exchange in shaping the evolution of very metal-poor massive stars \citep{Shenar2020}. In addition, the evolutionary channels that lead to \mbox{$\sim$ 30~\Msun}~stellar-size black holes are also most likely related to the extremely low metallicity regimes \citep{MandelMink2016, Marchant2016,vanSon2020}.

The spectroscopic studies of very metal-poor massive stars are greatly hampered by sensitivity issues. For decades, the Small Magellanic Cloud (SMC, \mbox{$Z$ = 1/5 \Zsun}) has represented both a metallicity and a distance frontier since metal-poorer environments are usually found in galaxies at the outer rims of the Local Group \citep{Garcia2021}. This situation is about to change with the advent of the Extremely Large Telescope (ELT) and its multi-object spectrographs HARMONI \citep{Thatte2016, Thatte2020} and MOSAIC \citep{Hammer2014, Evans2015}. They will enable exquisite quantitative spectroscopic analyses of extragalactic massive stars that so far have been reserved to the Magellanic Clouds \citep[e.g. ][]{Mokiem2007a, Ramirez-Agudelo2017, Sabin-Sanjulian2017}. However, the design of both instruments hinders their use as discoverers of massive stars in nearby very metal-poor galaxies. The typical diameter of these galaxies is a few arcminutes, while the field of view of HARMONI is very limited (\mbox{9'' $\times$ 6''} at most) and MOSAIC's spacing of the robot positioners is too broad (23'' diameter). As a consequence, the construction of catalogues of very metal-poor massive stars is still restricted to contemporary observing facilities.

In this work, we present the first major effort to mine the population of massive stars in the galaxy Sextans~A 
\citep[\mbox{10\h 11\m 00.\s 8} - \mbox{04\textdegree41$'$34$''$}, dIrr, \textit{aka} DDO 75, ][]{McConnachie2012}. 
Located in the outer edge of the Local Group (LG), 1.34 Mpc away, and presenting low foreground extinction \citep[\EBV~= 0.044, ][]{Tammann2011}, Sextans~A is a contender to become the next benchmark for low metallicity regimes. Its oxygen abundance derived from H~{\sc ii}~regions ranges \mbox{12 + log(O/H)} = 7.49–7.71 \citep[1/15-1/10 \Osun,][]{Skillman1989,Pilyugin2001,Kniazev2005,Berg2012} and the analysis of blue supergiant revealed the lowest abundance of Fe-group elements in the LG \citep[\mbox{$\langle$[(\FeII, \CrII)/H]$\rangle$ = -0.99 $\pm$ 0.06},][K04]{Kaufer2004}. Its poor Fe content has been confirmed by \citet{Garcia2017} with HST-COS spectroscopy of a selection of O stars. In addition, \citet{Kaufer2004} found that the [$\alpha$/Fe] abundance ratio of Sextans~A is consistent with the solar ratio.

With such promising low values for metallicity, this galaxy has sparked the interest of the community. \citet[][C16]{Camacho2016} were the first to confirm the presence of OB-type stars in Sextans~A. \citet[][G19]{Garcia2019} reported four young O stars on its outskirts, signalling massive star formation in low-density environments. In addition, \citet{Britavskiy2014, Britavskiy2015} unveiled the first red supergiants in the galaxy with optical spectroscopy of \textit{Spitzer}/IRAC candidates. Nevertheless, these works provide a deficient coverage of the most prominent star-forming regions of Sextans~A.

In this work, we present the results of four new observing campaigns at the 10.4-m Gran Telescopio Canarias (GTC) designed to cover all the UV bright sources of the galaxy and the massive stellar content of its H~{\sc ii} shells. Together with the sources from our first observing run \citep{Camacho2016}, the resulting sample will yield first-order information on the evolution of massive stars with sub-SMC metallicity and provide an extensive collection of spectra to characterise the physical properties of massive star populations at these low-$Z$ regimes.

The paper is structured as follows.
The spectroscopic observations and adopted target selection criteria are presented in Sect. \ref{sec:observations}. In Sect. \ref{sec:reduction}, we describe the data reduction process. The spectral classification, membership assessment and radial velocity measurement are described in Sect. \ref{sec:analysis}, along with an evaluation of the $Q$ pseudo-colour as a photometric method to select OB candidates. In Sect. \ref{sec:discussion}, we offer an overview of the entire sample and discuss the colour-magnitude diagram of Sextans~A, exploring the implications for its recent star formation history and stellar evolution at very low metallicity regimes. Finally, a summary and general conclusions are provided in Sect. \ref{sec:conclusions}.


\section{Description of observations}
\label{sec:observations}

Observations consisted of two mask multi-object spectroscopy (MOS) observing runs, GTC3-14AGOS (MOS1, PI A. Herrero) and GTC118-17B (MOS2, PI M. Garcia), and two long-slit (LSS) observations, GTC43-19B (LSS2, PI M. Garcia) and GTC58-20B (LSS3, PI M. Lorenzo). All of them were obtained with the Optical System for Imaging and low-Intermediate-Resolution Integrated Spectroscopy (OSIRIS) installed at the 10.4-m Gran Telescopio Canarias (GTC). Our data were complemented with long-slit spectroscopy published in \citet{Camacho2016}, denoted as LSS1. The observing logs are provided in Appendix~\ref{sec:appx_logs}. 

The MOS1 programme consisted of 13 hours of dark and grey sky and seeing lower than 1.2''. It was broken into one-hour-long observing blocks distributed in 2014 and 2015, and taken in service mode (see Table~\ref{table:mos1_log}). 
In this data set, targets were selected from \citet{Massey2007}'s photometric catalogue, with \mbox{19.0 $\leq$ $V$ $\leq$ 21.0} and a reddening-free pseudo-colour \mbox{$Q = (U-B)-0.72(B-V)\leq -0.8$} (see Section \ref{sec:SuccessRate}), following the criteria described in \citet{GarciaHerrero2013} and \citet{Camacho2016}. 
The observations of MOS2 consisted of 7 hours divided into two dark nights carried out in visitor mode (see Table~\ref{table:mos2_log}). The primary targets of this proposal were UV sources from \citet{Bianchi2012} with \mbox{$m_\mathrm{F170W}$ $\leq$ 17.6} (HST-WFPC2's filter F170W, \mbox{$\lambda_\mathrm{c}$ = 1730 \AA{}}) and without previous optical coverage. 
The long-slit LSS2 observations were designed to collect all the UV bright sources not covered by the MOS2 run. A total of 4 hours in 2020 \textit{via} service mode was devoted to the run.
Lastly, the LSS3 programme was granted 42 hours of dark sky in visitor mode to mine the population of massive stars in the giant H~{\sc ii} shells of the southeast of the galaxy. However, due to bad weather conditions, only 26.2 hours were observed
over the first semester of 2021.
Targets with 18.0 $\leq$ $V$ $\leq$ 21.4 and $Q \leq -0.6$ were included.

Both the MOS and the LSS2 runs used the R2000B VPH filter, 2$\times$2 binning and 1.2'' slits, which granted a resolving power of \mbox{$R$ = $\lambda / \Delta \lambda \sim$ 1000} in the \mbox{4000 - 5500 \AA{}} range. For MOS data sets, the exact wavelength coverage depends on the position of the slit in the mask. Some slits (5, 8 and 13 from MOS1, and 2, 20 and 24 from MOS2) were tilted (80\textdegree, 49\textdegree, -25\textdegree, 5\textdegree, 7.1\textdegree{} and 8\textdegree{} from the North-South axis, respectively) in order to accommodate a larger amount of stars in the sample. This inclination decreased the spatial and spectral resolution of their corresponding spectra. 
Observing run LSS3 was carried out in remote-visitor mode, and different slit widths could be used to match the seeing conditions, ranging from 0.8'' to 1.5''. This resulted in different spectral resolutions within the sample. Both the tilt and the width of the slits are noted for the corresponding stars in Tables \ref{tab:cat_OB_highQ}-\ref{tab:cat_late}.


\section{Data reduction}
\label{sec:reduction}

The MOS and the LSS data sets were reduced using different procedures. For the long-slit observations, we followed the routine described in \citet{GarciaHerrero2013}.
The reduction of the two MOS data sets was carried out using the \textsc{gtcmos} package, an \textsc{iraf}\footnote{\textsc{iraf} is distributed by the National Optical Astronomy Observatory, operated by the Association of Universities for Research in Astronomy (AURA) under agreement with the National Science Foundation.} pipeline developed by Dr. Divakara Mayya to reduce OSIRIS-MOS spectra \citep{divakara}. The pipeline delivers a 2D wavelength calibrated image where we applied \textsc{iraf}'s \textit{apall} task to extract the spectra and perform background subtraction. Finally, we used our own IDL routines to perform heliocentric correction, align the spectra in velocity space, coadd the selected observing blocks, and generate the final spectrum. The spectra were then normalized with \textsc{iraf}'s \textit{continuum} task. Further details on relevant aspects of the reduction steps are given below.

In order to maximise the final quality of the spectra obtained with MOS observations and to define a reduction protocol for future MOS runs, we tested three different approaches in the MOS1 run: (i) reducing the data sets obtained in each observing block (OB) individually; (ii) creating super-observing blocks (sOB), i.e. combining exposures that were carried out the same night and shared the calibration images (sOB03 = OB03 + OB04 and sOB13 = OB12 + OB13 + OB14); and (iii) coadding all raw images prior to reduction. The latter idea was dismissed due to the spatial shift between the location of the stars in the data sets of OBs taken in different years, hence coadding all data would result in severely decreased spatial resolution. A similar shift between OBs can be detected within the same night if the mask is not frequently re-aligned. In Table~\ref{tab:OBosOB}, we list the OBs and sOBs used to generate the final spectrum of each star observed in MOS1. We find that both procedures return very similar results, although reducing the spectra individually prior to coaddition provided slightly better results for most of the stars. For the MOS2 data set, we coadded all exposures of Night-1 and Night-2 separately.

We did not apply Flat Field correction to MOS runs since the corresponding \textsc{gtcmos} routine would only offer a solution in the common wavelength intervals covered by all the slits of a given MOS data set. Hence its implementation would lead to losing the extremes of the spectral range. Our tests showed little signal-to-noise ratio (S/N) improvement if the Flat Field correction was applied.

For the reduction of all the observing runs, sky subtraction was carried out at the step of \textit{apall} spectral extraction. 
In this step, we tried to minimize the nebular contribution in the extracted stellar spectrum to avoid compromising the spectral classification. To this end,  we estimated the background level from regions as close as possible to the source.
This method was preferred since it preserves spatial information of the nebular emission at the vicinity of the stars. Yet, nebular subtraction is not always complete as evinced by the presence of \hb~and [O~{\sc iii}] lines in emission in the final spectra of the sample stars (Figs. \ref{fig:OB_highQ_0}-\ref{fig:late_1}). In other cases, the nebular contribution is overcorrected, and the spectra show
these lines in absorption, or in some rare occasions, \hb~in strong absorption and [O~{\sc iii}] in emission.
We note that it is not possible to perform a complete correction of the nebular emission since it would require to model the variation of the nebular emissivity along the slit in each case.
Moreover, the spatial resolution of our observations is likely insufficient to adequately sample the ionization structure of the H~{\sc ii} shells, which translates into an incomplete nebular subtraction. During the reduction of the long-slit runs, we tried to improve the sky subtraction by selecting the sky boxes at different distances from the targets at \textit{apall}, but the results were again unsatisfactory.

In MOS observations, we detected dark columns on the right part of some slits after applying the \textsc{gtcmos} \textit{omreduce} task. \textit{Omreduce} trims the portions of the image corresponding to individual slits, applies the wavelength calibration and finally returns a reconstructed MOS image with the slits aligned in wavelength.
The dark columns arise when the slits are too close to each other and the extraction regions assigned by the pipeline overlap. 
The overlap can sometimes create an artificial emission that may be mistaken for a stellar source.
At the step of \textit{apall} extraction, we selected the stellar apertures and defined their sky boxes avoiding these artefacts. A future recommendation to prevent this issue is designing the MOS masks with enough space between slits. We recommend leaving a projected distance larger than 5'' in the horizontal direction if \textsc{gtcmos} will be used for reduction. We also suggest using slits with a length seven times greater than the expected seeing in the observations to allow a proper selection of the sky boxes during sky subtraction.

Once the data of all individual observing blocks or super-observing blocks had been reduced and extracted with \textit{apall}, all the spectra extracted for a given star were co-added with a semi-automatic IDL script. This script sets the spectra to the heliocentric standard of rest and combines the exposures weighing by their standard deviation. Observing blocks with too poor spectral quality were discarded.

Some stars were observed in more than one programme and even more than once within an LSS programme. To improve the S/N of the data, we combined the observations after matching their spectral resolution and correcting for radial velocity variations when necessary. Those stars suffering from radial velocity shifts between exposures are marked in Tables \ref{tab:cat_OB_highQ}-\ref{tab:cat_late} as "SB1", 
as they suggest a possible binary nature.

The final reduced spectra are shown in Figs.~\ref{fig:OB_highQ_0}-\ref{fig:late_1} in Appendix \ref{sec:appx_Spectra}, arranged by spectral type. Identification numbers, coordinates and photometric measurements are collected in Tables~\ref{tab:cat_OB_highQ}-\ref{tab:cat_late}, along with the spectral types and radial velocities derived in this work.

\section{Analysis}
\label{sec:analysis}


\subsection{Spectral classification}
\label{sec:SpTclas}

The spectral classification of the OB-type stars was carried out following the criteria outlined in \citet{Castro2008} for metal-poor regimes ($Z$ $\sim$ 0.5 \Zsun), adapted to the metallicity of Sextans~A by \citet{Camacho2016_t}.
We also used the guidelines given by \citet{Sota2011} for late O-types and the ones from \citet{Walborn2002} for stars earlier than O5. The main diagnostics for O stars are based on the relative intensity of \HeI~and \HeII~lines. The luminosity class of O-type stars was assigned based on the width of the Balmer lines and on whether the \mbox{\HeII~4686} line was in absorption or emission. We note the latter diagnostic may be affected by the weak winds of the targets. 
The spectral types of B stars are indicated by the presence and strength of Si transitions in different ionization stages, \mbox{\MgII~4481} and \HeI~lines. Their luminosity class was assigned from the width of the Balmer lines.

We also used the absolute magnitude $M_{\mathrm{V}}$ as an additional indicator of luminosity class for some sources. We estimated $M_{\mathrm{V}}$ using the distance modulus of the galaxy \citep[\mbox{$\mu_0$ = 25.63}, ][]{Tammann2011} and the intrinsic colour $(B-V)_{\rm{0}}$ calculated from the pseudo-colour $Q$ \citep[$(B-V)_0 = -0.005+0.317 Q$, ][]{Massey2000}.

Later spectral types (A-types and later) were classified following \citet{GrayCorbally2009}. The spectral classification of these late stars may be affected by the low metallicity of the environment since metallic lines and molecular bands are the main diagnostics. 

Spectral types are compiled in Tables~\ref{tab:cat_OB_highQ}-\ref{tab:cat_late}. In addition, we provide brief notes on the spectral classification of targets in Appendix~\ref{sec:appx_IndividualNotes}. 

\clearpage
\onecolumn
\begin{landscape}
\small
\setlength\LTleft{-15pt}
\setlength\LTright{-20pt}
\setlength\LTcapwidth{\linewidth}
\captionsetup[longtable]{labelfont=bf, font = footnotesize, labelsep=period}
\begin{longtable}{lcccccccccl}
\caption{Blue type stars (from O to F types) confirmed by spectroscopy in this work. Identification codes (ID), spectral types (SpT), radial velocities (\vrad) and signal-to-noise ratio per element of resolution (S/N, calculated in a region free of spectral lines around 4800 \AA{}) are provided. Stars whose radial velocity was determined manually are marked with $^{(*)}$. Identification numbers, coordinates and photometry by \citet{Massey2007} are also included when available, otherwise we provide the ones from \citet{Bianchi2012}\textquotesingle s or \citet{Holtzman2006}\textquotesingle s catalogues, and mark the sources with $^{(\dagger)}$ or $^{(\ddagger)}$, respectively. We include additional information in column \textit{Notes}.}
\label{tab:cat_OB_highQ}\\
\toprule
\toprule
                ID &             Alt. ID & RA (J2000.0) & DEC (J2000.0) &             SpT &            $V$ &  $B-V$ &    $Q$ &  \vrad~(\kms) &  S/N &                                                             Notes \\
\midrule
\endfirsthead
\caption[]{continued.} \\
\toprule
\toprule
                ID &             Alt. ID & RA (J2000.0) & DEC (J2000.0) &             SpT &            $V$ &  $B-V$ &    $Q$ &  \vrad~(\kms) &  S/N &                                                             Notes \\
\midrule
\endhead
\midrule
\multicolumn{11}{r}{{\textit{continued on next page}}} \\

\endfoot

\bottomrule
\endlastfoot
s001*$^{\ddagger}$ &                   - &  10:11:05.29 &   -04:42:41.7 &         O3.5 V  & F555W = 21.851 &      - &      - & 460 $\pm$ 100 &   55 &                                                  1.0'' slit. SB2? \\
              s002 & J101105.28-044238.3 &  10:11:05.28 &   -04:42:38.3 &          O4 Vz  &         20.369 & -0.279 & -0.937 &  275 $\pm$ 30 &  111 &                                                       1.0'' slit. \\
             s003* & J101058.59-044328.9 &  10:10:58.59 &   -04:43:28.9 &       O3-O5 Vz  &         20.804 & -0.095 & -1.120 &  425 $\pm$ 60 &   83 &                                        \citetalias{Garcia2019}-s2 \\
              s004 & J101057.89-044310.2 &  10:10:57.89 &   -04:43:10.2 &         O5 III  &         20.917 & -0.277 & -1.006 &  310 $\pm$ 30 &   91 &                                  \citetalias{Garcia2019}-s4. Bin? \\
              s005 & J101053.95-044057.4 &  10:10:53.95 &   -04:40:57.4 &     O5 V((fc))  &         21.751 & -0.364 & -0.800 &  375 $\pm$ 40 &   30 &                                                       0.8'' slit. \\
             s006* & J101104.74-044206.1 &  10:11:04.74 &   -04:42:06.1 &           O6 I  &         20.721 & -0.068 & -0.416 &  500 $\pm$ 50 &   29 &                                                                   \\
              s007 & J101106.48-044237.1 &  10:11:06.48 &   -04:42:37.1 &           O6 V  &         20.515 & -0.280 & -0.965 &  395 $\pm$ 35 &  105 &                                                              SB2? \\
             s008* & J101106.65-044213.0 &  10:11:06.65 &   -04:42:13.0 &         O6.5 V  &         21.071 & -0.199 & -0.985 &  300 $\pm$ 50 &   71 &                                                1.2'' slit. Blend. \\
              s009 & J101108.48-044149.2 &  10:11:08.48 &   -04:41:49.2 &           O7 V  &         20.659 & -0.285 & -0.948 &  320 $\pm$ 35 &   66 &                                                                   \\
             s010* & J101059.19-044308.2 &  10:10:59.19 &   -04:43:08.2 &           O7 V  &         21.003 & -0.134 & -0.549 &  250 $\pm$ 50 &   30 &                                                       1.5'' slit. \\
              s011 & J101103.70-044234.2 &  10:11:03.70 &   -04:42:34.2 & O7.5 III + B0 I &         20.191 & -0.220 & -1.019 &  315 $\pm$ 30 &   80 &                                                1.0'' slit. Blend. \\
              s012 & J101054.23-044115.3 &  10:10:54.23 &   -04:41:15.3 &       O7.5 III  &         20.115 & -0.181 & -0.835 & 290 $\pm$ 155 &   52 &                                                     SB2 or blend? \\
             s013* & J101100.09-044336.0 &  10:11:00.09 &   -04:43:36.0 &       O7.5 III  &         21.532 & -0.295 & -0.914 &  370 $\pm$ 50 &   43 &                                                                   \\
              s014 & J101053.81-044113.0 &  10:10:53.81 &   -04:41:13.0 &  O7.5 III((f))  &         20.688 & -0.265 & -1.015 &  330 $\pm$ 35 &   49 &                         \citetalias{Camacho2016}-OB326. Off-slit. \\
             s015* & J101105.47-044236.8 &  10:11:05.47 &   -04:42:36.8 &     O7.5 V + B? &         22.259 & -0.410 & -1.010 &  225 $\pm$ 50 &   35 &                                                  1.0'' slit. SB2? \\
              s016 & J101056.28-044253.0 &  10:10:56.28 &   -04:42:53.0 &          O8 II  &         20.639 & -0.192 & -1.065 &  285 $\pm$ 95 &   72 &                                                                   \\
             s017* & J101106.50-044139.0 &  10:11:06.50 &   -04:41:39.0 &          O8 II  &         21.399 &  0.198 & -1.281 &  330 $\pm$ 50 &   18 &                                                         Off-slit. \\
             s018* & J101053.61-044246.3 &  10:10:53.61 &   -04:42:46.3 &         O8 III  &         21.092 & -0.321 & -0.894 &  210 $\pm$ 50 &   41 &                                                                   \\
  s019$^{\dagger}$ &                3485 &  10:11:05.84 &   -04:42:12.3 &         O8 III  &         21.930 & -0.250 & -0.850 &  270 $\pm$ 35 &   32 &                                                0.8'' slit. Blend. \\
              s020 & J101107.33-044205.6 &  10:11:07.33 &   -04:42:05.6 &           O8 V  &         20.544 & -0.196 & -0.968 &  315 $\pm$ 35 &  125 &                                         1.5'' slit. SB2 or blend? \\
              s021 & J101104.79-044220.9 &  10:11:04.79 &   -04:42:20.9 &           O8 V  &         20.598 & -0.468 & -0.616 &  335 $\pm$ 35 &   95 &                          0.63'' slit. Tilted 3\textdegree. Blend. \\
              s022 & J101105.38-044240.1 &  10:11:05.38 &   -04:42:40.1 &           O8 V  &         19.459 & -0.259 & -0.950 &  370 $\pm$ 30 &   95 &                            \citetalias{Camacho2016}-OB521. Blend. \\
              s023 & J101105.21-044240.5 &  10:11:05.21 &   -04:42:40.5 &           O8 V  &         20.288 & -0.299 & -0.927 &  235 $\pm$ 35 &   82 &                                                       1.0'' slit. \\
              s024 & J101105.90-044210.3 &  10:11:05.90 &   -04:42:10.3 &           O8 V  &         20.810 & -0.162 & -1.044 &  280 $\pm$ 35 &   77 &                                                                   \\
              s025 & J101110.14-044129.7 &  10:11:10.14 &   -04:41:29.7 &           O8 V  &         21.060 & -0.265 & -0.976 &  350 $\pm$ 45 &   52 &                                                                   \\
  s026$^{\dagger}$ &                3935 &  10:11:07.43 &   -04:42:04.5 &           O8 V  &         22.500 & -0.060 & -1.007 & 325 $\pm$ 110 &   51 &                                                0.8'' slit. Blend. \\
  s027$^{\dagger}$ &                2398 &  10:11:03.39 &   -04:42:37.1 &           O8 V  &         21.960 & -0.220 & -0.852 &  280 $\pm$ 35 &   20 &                                                       1.5'' slit. \\
              s028 & J101107.34-044231.7 &  10:11:07.34 &   -04:42:31.7 &       O8.5 III  &         20.568 & -0.228 & -1.033 &  300 $\pm$ 35 &   83 &                                                                   \\
              s029 & J101058.19-044318.4 &  10:10:58.19 &   -04:43:18.4 &       O8.5 III  &         20.803 & -0.247 & -1.005 &  285 $\pm$ 35 &   74 &                                  \citetalias{Garcia2019}-s3. SB1. \\
              s030 & J101109.39-044055.5 &  10:11:09.39 &   -04:40:55.5 &       O8.5 III  &         21.008 & -0.322 & -0.848 &  380 $\pm$ 35 &   56 &                                            Tilted 7.1\textdegree. \\
             s031* & J101057.33-044016.6 &  10:10:57.33 &   -04:40:16.6 &         O8.5 V  &         21.229 & -0.181 & -1.069 &  475 $\pm$ 50 &  115 &                                              Tilted 8\textdegree. \\
              s032 & J101109.35-044057.8 &  10:11:09.35 &   -04:40:57.8 &         O8.5 V  &         21.216 & -0.326 & -0.776 &  365 $\pm$ 30 &   67 &                                     Tilted 7.1\textdegree. Blend. \\
              s033 & J101105.64-044215.6 &  10:11:05.64 &   -04:42:15.6 &         O8.5 V  &         21.654 & -0.271 & -0.960 &  295 $\pm$ 75 &   40 &                                                       0.8'' slit. \\
              s034 & J101058.05-044021.8 &  10:10:58.05 &   -04:40:21.8 &         O8.5 V  &         20.923 & -0.168 & -0.548 &  250 $\pm$ 30 &   26 &                                                       1.5'' slit. \\
             s035* & J101106.50-044212.4 &  10:11:06.50 &   -04:42:12.4 &         O8.5 V  &         20.683 & -0.089 & -0.498 & 170 $\pm$ 100 &   24 &                                                            Blend. \\
              s036 & J101105.07-044214.6 &  10:11:05.07 &   -04:42:14.6 &           O9 I  &         19.738 & -0.237 & -0.931 &  360 $\pm$ 30 &  200 &                                                                   \\
              s037 & J101104.78-044224.1 &  10:11:04.78 &   -04:42:24.1 &           O9 I  &         20.681 & -0.234 & -1.001 &  345 $\pm$ 30 &  110 &                                   \citetalias{Camacho2016}-OB623. \\
              s038 & J101106.05-044211.4 &  10:11:06.05 &   -04:42:11.4 &      O9 I((f))  &         19.492 & -0.231 & -1.013 &  450 $\pm$ 35 &   91 &                                   \citetalias{Camacho2016}-OB523. \\
              s039 & J101059.96-044332.9 &  10:10:59.96 &   -04:43:32.9 &         O9 III  &         21.084 & -0.171 & -0.502 &  340 $\pm$ 35 &   49 &                                                       1.0'' slit. \\
             s040* & J101106.56-044217.1 &  10:11:06.56 &   -04:42:17.1 & O9 III + mid-B? &         21.285 & -0.311 & -0.900 &  250 $\pm$ 30 &   29 &                                                              SB2? \\
              s041 & J101058.53-044414.4 &  10:10:58.53 &   -04:44:14.4 &        O9.5 II  &         20.877 & -0.029 & -1.071 &  290 $\pm$ 30 &   73 &                                        \citetalias{Garcia2019}-s1 \\
              s042 & J101105.44-044221.4 &  10:11:05.44 &   -04:42:21.4 &       O9.5 III  &         20.642 & -0.226 & -0.939 &  275 $\pm$ 30 &  108 &                                                            Blend. \\
              s043 & J101105.03-044224.8 &  10:11:05.03 &   -04:42:24.8 &         O9.5 V  &         20.863 &  0.250 & -0.607 &  305 $\pm$ 30 &   92 &                                                1.5'' slit. Blend. \\
              s044 & J101110.45-044132.9 &  10:11:10.45 &   -04:41:32.9 &         O9.5 V  &         20.800 & -0.255 & -0.966 &  270 $\pm$ 95 &   60 &                                                              SB2? \\
              s045 & J101105.93-044208.0 &  10:11:05.93 &   -04:42:08.0 &         O9.5 V  &         20.511 & -0.169 & -0.430 &  250 $\pm$ 35 &   60 &                                                       0.8'' slit. \\
             s046* & J101104.87-044216.0 &  10:11:04.87 &   -04:42:16.0 &         O9.5 V  &         20.748 & -0.214 & -0.872 &  325 $\pm$ 30 &   45 &                                                0.8'' slit. Blend. \\
              s047 & J101104.58-044213.0 &  10:11:04.58 &   -04:42:13.0 &         O9.5 V  &         20.668 &  0.267 & -0.274 & 335 $\pm$ 100 &   31 &                                   \citetalias{Camacho2016}-OB625. \\
 s048*$^{\dagger}$ &                3691 &  10:11:06.34 &   -04:42:28.2 &        O9.5 Vn  &         22.180 & -0.240 & -0.797 &  200 $\pm$ 60 &   20 &                                                       1.0'' slit. \\
              s049 & J101105.30-044210.1 &  10:11:05.30 &   -04:42:10.1 &         O9.7 I  &         19.976 & -0.268 & -0.929 &  350 $\pm$ 30 &   91 &                                                       SB1? Blend. \\
              s050 & J101100.66-044044.3 &  10:11:00.66 &   -04:40:44.3 &         O9.7 I  &         19.609 & -0.248 & -0.960 &  295 $\pm$ 30 &   90 &                      \citetalias{Camacho2016}-OB321. 0.63'' slit. \\
              s051 & J101058.27-044230.6 &  10:10:58.27 &   -04:42:30.6 &         O9.7 I  &         20.774 & -0.277 & -0.875 &  335 $\pm$ 30 &   70 &                                                                   \\
              s052 & J101106.46-044207.3 &  10:11:06.46 &   -04:42:07.3 &         O9.7 I  &         20.320 & -0.166 & -0.690 &  345 $\pm$ 30 &   62 &                                                              SB2? \\
             s053* & J101106.13-044208.3 &  10:11:06.13 &   -04:42:08.3 &         O9.7 I  &         21.255 & -0.269 & -0.904 &  325 $\pm$ 50 &   60 &                                                                   \\
              s054 & J101104.53-044219.3 &  10:11:04.53 &   -04:42:19.3 &       O9.7 III  &         20.511 & -0.202 & -0.951 &  330 $\pm$ 30 &  125 &                                                                   \\
              s055 & J101053.90-044111.0 &  10:10:53.90 &   -04:41:11.0 &       O9.7 III  &         20.323 & -0.103 & -0.920 &  300 $\pm$ 30 &  103 &                                                            Blend. \\
              s056 & J101100.94-044003.2 &  10:11:00.94 &   -04:40:03.2 &         O9.7 V  &         20.914 & -0.147 & -0.922 &  335 $\pm$ 30 &   98 &                                                               Oe? \\
              s057 & J101059.58-044135.4 &  10:10:59.58 &   -04:41:35.4 &         O9.7 V  &         20.697 & -0.110 & -0.393 &  355 $\pm$ 35 &   59 &                                                                   \\
             s058* & J101104.05-044226.1 &  10:11:04.05 &   -04:42:26.1 &     O9.7 V + B? &         20.957 & -0.204 & -0.932 &  350 $\pm$ 50 &   40 &                                                              SB2? \\
              s059 &                   - &  10:10:57.09 &   -04:44:46.3 &        O9.7 Vn  &              - &      - &      - &  355 $\pm$ 30 &   36 &                                                                   \\
             s060* & J101107.27-044032.4 &  10:11:07.27 &   -04:40:32.4 &         O9.7 V  &         21.222 & -0.065 & -1.121 & 400 $\pm$ 100 &   26 &                                         Blend or bin? Broad \HeI. \\
              s061 & J101057.32-044055.4 &  10:10:57.32 &   -04:40:55.4 &           B0 I  &         19.915 & -0.247 & -0.899 &  290 $\pm$ 30 &  159 &                                                                   \\
              s062 & J101103.80-044212.8 &  10:11:03.80 &   -04:42:12.8 &           B0 I  &         20.358 & -0.269 & -0.872 &  315 $\pm$ 30 &   76 &                                                                   \\
              s063 & J101107.43-044025.4 &  10:11:07.43 &   -04:40:25.4 &         B0 III  &         20.727 & -0.295 & -0.885 &  305 $\pm$ 35 &  133 &                                                                   \\
             s064* & J101108.49-043905.9 &  10:11:08.49 &   -04:39:05.9 &         B0 III  &         21.223 & -0.268 & -0.879 &  325 $\pm$ 50 &   44 &                                                              SB2? \\
             s065* & J101107.31-044205.3 &  10:11:07.31 &   -04:42:05.3 &           B0 V  &         20.889 & -0.601 & -0.595 &  100 $\pm$ 50 &   63 &                                                          SB2? Be? \\
              s066 & J101058.01-044027.8 &  10:10:58.01 &   -04:40:27.8 &         B0.5 I  &         20.041 & -0.225 & -0.837 &  310 $\pm$ 40 &  124 &                                                                   \\
              s067 & J101059.29-044343.8 &  10:10:59.29 &   -04:43:43.8 &         B0.5 I  &         20.342 & -0.137 & -0.464 &  360 $\pm$ 40 &   82 &                                                                   \\
             s068* & J101106.48-044210.5 &  10:11:06.48 &   -04:42:10.5 &         B0.5 I  &         20.521 &  0.272 & -0.344 &    0 $\pm$ 50 &   69 &                  = \Remark{LSS3.OB18203}. 0.8'' slit. Bin. Blend. \\
              s069 & J101107.59-044159.9 &  10:11:07.59 &   -04:41:59.9 &    B0.5 II + O? &         19.977 & -0.214 & -0.941 &  390 $\pm$ 40 &  109 &                                                              SB2? \\
              s070 & J101106.03-044209.1 &  10:11:06.03 &   -04:42:09.1 &       B0.5 III  &         20.834 & -0.272 & -0.907 &  270 $\pm$ 45 &   82 &                                    \citetalias{Camacho2016}-OB524 \\
              s071 & J101105.69-044213.6 &  10:11:05.69 &   -04:42:13.6 &           B1 I  &         19.699 & -0.236 & -0.903 &  265 $\pm$ 35 &  256 &                                        Tilted 49\textdegree. SB1? \\
              s072 & J101059.21-043948.1 &  10:10:59.21 &   -04:39:48.1 &           B1 I  &         19.329 & -0.192 & -0.880 &  340 $\pm$ 30 &  211 &                              \citetalias{Camacho2016}-OB421. SB1? \\
              s073 & J101103.98-044232.5 &  10:11:03.98 &   -04:42:32.5 &           B1 I  &         20.069 & -0.199 & -0.790 &  285 $\pm$ 30 &  101 &                                                       1.5'' slit. \\
              s074 & J101054.62-044103.0 &  10:10:54.62 &   -04:41:03.0 &           B1 I  &         19.739 & -0.218 & -0.942 &  330 $\pm$ 30 &   42 &                                   \citetalias{Camacho2016}-OB422. \\
             s075* & J101104.02-044222.9 &  10:11:04.02 &   -04:42:22.9 &       B1 I + O? &         21.220 & -0.023 & -0.107 &  200 $\pm$ 50 &   34 &                                                  0.8'' slit. SB2? \\
  s076$^{\dagger}$ &                2990 &  10:11:04.66 &   -04:42:23.6 &           B1 I  &         21.940 & -0.070 & -0.920 &  335 $\pm$ 40 &   29 &                                                                   \\
              s077 & J101054.08-044111.5 &  10:10:54.08 &   -04:41:11.5 &         B1 III  &         19.487 & -0.217 & -0.875 &  340 $\pm$ 30 &   92 &                       \citetalias{Camacho2016}-OB323. SB1. Blend. \\
              s078 & J101102.18-044315.4 &  10:11:02.18 &   -04:43:15.4 &         B1 III  &         21.290 & -0.137 & -0.915 &  290 $\pm$ 30 &   54 &                                                                   \\
             s079* & J101105.82-044213.0 &  10:11:05.82 &   -04:42:13.0 &         B1 III  &         21.774 & -0.163 & -0.984 &  325 $\pm$ 50 &   50 &                                                            Blend. \\
             s080* & J101054.97-044116.2 &  10:10:54.97 &   -04:41:16.2 &         B1 III  &         21.471 & -0.292 & -0.983 &  250 $\pm$ 50 &   34 &                                                                   \\
  s081$^{\dagger}$ &                3646 &  10:11:06.21 &   -04:42:25.7 &           B1 V  &         23.250 &  0.000 & -1.260 &  320 $\pm$ 40 &   44 &                                                            Blend. \\
              s082 & J101105.65-044038.9 &  10:11:05.65 &   -04:40:38.9 &         B1.5 I  &         19.896 & -0.186 & -0.822 &  400 $\pm$ 30 &  160 &                                                                   \\
              s083 & J101055.35-044106.1 &  10:10:55.35 &   -04:41:06.1 &           B2 I  &         19.067 & -0.171 & -1.007 &  280 $\pm$ 30 &  183 &                              \citetalias{Camacho2016}-OB222. SB1. \\
              s084 & J101104.33-044220.8 &  10:11:04.33 &   -04:42:20.8 &           B2 I  &         20.552 &  0.061 & -1.017 &  265 $\pm$ 35 &   72 &                                                       0.8'' slit. \\
              s085 & J101104.69-044228.8 &  10:11:04.69 &   -04:42:28.8 &           B2 I  &         20.850 &  0.254 &      - &  320 $\pm$ 30 &   62 &                                                  1.5'' slit. SB1? \\
             s086* & J101104.97-044218.9 &  10:11:04.97 &   -04:42:18.9 &         B2 III  &         20.509 &  0.107 &  0.014 &  100 $\pm$ 50 &  126 &                                                              SB1? \\
              s087 & J101053.94-044110.1 &  10:10:53.94 &   -04:41:10.1 &     B2 III + O? &         20.526 & -0.230 & -0.921 &  255 $\pm$ 35 &   79 &                                                            Blend. \\
             s088* & J101110.46-044131.5 &  10:11:10.46 &   -04:41:31.5 &     B2 III + O? &         21.265 & -0.265 & -0.435 &  325 $\pm$ 50 &   29 &                                                  0.8'' slit. SB2? \\
              s089 & J101102.38-044014.6 &  10:11:02.38 &   -04:40:14.6 &         B2.5 I  &         19.581 & -0.099 & -0.683 &  340 $\pm$ 30 &  150 &     \citetalias{Camacho2016}-OB622. Tilted -25\textdegree. Bin?.  \\
              s090 & J101106.11-044229.8 &  10:11:06.11 &   -04:42:29.8 &         B2.5 I  &         20.358 & -0.133 & -0.507 &  295 $\pm$ 30 &   87 &                                                1.0'' slit. Blend. \\
              s091 & J101058.27-044257.8 &  10:10:58.27 &   -04:42:57.8 &         B2.5 I  &         19.176 & -0.074 & -0.706 &  330 $\pm$ 30 &   83 &                                   \citetalias{Camacho2016}-OB221. \\
              s092 & J101105.50-044128.8 &  10:11:05.50 &   -04:41:28.8 &         B2.5 I  &         20.155 & -0.141 & -0.671 &  325 $\pm$ 30 &   77 &                                                                   \\
             s093* & J101106.48-044210.5 &  10:11:06.48 &   -04:42:10.5 &         B2.5 I  &         20.521 &  0.272 & -0.344 &  450 $\pm$ 50 &   51 &                  = \Remark{LSS3.OB17203}. 0.8'' slit. Bin. Blend. \\
             s094* & J101058.58-044423.5 &  10:10:58.58 &   -04:44:23.5 &     B2.5 I + O? &         21.621 &  0.205 & -0.966 &  335 $\pm$ 50 &   37 &                               Off-slit. Tilted 5\textdegree. SB2? \\
              s095 & J101105.96-044221.4 &  10:11:05.96 &   -04:42:21.4 &       B2.5 III  &         21.672 & -0.188 & -0.730 &  320 $\pm$ 30 &   47 &                                                                   \\
             s096* & J101056.64-044102.1 &  10:10:56.64 &   -04:41:02.1 &       B2.5 III  &         21.062 &  0.263 & -0.056 &  200 $\pm$ 50 &   34 &                                                       0.8'' slit. \\
              s097 & J101107.62-044203.6 &  10:11:07.62 &   -04:42:03.6 &           B3 I  &         20.085 & -0.175 & -0.662 &  300 $\pm$ 30 &  158 &                                                       O.8'' slit. \\
              s098 & J101106.12-044219.9 &  10:11:06.12 &   -04:42:19.9 &           B3 I  &         19.782 & -0.163 & -0.789 &  335 $\pm$ 30 &  136 &                                                                   \\
              s099 & J101102.75-044047.6 &  10:11:02.75 &   -04:40:47.6 &           B3 I  &         20.201 & -0.141 & -0.657 &  275 $\pm$ 40 &  121 &                                                                   \\
              s100 & J101051.62-044326.4 &  10:10:51.62 &   -04:43:26.4 &         B3 III  &         20.997 & -0.263 & -0.966 &  320 $\pm$ 35 &   62 &                                                                   \\
              s101 & J101054.45-044114.1 &  10:10:54.45 &   -04:41:14.1 &         B3 III  &         21.254 & -0.328 & -0.897 &  255 $\pm$ 35 &   43 &                                                                   \\
              s102 & J101104.59-044222.2 &  10:11:04.59 &   -04:42:22.2 &           B7 I  &         20.189 & -0.074 & -0.463 &  280 $\pm$ 35 &   94 &                                                                   \\
             s103* & J101103.93-044214.0 &  10:11:03.93 &   -04:42:14.0 &           B7 I  &         20.437 & -0.069 & -0.386 &  400 $\pm$ 40 &   93 &                                                            Blend. \\
              s104 & J101105.41-044210.9 &  10:11:05.41 &   -04:42:10.9 &       B7 I + O? &         21.153 & -0.206 & -0.751 &  250 $\pm$ 40 &   90 &                                                1.5'' slit. Blend? \\
              s105 & J101104.42-044232.4 &  10:11:04.42 &   -04:42:32.4 &           B7 I  &         19.227 & -0.075 & -0.554 &  270 $\pm$ 35 &   86 &                                                                   \\
              s106 & J101106.02-044214.2 &  10:11:06.02 &   -04:42:14.2 &          B7 II  &         18.771 &  0.054 & -0.277 &  250 $\pm$ 30 &  102 &                       \citetalias{Camacho2016}-OB121. SB1. Blend. \\
             s107* & J101059.81-044305.5 &  10:10:59.81 &   -04:43:05.5 &         B7 III  &         20.575 &  0.126 & -0.011 & 200 $\pm$ 100 &   77 &                                                       1.5'' slit. \\
              s108 & J101105.86-044232.0 &  10:11:05.86 &   -04:42:32.0 &     B7 III + B? &         21.560 & -0.119 & -0.464 &  320 $\pm$ 55 &   44 &                                                  1.0'' slit. SB2? \\
             s109* & J101107.14-044131.4 &  10:11:07.14 &   -04:41:31.4 &     B7 III + O? &         21.538 & -0.268 & -0.873 &  350 $\pm$ 50 &   28 &                                                              SB2? \\
              s110 & J101105.98-044217.9 &  10:11:05.98 &   -04:42:17.9 &           B8 I  &         20.008 & -0.048 & -0.358 &  260 $\pm$ 55 &  111 &                                                       1.5'' slit. \\
              s111 & J101059.13-044051.0 &  10:10:59.13 &   -04:40:51.0 &          B8 II  &         20.486 & -0.069 & -0.412 &  340 $\pm$ 35 &   54 &                         \citetalias{Camacho2016}-OB324. Off-slit. \\
             s112* & J101104.08-044224.2 &  10:11:04.08 &   -04:42:24.2 &         B8 III  &         21.104 &  0.038 & -0.144 &  270 $\pm$ 30 &   67 &                                                       0.8'' slit. \\
              s113 & J101059.15-044215.1 &  10:10:59.15 &   -04:42:15.1 &         B8 III  &         21.514 & -0.111 & -0.281 &  265 $\pm$ 35 &   38 &                                                       1.0'' slit. \\
              s114 & J101056.30-044105.7 &  10:10:56.30 &   -04:41:5.70 &         B8 III  &         21.478 & -0.228 & -0.719 &  340 $\pm$ 50 &   23 &                                                            Blend. \\
              s115 & J101105.17-044236.0 &  10:11:05.17 &   -04:42:36.0 &           B9 I  &         19.360 & -0.044 & -0.427 & 295 $\pm$ 250 &   95 &                                                       0.8'' slit. \\
              s116 & J101054.89-044112.1 &  10:10:54.89 &   -04:41:12.1 &           B9 I  &         18.480 & -0.010 & -0.529 &  375 $\pm$ 30 &   75 &                            \citetalias{Camacho2016}-OB122. Blend. \\
              s117 & J101058.86-044126.2 &  10:10:58.86 &   -04:41:26.2 &         B9 III  &         20.169 &  0.020 & -0.040 &  325 $\pm$ 45 &   46 &                                                       0.8'' slit. \\
 s118*$^{\dagger}$ &                3713 &  10:11:06.39 &   -04:42:16.4 &           B9 V  &         21.980 & -0.120 & -0.184 & 300 $\pm$ 100 &   35 &                                                                   \\
             s119* & J101106.38-044155.9 &  10:11:06.38 &   -04:41:55.9 &          A0 II  &         20.519 & -0.043 & -0.155 &  350 $\pm$ 40 &   41 &                                   \citetalias{Camacho2016}-OB525. \\
              s120 & J101107.24-044219.7 &  10:11:07.24 &   -04:42:19.7 &           A2 I  &         18.873 &  0.081 & -0.009 &  370 $\pm$ 30 &  114 &                                                                   \\
             s121* & J101054.71-044115.6 &  10:10:54.71 &   -04:41:15.6 &           A2 I  &         19.211 &  0.129 & -0.019 &  270 $\pm$ 30 &   83 &                                                                   \\
              s122 & J101057.50-044208.9 &  10:10:57.50 &   -04:42:08.9 &           A2 I  &         19.499 & -0.011 & -0.259 &  275 $\pm$ 40 &   60 &                                                       1.0'' slit. \\
              s123 & J101056.97-044206.6 &  10:10:56.97 &   -04:42:06.6 &         A2 III  &         20.449 &  0.172 & -0.017 &  320 $\pm$ 40 &   32 &                                                       1.0'' slit. \\
             s124* & J101104.97-044233.8 &  10:11:04.97 &   -04:42:33.8 &           F5 I  &         17.477 &  0.174 & -0.188 &  350 $\pm$ 40 &  132 & \citetalias{Camacho2016}-OB621. \citetalias{Kaufer2004}-SexA-513. \\
 s125*$^{\dagger}$ &   J10111837-0440480 &  10:11:18.37 &   -04:40:48.1 &           F8 V  &              - &      - &      - &    0 $\pm$ 30 &   74 &                                                       Foreground. \\
\end{longtable}
\end{landscape}
\begin{landscape}
\small
\setlength\LTleft{-15pt}
\setlength\LTright{-20pt}
\setlength\LTcapwidth{\linewidth}
\captionsetup[longtable]{labelfont=bf, font = footnotesize, labelsep=period}
\begin{longtable}{lcccccccccl}
\caption{Blue type stars for which a spectral subtype could not be assigned. The description of columns is the same as Table \ref{tab:cat_OB_highQ}.}
\label{tab:cat_OB_lowQ}\\
\toprule
\toprule
               ID &             Alt. ID & RA (J2000.0) & DEC (J2000.0) &         SpT &       $V$ &  $B-V$ &    $Q$ &  \vrad~(\kms) &  S/N &                Notes \\
\midrule
\endfirsthead
\caption[]{continued.} \\
\toprule
\toprule
               ID &             Alt. ID & RA (J2000.0) & DEC (J2000.0) &         SpT &       $V$ &  $B-V$ &    $Q$ &  \vrad~(\kms) &  S/N &                Notes \\
\midrule
\endhead
\midrule
\multicolumn{11}{r}{{\textit{continued on next page}}} \\

\endfoot

\bottomrule
\endlastfoot
s126*$^{\dagger}$ &                3545 &  10:11:06.00 &   -04:42:28.4 &        O V  & B = 23.04 &      - &      - &  230 $\pm$ 50 &   48 &               Blend. \\
            s127* & J101100.30-044330.7 &  10:11:00.30 &   -04:43:30.7 &   O V + gal &    21.398 & -0.071 & -0.263 & 250 $\pm$ 100 &   84 &                      \\
            s128* &                   - &  10:11:06.64 &   -04:41:33.5 &        O V  &         - &      - &      - & 235 $\pm$ 100 &    8 &                      \\
             s129 & J101104.28-044238.2 &  10:11:04.28 &   -04:42:38.2 &   O V + neb &    21.237 & -0.310 & -1.000 &  330 $\pm$ 30 &   15 &                      \\
             s130 & J101106.30-044140.8 &  10:11:06.30 &   -04:41:40.8 &         O   &    20.934 & -0.075 & -0.473 &  380 $\pm$ 45 &   15 &            Off-slit. \\
             s131 & J101106.32-043914.0 &  10:11:06.32 &   -04:39:14.0 &    O  + neb &    21.277 & -0.140 & -1.003 &  390 $\pm$ 35 &   31 &                      \\
            s132* & J101101.33-044150.9 &  10:11:01.33 &   -04:41:50.9 &   O V + neb &    21.347 & -0.093 & -1.017 &  490 $\pm$ 70 &   23 &            Off-slit. \\
             s133 & J101100.17-043939.2 &  10:11:00.17 &   -04:39:39.2 &    O  + neb &    21.219 & -0.098 & -1.038 &  225 $\pm$ 30 &   74 &                      \\
            s134* & J101101.20-044337.2 &  10:11:01.20 &   -04:43:37.2 &        O V  &    21.097 & -0.056 & -0.447 &  325 $\pm$ 50 &   33 &                      \\
            s135* & J101105.28-044223.4 &  10:11:05.28 &   -04:42:23.4 &        O V  &    20.689 &  0.011 & -0.239 &  325 $\pm$ 50 &   50 &          1.5'' slit. \\
             s136 & J101109.27-044147.7 &  10:11:09.27 &   -04:41:47.7 &    O  + neb &    20.639 & -0.062 & -1.111 &  435 $\pm$ 80 &   31 &                      \\
            s137* & J101108.58-044125.7 &  10:11:08.58 &   -04:41:25.7 &    Be  + O? &    21.007 & -0.096 & -1.108 &  325 $\pm$ 50 &   63 &                 SB2? \\
            s138* & J101100.56-044327.6 &  10:11:00.56 &   -04:43:27.6 & B III + gal &    20.406 &  0.026 & -0.513 &  175 $\pm$ 50 &   69 &     1.0'' slit. SB1? \\
s139*$^{\dagger}$ &                3988 &  10:11:07.24 &   -04:41:50.3 &     B  + O? &    22.190 & -0.180 & -0.820 &  200 $\pm$ 50 &   18 &                 SB2? \\
            s140* & J101105.48-044202.9 &  10:11:05.48 &   -04:42:2.90 &      B III  &    20.907 & -0.266 & -0.837 &  250 $\pm$ 50 &   15 &            Off-slit. \\
             s141 & J101059.21-044348.8 &  10:10:59.21 &   -04:43:48.8 &         B   &    21.159 & -0.138 & -0.576 &  255 $\pm$ 85 &   16 &            Off-slit. \\
             s142 & J101055.14-044117.0 &  10:10:55.14 &   -04:41:17.0 &         B   &    21.422 &  0.230 & -0.914 &  315 $\pm$ 50 &   26 &               Blend. \\
             s143 &                   - &  10:11:06.17 &   -04:42:13.7 &         B   &         - &      - &      - &  270 $\pm$ 40 &   27 &          0.8'' slit. \\
             s144 & J101105.90-044155.8 &  10:11:05.90 &   -04:41:55.8 &         B   &    21.535 &  0.043 &  0.069 &  305 $\pm$ 40 &   14 &            Off-slit. \\
s145*$^{\dagger}$ &                2432 &  10:11:03.50 &   -04:42:27.4 &         B   &    22.380 &  1.020 &      - &  325 $\pm$ 50 &   24 &          0.8'' slit. \\
             s146 & J101101.76-044314.0 &  10:11:01.76 &   -04:43:14.0 &         B   &    21.678 & -0.006 & -0.306 &  345 $\pm$ 35 &   18 &                      \\
            s147* & J101101.90-044241.1 &  10:11:01.90 &   -04:42:41.1 &         B   &    21.001 & -0.095 & -0.393 &  150 $\pm$ 50 &   23 &          0.8'' slit. \\
            s148* & J101101.93-044147.1 &  10:11:01.93 &   -04:41:47.1 &        OB   &    20.904 &  0.338 & -0.216 & 325 $\pm$ 100 &   40 &          0.8'' slit. \\
            s149* & J101103.95-044225.0 &  10:11:03.95 &   -04:42:25.0 &        OB   &    21.163 &  0.684 &      - & 325 $\pm$ 100 &   36 &          0.8'' slit. \\
s150*$^{\dagger}$ &                3416 &  10:11:05.71 &   -04:42:10.6 &        OB   &    22.190 & -0.260 & -0.793 & 400 $\pm$ 100 &   32 &          0.8'' slit. \\
             s151 & J101105.55-044221.9 &  10:11:05.55 &   -04:42:21.9 &        OB   &    20.376 & -0.002 & -0.062 &  405 $\pm$ 85 &   29 &               Blend. \\
s152*$^{\dagger}$ &                2392 &  10:11:03.37 &   -04:42:28.9 &        OB   &    21.590 &  1.190 &      - &  325 $\pm$ 50 &   28 &          0.8'' slit. \\
            s153* & J101103.34-044156.2 &  10:11:03.34 &   -04:41:56.2 &        OB   &    21.589 & -0.164 & -0.440 & 245 $\pm$ 100 &   26 &          0.8'' slit. \\
            s154* & J101100.55-044352.8 &  10:11:00.55 &   -04:43:52.8 &        OB   &    21.314 & -0.115 & -0.234 &  450 $\pm$ 50 &   22 &                      \\
            s155* & J101104.54-044229.7 &  10:11:04.54 &   -04:42:29.7 &        OB   &    22.006 & -0.212 & -0.142 &  325 $\pm$ 50 &   22 &          0.8'' slit. \\
s156*$^{\dagger}$ &                3509 &  10:11:05.95 &   -04:41:54.0 &        OB   &    22.180 & -0.170 & -0.848 &  205 $\pm$ 50 &   16 &                      \\
            s157* &                   - &  10:11:07.12 &   -04:42:09.6 &        OB   &         - &      - &      - & 260 $\pm$ 100 &   13 &          1.5'' slit. \\
            s158* & J101057.38-044019.0 &  10:10:57.38 &   -04:40:19.0 &       OBA   &    21.217 & -0.075 & -0.284 &  460 $\pm$ 50 &   34 & Tilted 8\textdegree. \\
            s159* & J101105.64-044224.0 &  10:11:05.64 &   -04:42:24.0 &       OBA   &    21.023 & -0.204 & -0.528 & 325 $\pm$ 100 &   32 &               Blend. \\
            s160* & J101101.83-044117.6 &  10:11:01.83 &   -04:41:17.6 &       OBA   &    20.706 & -0.032 &  0.039 &  160 $\pm$ 50 &   27 &          1.5'' slit. \\
             s161 & J101108.21-044153.5 &  10:11:08.21 &   -04:41:53.5 &       OBA   &    21.899 & -0.153 & -1.048 &  440 $\pm$ 55 &   25 &                      \\
             s162 & J101105.15-044212.8 &  10:11:05.15 &   -04:42:12.8 &       OBA   &    21.676 & -0.080 & -0.357 & 475 $\pm$ 230 &   24 &                      \\
            s163* & J101103.49-044157.2 &  10:11:03.49 &   -04:41:57.2 &       OBA   &    21.063 &  0.263 & -0.123 & 290 $\pm$ 100 &   20 &          0.8'' slit. \\
            s164* & J101106.92-044217.8 &  10:11:06.92 &   -04:42:17.8 &       OBA   &    22.253 & -0.436 & -0.616 & 325 $\pm$ 150 &   19 &                      \\
            s165* & J101057.66-044130.2 &  10:10:57.66 &   -04:41:30.2 &       OBA   &    21.497 & -0.107 & -0.130 & 325 $\pm$ 100 &   19 &                      \\
            s166* & J101104.51-044225.9 &  10:11:04.51 &   -04:42:25.9 &       OBA   &    20.934 & -0.201 & -0.925 & 350 $\pm$ 100 &   17 &                      \\
\end{longtable}
\end{landscape}
\begin{landscape}
\small
\setlength\LTleft{-15pt}
\setlength\LTright{-20pt}
\setlength\LTcapwidth{\linewidth}
\captionsetup[longtable]{labelfont=bf, font = footnotesize, labelsep=period}
\begin{longtable}{lccccccccl}
\caption{Late type stars observed in this work. Identification codes (ID), spectral types (SpT) and radial velocities (\vrad) are included. Stars whose radial velocity was determined manually are marked with $^{(*)}$. Identifications numbers and coordinates by \citet{Massey2007} are provided when available,  otherwhise we give the ones from the 2MASS catalogue \citep{Cutri2003} and mark the sources with $^{(\dagger)}$. Column \textit{Memb.} flags the membership to Sextans~A.}
\label{tab:cat_late}\\
\toprule
\toprule
               ID &             Alt. ID & RA (J2000.0) & DEC (J2000.0) &         SpT &    $V$ & $B-V$ & \vrad~(\kms) & Memb. &                                    Notes \\
\midrule
\endfirsthead
\caption[]{continued.} \\
\toprule
\toprule
               ID &             Alt. ID & RA (J2000.0) & DEC (J2000.0) &         SpT &    $V$ & $B-V$ & \vrad~(\kms) & Memb. &                                    Notes \\
\midrule
\endhead
\midrule
\multicolumn{10}{r}{{\textit{continued on next page}}} \\

\endfoot

\bottomrule
\endlastfoot
            s167* & J101106.39-044218.1 &  10:11:06.39 &   -04:42:18.1 &       G5 I  & 19.588 & 1.446 & 300 $\pm$ 30 &   yes & \citetalias{Britavskiy2019}-Sextans A 6. \\
            s168* & J101106.00-044212.5 &  10:11:06.00 &   -04:42:12.5 & G8 I + B7 I & 19.453 & 1.689 & 350 $\pm$ 30 &   yes &                       0.8'' slit. Blend. \\
            s169* & J101103.95-044228.3 &  10:11:03.95 &   -04:42:28.3 &       G8 I  & 20.031 & 1.768 & 250 $\pm$ 30 &   yes & \citetalias{Britavskiy2019}-Sextans A 4. \\
            s170* & J101102.21-044149.4 &  10:11:02.21 &   -04:41:49.4 &       K0 V  & 16.785 & 0.747 &  40 $\pm$ 30 &    no &                              0.8'' slit. \\
            s171* & J101116.62-044014.4 &  10:11:11.78 &   -04:41:18.6 &       K2 V  & 20.144 & 1.349 &   0 $\pm$ 30 &    no &                                          \\
            s172* & J101104.78-044238.8 &  10:11:04.78 &   -04:42:38.8 &     K5 III  & 20.266 & 1.423 & 300 $\pm$ 40 &   yes &                              1.0'' slit. \\
s173*$^{\dagger}$ &   J10110861-0440473 &  10:11:08.60 &   -04:40:47.0 &       K7 V  &      - &     - &   0 $\pm$ 30 &    no &                                Off-slit. \\
            s174* & J101115.72-044441.0 &  10:11:15.72 &   -04:44:41.0 &       M3 V  & 19.797 & 1.714 &   0 $\pm$ 30 &    no &                              1.5'' slit. \\
\end{longtable}
\end{landscape}
\clearpage
\twocolumn

\subsubsection{Spectral content of the catalogue}
\label{sec:SpTcontent}

Although our observations were designed to find OB stars in Sextans~A, we also observed other types of sources.
Here we briefly describe the objects classes unveiled by our observations.

\textit{\textbf{O stars.}}
There are \Ostars~O-type stars in the sample. Among them, we find the stars with the earliest spectral types observed at metallicities lower than 1/5~\Zsun. 
Some of the sources classified as O-types are candidates to undergo chemically homogeneous evolution or to host a stripped star. 
This will be discussed in more detail in \mbox{Section \ref{sec:EvolutionaryStatus}}. 


\textit{\textbf{BA supergiants.}}
Our sample contains \BAI~B- and A-type supergiants. 
This collection will allow us to better constrain the current chemical composition of Sextans~A and to study the [$\alpha$/Fe] abundance ratio. 
The quantitative analysis of the spectra of \EarlyBI~early-B supergiants will yield the abundances of C, N, O and Si.
Strikingly, the silicon lines in the observed spectra appear weaker than predicted for their corresponding types and the metallicity of the galaxy. 
The abundance of magnesium and iron-group elements, can be derived from analyses of late-B and A supergiants.
These analyses will be carried out as future work.


\textit{\textbf{Red stars.}}
There are \LateStarsSexA~evolved stars from Sextans~A in our catalogue, two of them already reported by \citet[][B19]{Britavskiy2019}. These two are reclassified from K1-3 to G types. 
This update reinforces the observed trend of the average spectral types of red supergiants (RSGs) shifting towards earlier types at lower host galaxy metallicity \citep{Elias1985, MasseyOlsen2003, LevesqueMassey2012}.
A plausible explanation for this observed pattern is the dependence of the molecular bands, main indicators of spectral type in evolved stars, on the metallicity of the environment. However, this trend may also reflect the shift of the Hayashi limit to higher effective temperatures as the metallicity decreases \citep{Elias1985}.


\textit{\textbf{Foreground stars.}}
\ForegroundStars~foreground stars landed in our programmed slits. Their identification as non-members of Sextans~A is justified in Section \ref{sec:membership}.


\textit{\textbf{Binary candidates.}}
A total of \SBStars~stars show some signatures of binarity in their spectra.
Some targets present radial velocity variations between different observations. These are marked as "SB1" in the \textit{Notes} column of \mbox{Tables \ref{tab:cat_OB_highQ}-\ref{tab:cat_late}}, along with a question mark when the variation was smaller than \mbox{50~\kms}~or the S/N of the data was poor. Those sources whose spectra show doubled-peaked lines or exhibit lines of a second component are marked as "SB2". The ones that met other binary diagnostics, such as variation in their spectral features, are labelled as "Bin".  All of these cases are discussed individually in \mbox{Appendix. \ref{sec:appx_IndividualNotes}}. We note that this constitutes the first sample of candidate binary systems ever reported in Sextans~A.


\subsection{Membership}
\label{sec:membership}

To assess the membership of the targets to Sextans~A, 
we inspect whether they have registered parallaxes or proper motions in the Gaia Early Data Release 3 \citep[Gaia EDR3, ][]{GaiaEDR3}. Given the distance to the galaxy,
the majority of the catalogue stars are too faint to have an entry in Gaia. For those that are included, we expect null proper motions and parallaxes. Most of our sample stars follow this trend.

We find 12 stars that deviate from the expected parameters. Two of them had already been identified as foreground based on their radial velocity (Sect. \ref{sec:vrad}). 
The rest of the sources lie within 2$\sigma$ of the mean of the weighted normal distribution of the Gaia entries for Sextans~A, for both parallax and proper motions. We thus consider that they follow the general trend observed in the galaxy and discard them as outliers.

We also checked that the absolute magnitudes $M_V$ of all catalogue stars (calculated as described in Section \ref{sec:SpTclas}) are consistent with their spectral type at the distance of Sextans~A. There are only small discrepancies with respect to the calibrated values, which are discussed in Sect. \ref{sec:EvolutionaryStatus}.

Finally, we used radial velocity ($v_{\text{rad}}$) as an additional membership check for all of our sample stars. 
A group of \ForegroundStars~late type stars with \mbox{\vrad~$\leq$ 40~\kms}~was identified as foreground.
For the rest of our survey, the majority presents radial velocities consistent with the systemic velocity and radial velocity curve of Sextans~A, as discussed in Section \ref{sec:vrad}, which further supports their membership to the galaxy.


\subsection{Radial velocity}
\label{sec:vrad}

\begin{figure}
    \centering
    \includegraphics[width=\hsize]{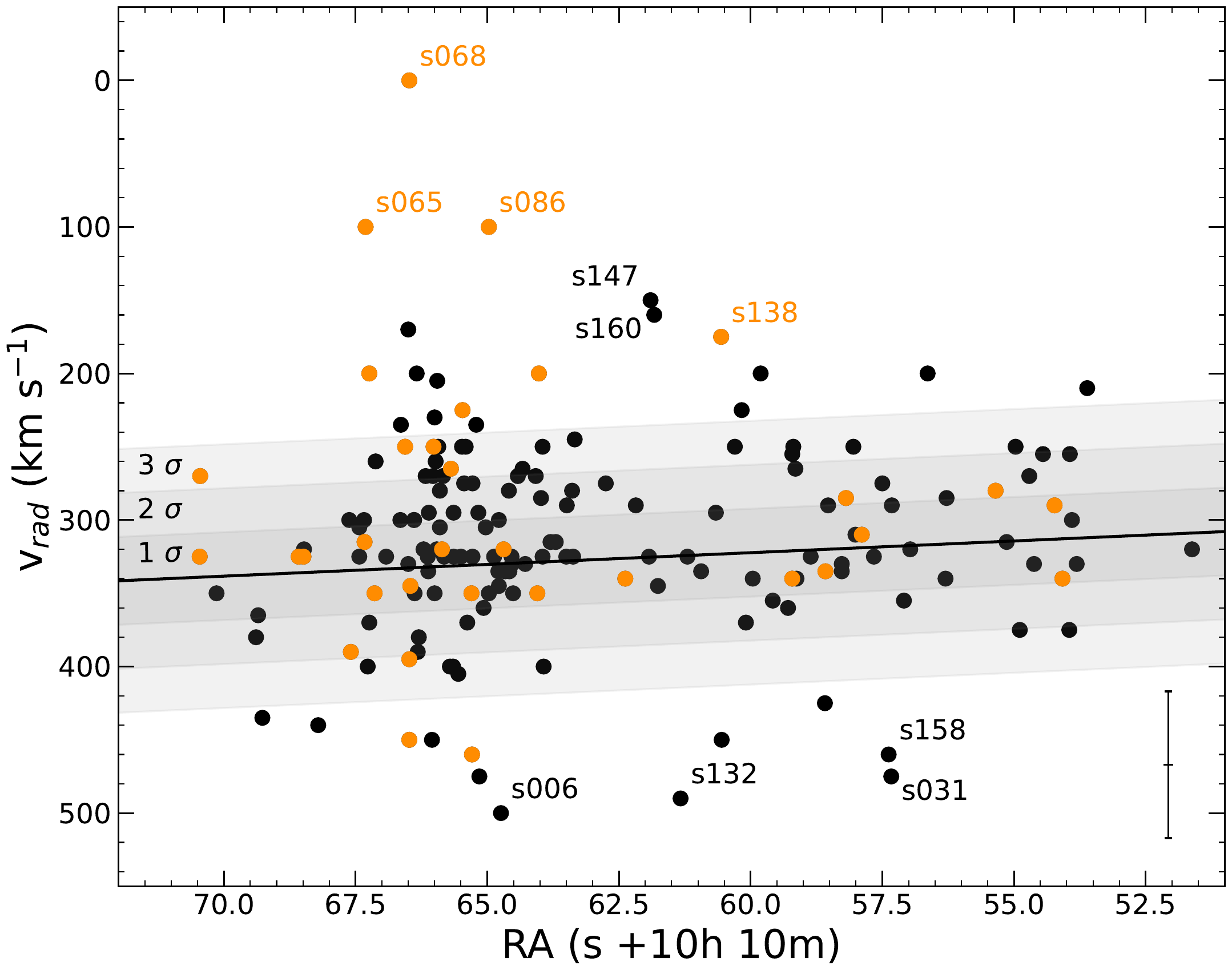}
    \caption{Comparison of the radial velocities obtained for the sample stars and the H~{\sc i}~rotation curve of Sextans~A derived by \citet{Skillman1989}. We excluded from the figure stars identified as foreground. The shaded area represents the 1$\sigma$, 2$\sigma$ and 3$\sigma$ uncertainty of the measured rotation curve, being \mbox{$\sigma$ = 30~\kms}. Stars whose IDs are provided depart more than 3$\sigma$ from the radial velocity curve, after error bars have been considered. We also highlight in orange our list of binary candidates identified from spectral features. 
    In the lower-right corner of the figure, we draw a representative error bar illustrating the average radial velocity error of $\pm$~50~\kms~for a star observed with resolution \mbox{R = 1000}.}
    \label{fig:vradcurve}
\end{figure}

We estimated the radial velocity of each star by measuring the Doppler shift experienced by the hydrogen and helium lines. Using an automatic procedure, we determined the centroid of the lines by fitting their profile to a Gaussian function. To avoid biases introduced by nebular oversubtraction that may affect the core and introduce profile asymmetries, we imposed a higher weight on the line wings, defined as the points located between the continuum and a value set by the program.  The height of this threshold was selected according to the strength of the lines compared to the local continuum. We calculated the depth of the lines as the difference between the flux of their cores ($f_{\mathrm{min}}$) and the continuum, and positioned the threshold at 15\% and 30\% of the estimated depth for strong (\mbox{1 - $f_{\mathrm{min}}$ > 0.4}) and weak (\mbox{1 - $f_{\mathrm{min}}$  $\leq$ 0.4}) Balmer lines, respectively. These values stem from the fact that contrasts higher than \mbox{1 - $f_{\mathrm{min}}$ > 0.4} are likely caused by nebular oversubtraction. For He lines, where nebular oversubtraction is less prominent, we set the threshold at 50\% of the line depth for weak (\mbox{1 - $f_{\mathrm{min}}$ < 0.1}) lines and 80\% for strong (\mbox{1 - $f_{\mathrm{min}}$  $\geq$ 0.1}) lines.

After the program fits all lines automatically to Gaussian profiles, it then rejects results from some lines in two steps according to: (i) their depth with respect to the noise of the spectrum (S/N values are compiled in Tables~\ref{tab:cat_OB_highQ}-\ref{tab:cat_late}), and (ii) their line width, rejecting lines
when the width exceeded twice the median of all fitted lines of the spectrum. We also imposed a limit of 1000 km s$^{-1}$ to the resulting values of the radial velocity for individual lines and applied a $\sigma$-clipping algorithm using the median as a reference and rejecting by 2$\sigma$. Finally, we calculated the weighted average of the estimated $v_{\text{rad}}$ of the selected lines.

The uncertainty was calculated by averaging the error of the Gaussian fit to the selected lines and then dividing by the square root of the number of used lines. This result depends not only on the spectral S/N, but also on the number of lines suitable to measure the radial velocity.

The radial velocity of some stars in the sample could not be determined by our automatic measurement due to poor spectral S/N. In such cases, we defined the central wavelength of lines manually and estimated their corresponding Doppler shift.
The assigned uncertainty was also determined manually.
These stars are marked with $^{(*)}$ in Tables~\ref{tab:cat_OB_highQ}-\ref{tab:cat_late}.

In Fig.~\ref{fig:vradcurve}, we compare our results with the rotation curve of Sextans~A, derived by \citet{Skillman1989}, to assess the membership of our sample stars to the galaxy. Overall, our targets are consistent with Sextans~A's radial velocity. 
However, there are ten sources (identified in the Figure) whose radial velocity (and error bars) departs from the H~{\sc i}~rotation curve by more than 3$\sigma$, being $\sigma$ the estimated error of the curve ($\sigma$ = 30~\kms).
The \vrad~departures of the ten outliers could be explained by a possible runaway nature or their belonging to a multiple system. Yet, spectroscopic follow-up will be needed to confirm any of these hypotheses. We highlight in orange the stars identified as binary candidates from their spectra in Figure~\ref{fig:vradcurve}.

\subsection{The $Q$ pseudo-colour as an indicator of OB type stars}
\label{sec:SuccessRate}

\begin{figure}
    \centering
    \includegraphics[width=\hsize]{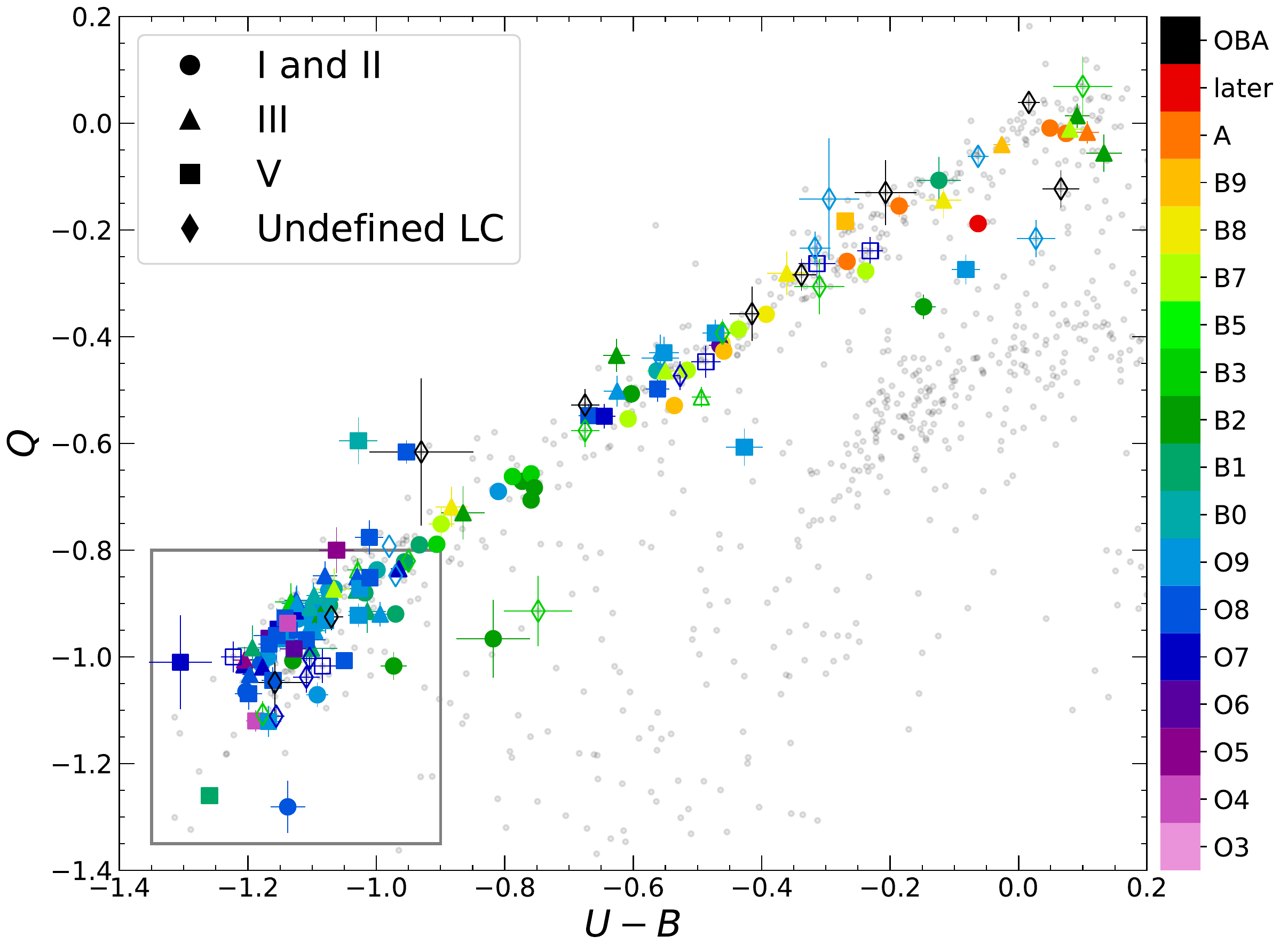}
\caption{$U-B$ vs $Q$ diagram of Sextans~A, showing \citet{Massey2007}'s catalogue in grey. 
The main criterion to select candidate O- and early B-type stars for our observing runs was \mbox{$Q$ $\leq$ -0.8}. 
The observed stars, excluding foreground sources, are colour-coded by spectral type and marked based on their luminosity classes (see legend). Stars without a defined spectral subtype are drawn with empty symbols. Most of the O-type stars confirmed by this work are found within the black box, which further supports the efficiency of the $Q$ pseudo-colour as a selection criterion of massive stars.}
    \label{fig:UBvsQ}
\end{figure}

Due to their degenerate colours in the optical range and the effect of extinction, O- and early B-type stars cannot be identified using photometry only.
However, \citet{Garcia2009} found that the OB stars of the galaxy IC~1613 lied in a particular locus of the \mbox{$U-B$ vs. $Q$} diagram (see their \mbox{Fig. 5}). This is a consequence of the biunivocal relation between the reddening-free pseudo-colour $Q = (U-B)-0.72(B-V)$ and spectral type that holds down to A0 types.
We have used this as the main criterion to select OB candidates in our subsequent discovery spectroscopic runs.
\citet{GarciaHerrero2013} used \mbox{$Q$ < -0.8} together with UV photometry, achieving a success rate of 70\% in the detection of O-type stars in IC~1613. With the same criteria, \citet{Camacho2016} had a 50\% rate in Sextans~A.

The stars in our more extended catalogue are distributed throughout a broad range of $Q$ values, which enables us to reassess the efficiency of this candidate selection criterion.

Roughly half of our catalogue, \BoxStars~stars, are found in the \mbox{$Q < -0.8$} box (see Figure \ref{fig:UBvsQ}). Of these, we find \OBoxStars~O-type stars, including the earliest of the sample, \BBoxStars~B stars, the majority earlier than B1.5, and \OBAboxStars~OB stars. The latter are not considered in the forthcoming statistics since their poor spectral S/N prevented us from assigning more definite spectral types.
The ratio of confirmed O-type stars and total observed stars within the \mbox{$Q < -0.8$} box is \PrecisionBox\%.
However, if we push this criterion further and reduce the candidate box to values of \mbox{$Q < -1.0$}, the number increases to \PrecisionMiniBox\%, with \MiniOBoxStars~O stars out of the \MiniBoxStarsWithoutOBA~stars with confirmed spectral type below this limit. Therefore, in the hunt for the high-mass stars in Sextans~A and other galaxies, we advise selecting candidates with very low $Q$ values.

Outside the \mbox{$Q < -0.8$} box, we found \OoutBoxStars~O stars. The reason for their locus in the $U-B$ vs. $Q$ diagram is unclear. For instance, in the case of the O star \Remark{LSS3.OB13208}, 
its colours are unreliable as the star overlaps with a foreground galaxy (see individual notes in Appendix~\ref{sec:appx_IndividualNotes}). 
Therefore, if we aim to seek completeness in the O star population of the galaxy, we must observe beyond the \mbox{$Q < -0.8$} box. This highlights the importance of untargeted surveys of galaxies to fully account for their population of massive stars.


\section{Discussion}
\label{sec:discussion}

We have obtained spectroscopy of \TotalStars~sources,
of which \OBStars~are OB stars (including \LowOBStars~whose spectral subtype could not be assigned due to poor spectral S/N), \AFStars~are A- and F-type stars, \RedI~are red giants/supergiants members of Sextans~A, and \ForegroundStars~are identified as foreground stars.
In addition, we provide the first list of binary candidates with metallicities of 1/10 \Zsun~and a list of candidates experiencing chemically homogeneous evolution or systems hosting a stripped star (see Sect. \ref{sec:EvolutionaryStatus}).

Few massive stars were previously known in very low metallicity environments, as they are located at far distances \citep{Garcia2021}.
The existing catalogues in IC~1613, WLM and NGC~3109, with \mbox{1/7 \Osun}~and \mbox{1/5 \Fesun}, do not exceed 
seventy stars \citep[e.g. ][]{Bresolin2006, Bresolin2007, Evans2007, GarciaHerrero2013}. In our target galaxy Sextans~A, with \mbox{1/10 \Zsun}, only 19 massive stars had been confirmed so far \citep{Camacho2016, Garcia2019}. At lower metallicities, only 2 OB stars in SagDIG \citep[\mbox{$\sim$ 1/20 \Osun,} ][]{Garcia2018} and one confirmed O-type star in Leo~P \citep[\mbox{$\sim$ 1/30 \Osun,} ][]{Evans2019} are known.

The sample of \OBStars~OB stars assembled in this work constitutes the largest census of massive stars at sub-SMC metallicities to date. This catalogue also enables us to study the recent star formation history of Sextans~A and provides a first glimpse into the evolution of \mbox{1/10 \Zsun}~massive stars. We discuss the two topics in this section.


\subsection{The recent star formation history of Sextans~A}
\label{sec:StarFormation}

\begin{figure*}
    \centering
    \includegraphics[width=\hsize]{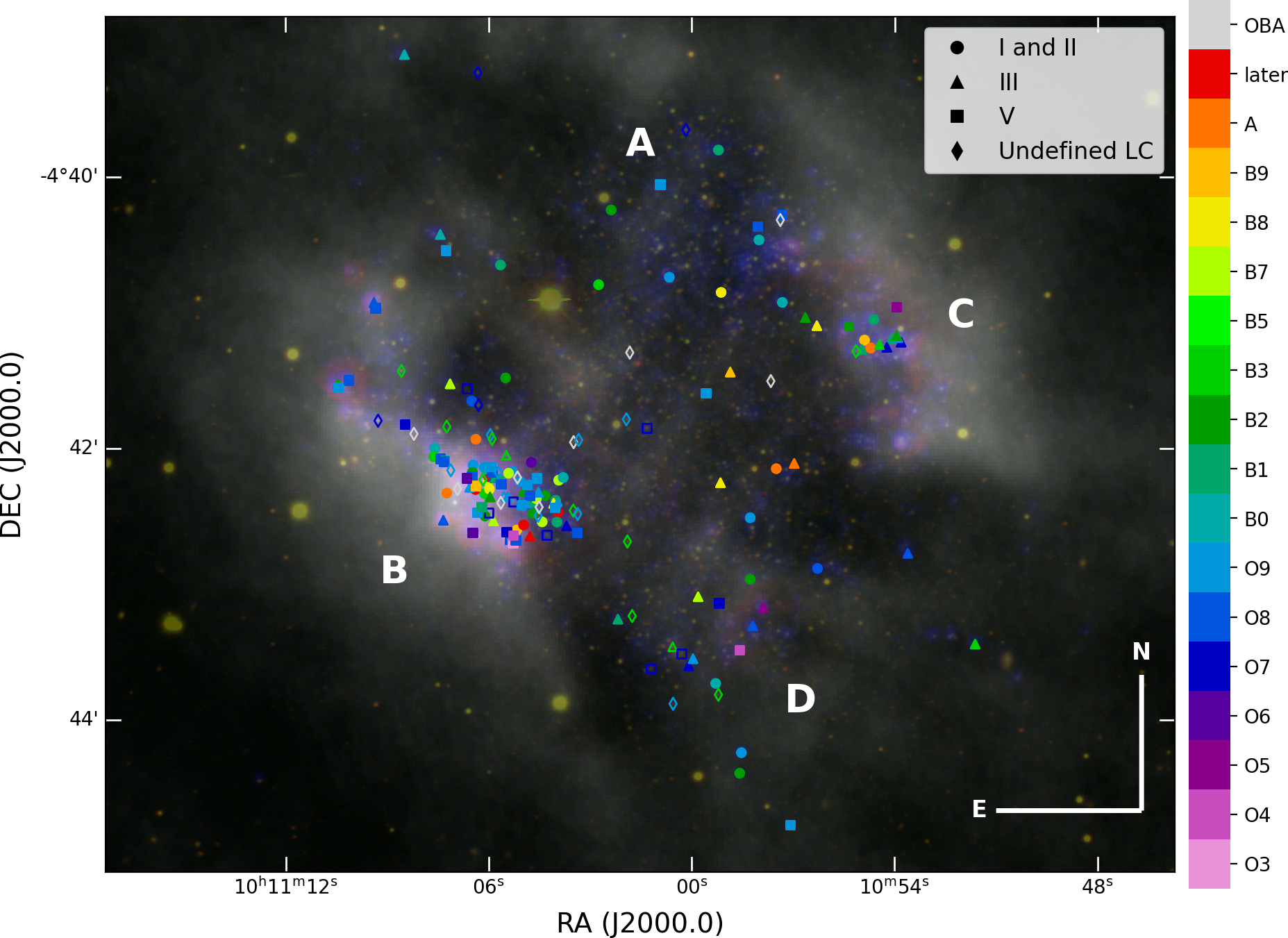}
\caption{Sextans~A RGB composite image made with \ha-- (red) and $V$--bands (green) from \citet{Massey2007}, and GALEX FUV (blue). The \textit{LITTLE THINGS} neutral hydrogen map \citep{Hunter2012} is overlaid in white. Regions with confirmed ongoing star formation are marked in the figure. We show the catalogue stars, excluding identified foreground stars, colour-coded according to their spectral type and with different symbols based on their luminosity class. Stars with an undefined spectral subtype are drawn with empty markers.}
    \label{fig:SexAmap}
\end{figure*}

\begin{figure*}
    \centering
    \includegraphics[width=\hsize]{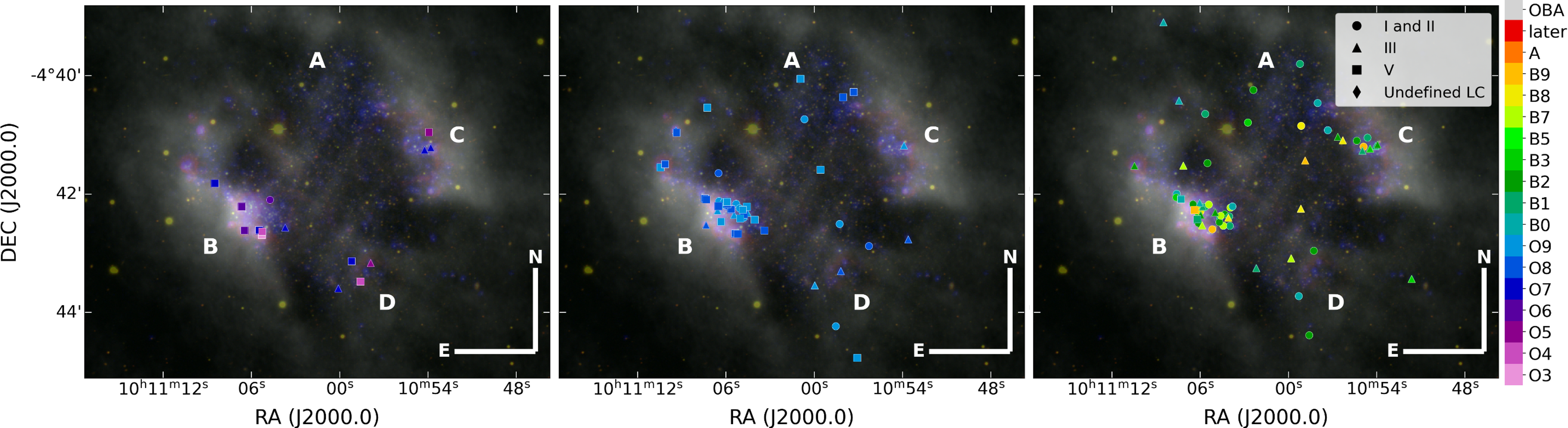}
\caption{Same as Fig. \ref{fig:SexAmap}. Each panel shows the sample OB stars in different intervals of spectral type. From left to right, we plotted targets with spectral types earlier than O8, stars with types between O8 and O9.7 (both included), and sources classified as B-type stars.
There is an apparent stratified pattern of spectral type in region--B, where the early O-type stars overlap with the high gas density areas, and later types are located at the inner rims of the H~{\sc i} cloud.
}
    \label{fig:SexAmap_mosaic}
\end{figure*}

After a quiescent period of \mbox{7.5 Gyr}, star formation resumed
\mbox{2.5 Gyr} ago in Sextans~A and has been growing continuously,
with an increase in the last \mbox{0.6 Gyr} \citep{Dolphin2003}. 
The galaxy hosts three major regions with confirmed ongoing star formation, qualitatively defined by \citet{Dohm-Palmer2002} and named region--A, --B and --C by \citet[][see also Fig. \ref{fig:SexAmap}]{Camacho2016}.
Region--A (in the north) has been forming stars for the past \mbox{400~Myr} and it is about to exhaust its local gas content.
The other two, regions--B (in the southeast) and --C (in the northwest), have been active for \mbox{200~Myr} and \mbox{20~Myr}, respectively, and overlap with the higher concentrations of ionized and neutral hydrogen that frame the visually bright part of the galaxy \citep{vanDyk1998,Dohm-Palmer2002}. Recently, \citet{Garcia2019} found young massive stars in the southern tip of Sextans~A and coined this area as region--D.

With typical ages under 50~Myr, massive stars signal ongoing star formation.
In Fig. \ref{fig:SexAmap}, we show the location of the sample stars in Sextans~A and use our \OBStars~OB sources as star formation tracers.

Most of the OB stars are located in regions--B and --C,  overlapping with the high column density areas of neutral gas and producing large structures and shells of ionized hydrogen.
Other preferred locations for hot stars are the northeastern and southern edges of the galaxy, where they ionize small H~{\sc ii} bubbles and arcs.

The stars holding the earliest types (see Fig. \ref{fig:SexAmap}) are located 
independently of the stellar density and star-forming rate of their host areas. Regions--B, --C, and --D correspond to regions--1, --3, and --4 from \citet{Shi2014}, who inferred  star formation rates using FUV photometry from GALEX of -2.66, -2.32 and -3.19~\mbox{log $\rm M_{\odot}yr^{-1}kpc^{-2}$}, respectively.
Finding such early types in these three areas indicates that star formation has been active until at least \mbox{5~Myr} ago \citep{Massey2003}. No supernova remnant that could help us further trace star formation is known.

Region--B presents more dispersion in spectral types, and we have found more early-type stars than in the other star-forming regions of the galaxy. 
This is in line with \citet{Camacho2016}'s suggestion that, although
activated longer ago, region--B is still undergoing star formation.

The distribution of massive stars in region--B appears to be layered by spectral type and correlated with the column density of the neutral gas (see Fig. \ref{fig:SexAmap_mosaic}). The early O-type stars overlap with the highest density of H~{\sc i}
(left panel of Fig. \ref{fig:SexAmap_mosaic}), while the later types are located at at the inner edges of the H~{\sc i} cloud where the column density is lower (middle and right panels of Fig. \ref{fig:SexAmap_mosaic}). The stratification might indicate a temporal sequence of star formation, which would proceed from the inner rim outwards, following the H~{\sc i} column-density gradient.
Region--C could show a similar layered pattern, although the catalogue coverage is more sparse in this area due to the way the observing runs were planned.

\begin{figure*}
    \centering
    \includegraphics[width=0.97\hsize]{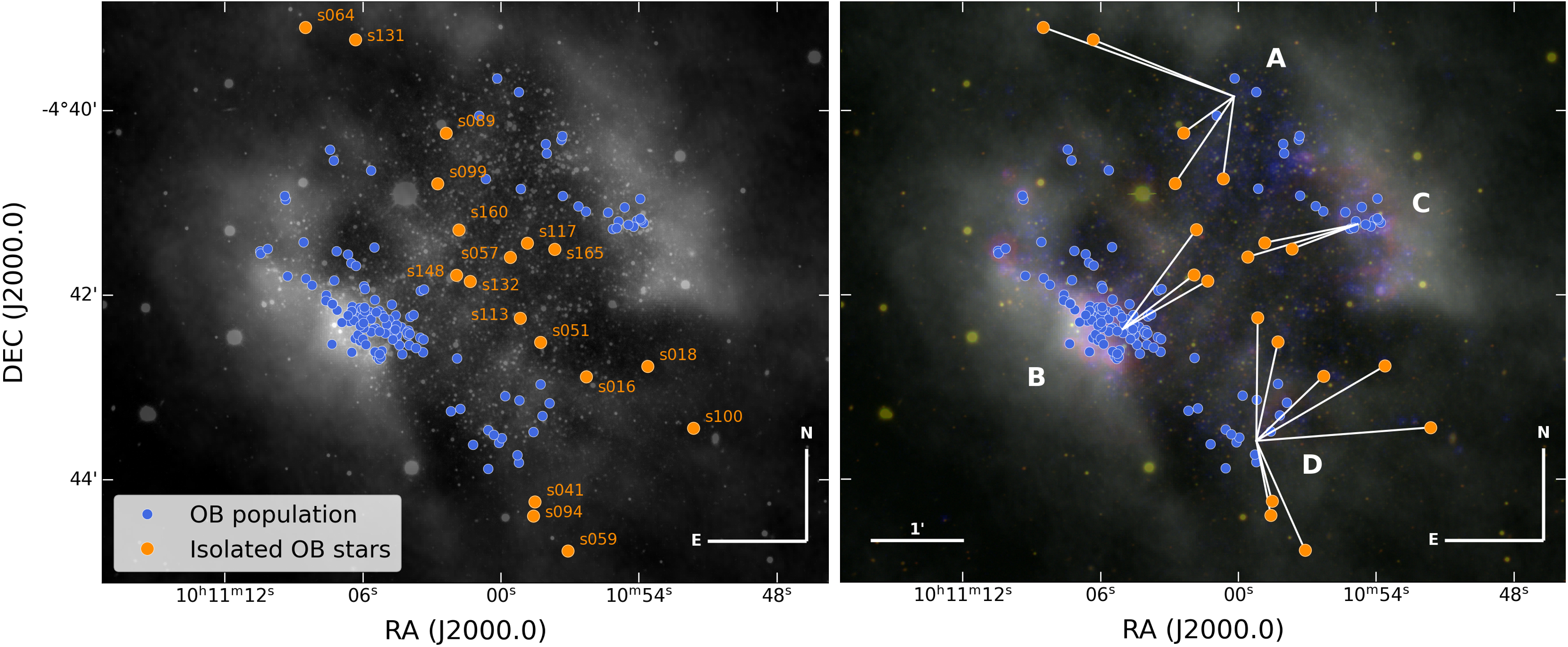}
\caption{Left: \textit{LITTLE THINGS} neutral hydrogen map \citep{Hunter2012} overlaid on a $V$-band image of Sextans~A \citep{Massey2007}. We show the population of OB stars in blue, and identify and highlight in orange the isolated OB stars. Right: Same RGB image as Fig. \ref{fig:SexAmap}. We again highlight the isolated OB stars in orange and show the distance these stars would have travelled from their nearest OB population if they were runaway stars (white lines). A projected angle of 1' corresponding to a distance of 390~pc in Sextans~A is plotted as reference.}
    \label{fig:SexAmap_Isolated}
\end{figure*}

A significant fraction of OB stars in our census are found on the outskirts of Sextans~A, at the edges of the neutral gas reservoir and with no associated H~{\sc ii} structures (see left panel of Fig. \ref{fig:SexAmap_Isolated}).
\citet{Garcia2019} already pointed out these unexpected star-forming sites with the detection of star \Remark{MOS1.s01} (their s1) in the southern region of the galaxy. Similar examples exist in other galaxies near and far. \citet{Garcia2010} spotted OB associations and UV sources outside the brightest centre of IC~1613. 
The detection of extended UV-disc galaxies \citep{Gildepaz2005} 
already suggested the existence of star formation in low-gas density environments on the galactic outskirts. Only the spectroscopic confirmation of OB stars at these sites has allowed us to directly associate some of the UV sources  with massive stars. 
In addition, we also find OB stars in the central part of the galaxy, where star formation was considered already finished. This recalls the massive star V39 detected by \citet{Herrero2010} in an inconspicuous region of IC~1613, and resembles the observed star formation in the outer fringes of the galaxy. Examples of star formation at low H~{\sc i} column density also occur in low-mass dwarf galaxies outside the Local Group \citep[e.g.][]{Teich2016}.

The existence of an underlying population supporting these apparently isolated stars cannot be discarded since no sufficiently deep IR observations of the galaxy exist. Yet, \citet{Garcia2019} examined archival WFC3-IR observations of region--D and found no additional stars around s1.
Moreover, \textit{Herschel} observations do not support the existence of large masses of dust at the location of most of these isolated stars.

Could these isolated stars be runaways?
Given the distance to Sextans A, 
we cannot infer any information from proper motions. A star crossing the entire galaxy in 10~Myr would exhibit proper motions of \mbox{$\sim$ 0.01 mas/yr}, significantly lower than the typical uncertainties of Gaia for the proper motions of stars with \mbox{$G$ = 20 mag} \citep[0.5 mas/yr, ][]{Lindegren2021}.

We have manually selected sources that are far from region--A, --B, --C and --D and highlighted them in Fig. \ref{fig:SexAmap_Isolated}.
The distances between the isolated sources and the approximate centre of
their nearest star-forming regions (see right panel of Fig. \ref{fig:SexAmap_Isolated}) range between 250 and 825~pc. Assuming lifetimes of 10~Myr and ejection at early stages of evolution, covering these distances would correspond to projected velocities of 25-80~\kms.
These values could be consistent with the high-end tail of the probability distribution of runaway stars produced by binary interaction found by \citet{Renzo2019}.
At the metallicity of Sextans~A, their calculations suggest that the velocity of the ejected stars would be $\sim$~10~\kms~higher on average. This is still significantly lower than the velocity of the alleged runaways. Moreover, the fraction of expected unbound binaries would be lower at low metallicity. Given the low expected probability of these cases \citep[$<$ 0.01\%,][]{Renzo2019}, we consider unlikely that these are runaway stars.

We therefore conclude that these sources were likely formed in situ, perhaps supported by compact, CO-dark gas clouds not signalled by H~{\sc i} maps. An alternative explanation would be that they are the products of mergers of two or more lower-mass sources, although this possibility does not solve the conundrum of the missing underlying population.


\subsection{Evolutionary status}
\label{sec:EvolutionaryStatus}

\begin{figure*}
    \centering
    \includegraphics[width=0.95\hsize]{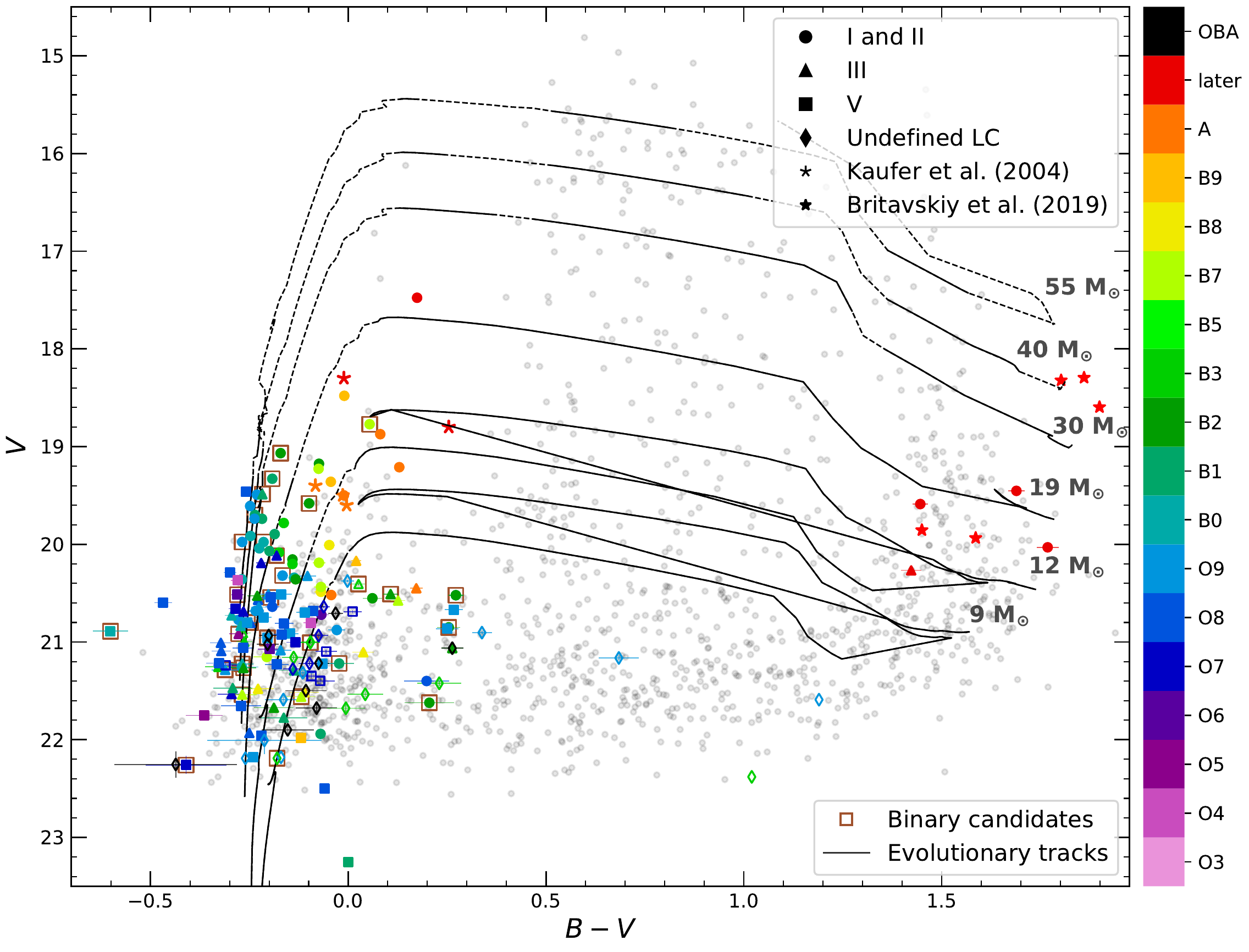}
\caption{CMD of Sextans~A, showing \citet{Massey2007}'s catalogue in grey and our sample sources, excluding foreground stars, colour-coded according to their spectral type. Those without a defined spectral subtype are drawn as empty symbols. Their luminosity classes are indicated by different symbols (see legends), and binary candidates are marked with empty brown squares. We include \citet{Kaufer2004} and \citet{Britavskiy2019} supergiants, labelled with distinct markers. 
Error bars for the photometry are shown; we note that errors for the $V$--band magnitudes are usually smaller than the symbol size. The computed photometry for \mbox{$Z$ = 0.10 \Zsun}, \mbox{\vsini~= 100~\kms} evolutionary tracks by \citet{Szecsi2022} are also provided (see Sect.~\ref{sec:EvolutionaryStatus}). Dashed lines mark the points where the interpolation of the theoretical atmosphere models to calculate the photometric magnitudes was incomplete. The tracks  have been shifted to account for Sextans~A's distance modulus and foreground extinction.}
    \label{fig:CMD_withoutcor}
\end{figure*}

Fig. \ref{fig:CMD_withoutcor} shows Sextans~A's colour-magnitude diagram (CMD) built with \citet{Massey2007}'s photometric catalogue, highlighting our sample stars.
We also incorporate blue and red supergiants from previous studies by \citet{Kaufer2004} and \citet{Britavskiy2019}.

We include evolutionary tracks by \citet{Szecsi2022} with Sextans~A's metallicity ($Z$ = 0.10 \Zsun = 0.00105) and an initial rotational velocity of \mbox{\vsini~= 100~\kms}, consistent with the average rotational velocity of OB stars in other low metallicity regimes \citep[e.g. ][]{Sabin-Sanjulian2017, Ramirez-Agudelo2017, Ramachandran2019}.
To compare observations with the evolutionary predictions, we computed the synthetic photometry of the evolutionary tracks using a grid of \textsc{tlusty} atmosphere models
with an iron abundance of \mbox{ $[\mathrm{Fe}/\mathrm{H}] = -1.0$ dex}. For each point of the tracks, defined by effective temperature (\Teff) and surface gravity ($\log g$), we assigned an atmosphere model by using a four-point linear interpolation. Some regions of the evolutionary paths lay outside the \textsc{tlusty} grid coverage, and only 2 or 3 points could be used to interpolate, leading to perhaps
non-realistic atmospheres. These regions are shown as dashed lines in Fig. \ref{fig:CMD_withoutcor}. Finally, the tracks are shifted to account for the distance modulus \citep[\mbox{$\mu_0$ = 25.63}, ][]{Tammann2011} and foreground extinction \citep[\mbox{\EBV = 0.044}, ][]{Skillman1989} of Sextans~A. No correction is applied to the photometry of the catalogue stars.

The OB stars of our catalogue are spread across the blue plume.
The observed dispersion in both spectral types and luminosity classes
impact the masses and ages 
that can be derived from the evolutionary tracks.
Although most of the earliest stars concentrate in the bluest regions of the plume, others overlap with stars of later spectral types. The spread can be partly explained by the intrinsic errors of the assigned spectral type, but it is mostly caused by internal interstellar extinction. The same effect has been observed in other LG dIrr galaxies like IC~1613 \citep{Garcia2009} and SagDIG \citep{Garcia2018}. 
A thorough characterisation of the internal extinction of Sextans~A will be published in a subsequent paper of this series.

The late-type stars of Sextans~A are mainly found at the red ends of the evolutionary tracks. They are concentrated in two different groups, a less massive one between 12 and 19~\Msun~and a second population with higher initial evolutionary masses ranging between 30 and 40~\Msun. 
Recent works \citep{Beasor2019, Britavskiy2019} speculate that the more luminous and massive group of RSGs are the evolved analogues of the massive blue straggler stars and might be products of mergers or binary mass transfer.

\begin{figure}
    \centering
    \includegraphics[width=\hsize]{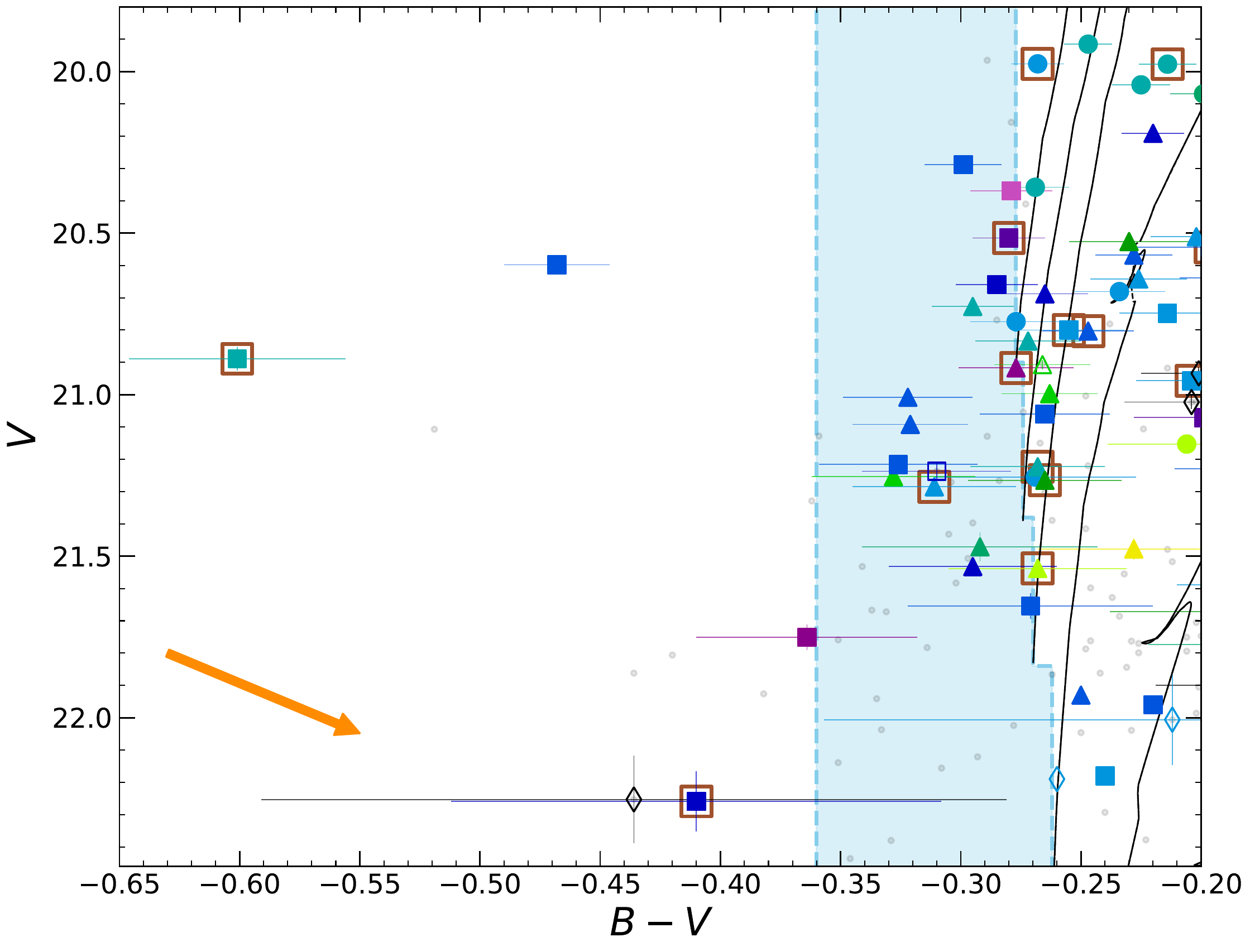}
\caption{Same as Fig. \ref{fig:CMD_withoutcor}, zooming into the main sequence and bluer colours. 
We shade in blue the area of bluer colours than the evolutionary tracks, with a lower boundary in $B-V$ defined by the intrinsic colour of a blackbody at 200~kK. This region contains the list of blue outliers listed in Table~\ref{tab:CHE}.
In the lower-left corner, we provide the reddening vector, adopting $R_V$ = 3.1. 
}
    \label{fig:CMD_zoom}
\end{figure}

We detect that \CHEStars~OB-type stars exhibit bluer colours than the evolutionary tracks (shaded region in Fig. \ref{fig:CMD_zoom}). This number may increase after internal reddening is corrected.
Slowly rotating single-star evolutionary tracks do not predict OB stars in this locus, as shown in Fig. \ref{fig:CMD_withoutcor}.
We imposed a lower limit on $B-V$ of -0.36~mag corresponding to the colour of a blackbody with \Teff~= 200~kK \citep[the highest effective temperature observed in Wolf-Rayet stars, ][]{Sander2012, Tramper2015}.
Photometric measurements bluer than this value are discarded as spurious. They might be due to strong emission lines contributing to the total spectral energy distribution and thus the measured colours.

A star might reach extreme hot temperatures, hence blue colours, if it evolves chemically homogeneously. Rapid rotation may lead a massive star to efficient mixing, dragging helium into its envelope as it is produced, and thus never establishing a chemical gradient \citep{MaederMeynet2000}. With hydrogen still burning in its core, the star will evolve to colours and effective temperatures hotter than the ZAMS, entering a Wolf-Rayet-like stage and remaining extremely hot throughout its lifetime \citep{YoonLanger2005}. This evolutionary path is more expected among very low-Z massive stars \citep{Brott2011}, since the reduced mass- and angular momentum-loss \textit{via} winds may allow for rapid rotation rates. However, there is still no observational confirmation for the existence of CHE stars.

The photometry and the assigned spectral types of some of the stars in the blue locus are consistent with them following chemically homogeneous evolution. However, this needs to be confirmed with measurements of high helium 
surface abundances and large rotational velocities \mbox{(\vsini~> 250~\kms)}. Unfortunately, the S/N and the resolution \mbox{($\sim$ 300~\kms)} of the current data set do not allow such analysis.

Alternatively, these sources could be post-binary interaction products, in particular, binary systems whose primary star has been removed of its hydrogen-rich envelope through the interaction with its companion. This loss produces very compact, hot helium stars \citep[e.g. ][]{Kippenhahn1969, Podsiadlowski1992}, known as stripped stars.
Theoretical atmosphere models predict that the spectra of these objects at their high-mass end show He and N lines \citep{Gotberg2017}, similar to the observed in nitrogen-rich Wolf-Rayet stars. Such spectral features have not been detected within our sample. 
However, at \mbox{$Z$ = 1/5 \Zsun}~and intermediate or low mass values, their spectra present typical features of early O-type stars with nitrogen enrichment \citep{Gotberg2018}. Some of our blue outliers exhibit this spectral morphology.
On the other hand, if the companion of the stripped star were to dominate the optical part of the spectrum, we would observe a typical OB spectrum, possible signals of multiplicity and UV and blue photometric excess.
The star would also experience higher rotational velocities after the mass and angular momentum transfer.
We detect \SBCHEStars~binary candidates within the \CHEStars~sources found at bluer colours than the evolutionary tracks.

We have assessed the different features that both CHE stars and systems hosting a stripped star would exhibit.
According to the theoretical predictions, CHE stars are hot and underluminous for their spectral type \citep{Kubatova2019}. They also hold high projected rotational velocities and helium abundances. On the other hand, systems hosting a stripped star would show signs of multiplicity and UV excess. If the stripped star dominates the optical spectrum, it would present a lower luminosity than a star with the same spectral classification \citep{Gotberg2018}.

\begin{figure}
    \centering
    \includegraphics[width=\hsize]{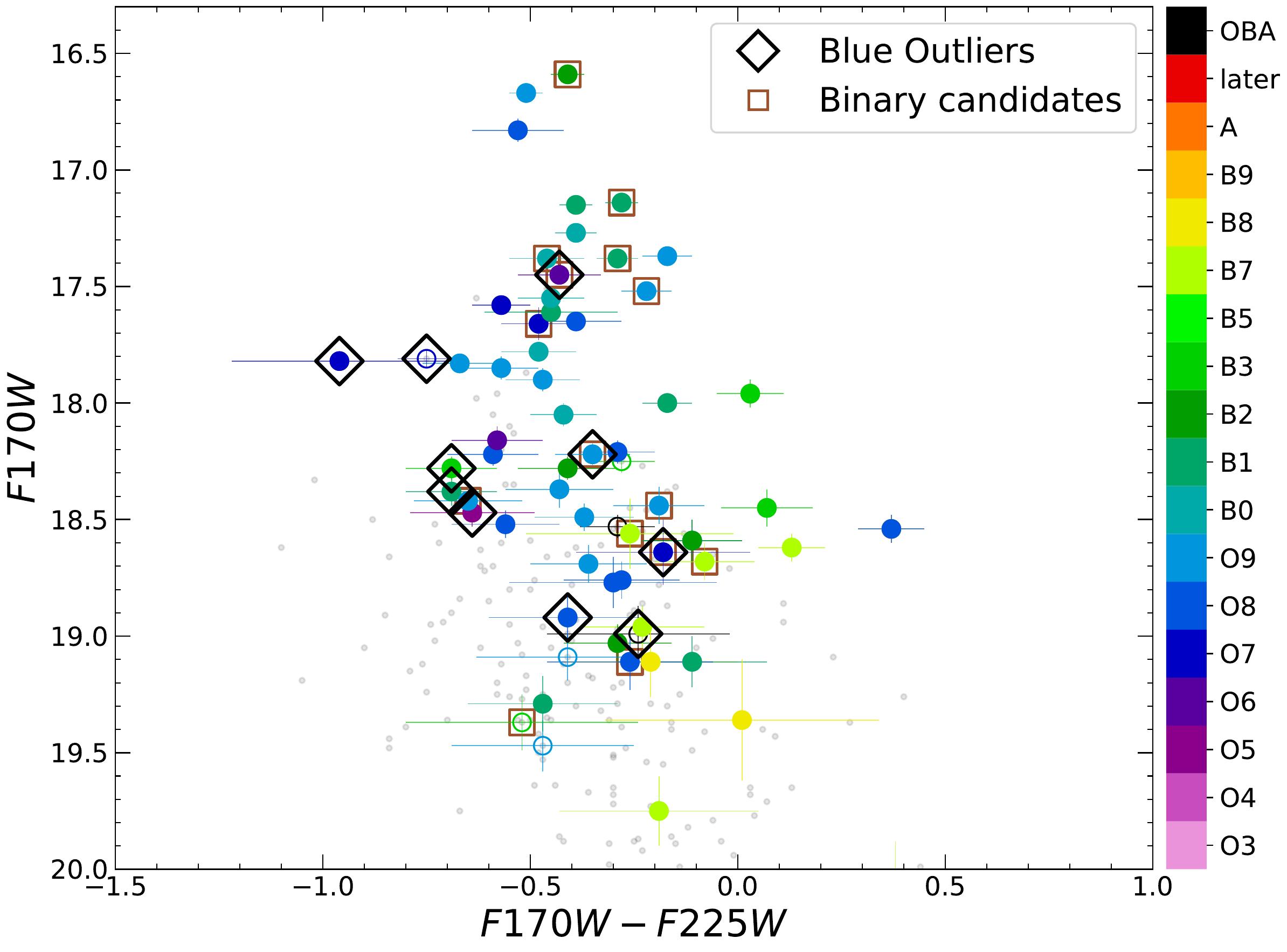}
\caption{UV CMD of Sextans A, displaying \citet{Bianchi2012}'s catalogue in grey and our sample OB stars colour-coded according to their spectral type. Binary candidates are marked with brown squares and the blue outliers with black diamonds.
}
    \label{fig:UVCMD}
\end{figure}
 
With this in mind, we have compiled a selection of properties for the list of blue outliers in Table \ref{tab:CHE}. 
We have analysed the location of these stars in the ultraviolet CMD using photometry by \citet{Bianchi2012} (see Fig. \ref{fig:UVCMD}), noting whether or not they are also blue outliers in this diagram, which would flag the existence of a UV excess.
In addition, we have derived the luminosity of these objects using Sextans~A's distance modulus \mbox{$\mu_0$ = 25.63} \citep{Tammann2011}, $(B-V)_0$ colours estimated from the pseudo-colour $Q$ \citep{Massey2000} and calibrations to estimate effective temperatures and bolometric corrections \citep{Balona1994, Martins2005, Markova2008}.
A luminosity lower than the calibrated values by  \citet{Martins2005, Ramachandran2019} for their spectral type would qualify the objects as candidates for CHE or stripped stars. 
Nine blue outliers are underluminous compared to the calibrations by a difference of $\geq$ 0.2 dex.

\begin{figure}
    \centering
    \includegraphics[width=\hsize]{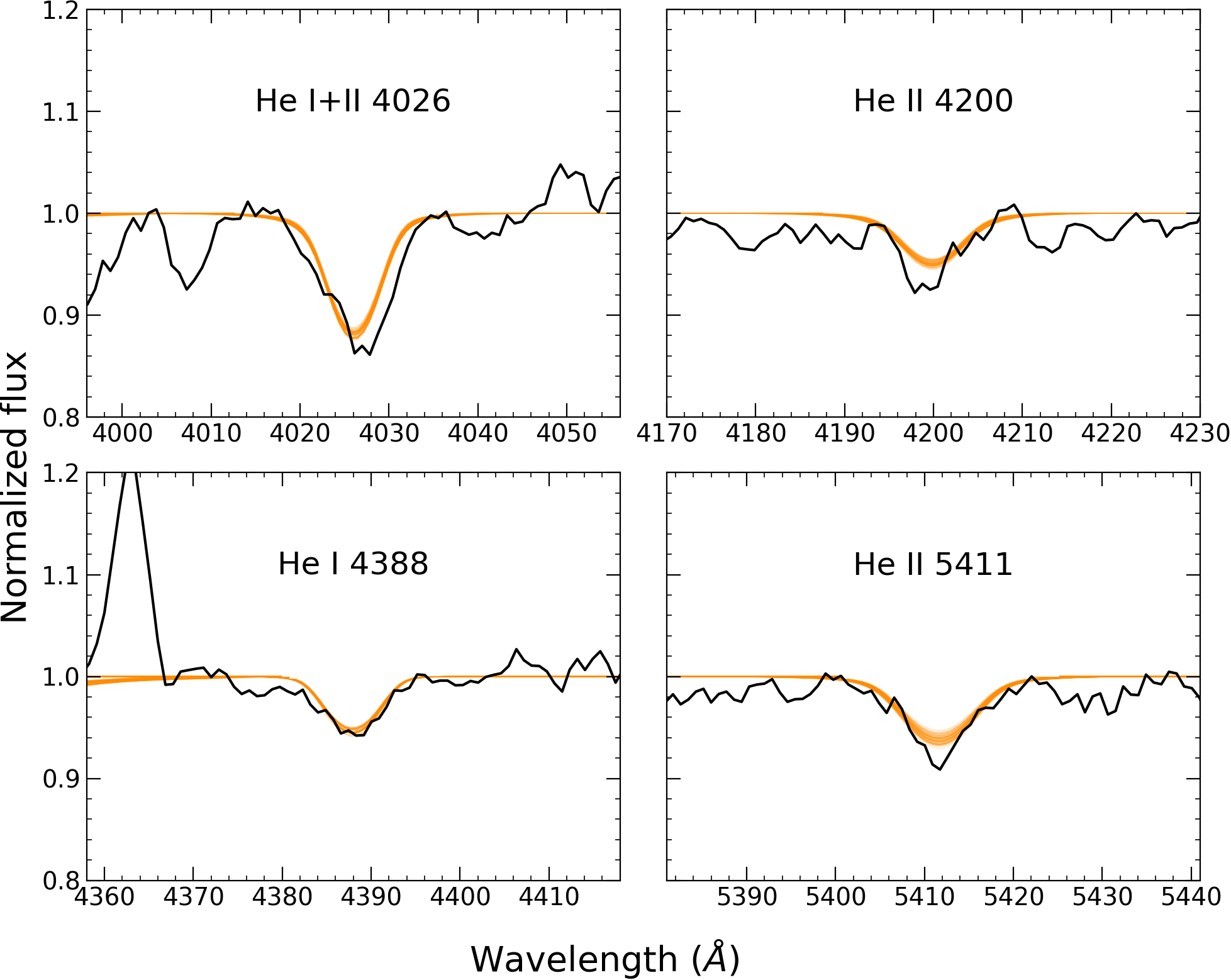}
\caption{Comparison of the spectrum of the blue outlier \Remark{LSS3.OB11202} (O8~Vz), in black and smoothed for clarity, with FASTWIND models compatible with O8-type dwarfs (Lorenzo in prep.), convolved to the resolution of the spectrum (\mbox{R $\sim$ 1300}) and with the projected rotational velocity \vsini~= 300~\kms (in orange). 
}
    \label{fig:CHElines}
\end{figure}

Finally, we have also studied whether the spectra are consistent with high projected rotational velocities to support their designation as CHE candidates or companions to stripped stars. We have used FASTWIND models compatible with the assigned spectral classification of the candidates (Lorenzo et al. in prep.) and helium abundance of \mbox{$\epsilon = 0.09$}, and convolved to the resolution of the respective observations.
Since the S/N and \mbox{$R$ $\sim$ 1000} of our observations prevent us from discerning \vsini~in the range of \mbox{75-300~\kms}, we have checked whether the line profiles of the observed spectra were broader than the synthetic lines with \mbox{\vsini~= 300~\kms}~(see Fig. \ref{fig:CHElines} for an example of the followed procedure). Results of the evaluation are collected in Table \ref{tab:CHE}. 
Six objects in our list of blue outliers present line profiles broader than fast-rotating models. 
We caution, however, that the broadening may also be due to blends or unresolved binary systems. 

We venture a classification for the blue outliers in Table \ref{tab:CHE}, based on their spectral type, luminosity, UV excess, tentative \vsini~and possible companion. Nonetheless, we remark that follow-up observations with higher resolution and S/N are strongly needed to measure rotational velocities and He abundances that may support any of these scenarios.

\begin{table*}
\centering
\caption{List of blue outliers in the CMD of Sextans~A.  Identification codes (ID), spectral types (SpT) and optical photometry from \citet{Massey2007} are provided. We also include the UV photometry from \citet{Bianchi2012} when available and flag the objects that are outliers in the UV CMD. The derived luminosities (calculated as indicated in the text) and the calibrated values for the corresponding spectral type are included. We also mark whether or not the source presents signatures of multiplicity in its spectra and if its line profiles are broader than models with \vsini~> 300 \kms. Armed with this information, we tentatively assign the objects into a category in column \textit{Suggested type}.}
\label{tab:CHE}
\resizebox{\textwidth}{!}{
\begin{tabular}{llccccccccccc}
\toprule
\toprule
 & & \multicolumn{2}{c|}{Optical phot.} & \multicolumn{4}{c|}{UV phot.} & & & \\ \cmidrule(lr){3-4} \cmidrule(lr){5-8}
  ID &             SpT &    $V$ &  $B-V$ & \multicolumn{1}{c|}{\begin{tabular}[c]{@{}c@{}}UV\\detection\end{tabular}} &  F170W & F170W - F225W & \multicolumn{1}{c|}{\begin{tabular}[c]{@{}c@{}}UV\\outlier\end{tabular}} & \multicolumn{1}{c|}{\begin{tabular}[c]{@{}c@{}}log (L/L$_{\mathbf{\odot}}$)\\This work\end{tabular}} & \multicolumn{1}{c|}{\begin{tabular}[c]{@{}c@{}}log (L/L$_{\mathbf{\odot}}$)\\Calibrated\end{tabular}} & \multicolumn{1}{c|}{\begin{tabular}[c]{@{}c@{}}Binary\\cand.\end{tabular}} & \multicolumn{1}{c|}{\begin{tabular}[c]{@{}c@{}}Broad\\lines\end{tabular}} & \multicolumn{1}{c|}{\begin{tabular}[c]{@{}c@{}}Suggested\\type\end{tabular}} \\
\midrule
s002 &          O4 Vz  & 20.369 & -0.279 &                                                                    no info &      - &             - &                                                                        - &                                                                                                 5.60 &                                                                                                  5.67 &                                                                         no &                                                                        no &                                                                            ? \\
s004 &         O5 III  & 20.917 & -0.277 &                                                                    no info &      - &             - &                                                                        - &                                                                                                 5.34 &                                                                                                  5.73 &                                                                        yes &                                                                        no &                                                                Stripped star \\
s005 &     O5 V((fc))  & 21.751 & -0.364 &                                                                        yes &  18.47 &         -0.64 &                                                                      yes &                                                                                                 4.83 &                                                                                                  5.49 &                                                                         no &                                                                        no &                                                                Stripped star \\
s007 &           O6 V  & 20.515 & -0.280 &                                                                        yes &  17.45 &         -0.43 &                                                                      yes &                                                                                                 5.44 &                                                                                                  5.32 &                                                                        yes &                                                                       yes &                                                    CHE or OB + stripped star \\
s009 &           O7 V  & 20.659 & -0.285 &                                                                        yes &  17.82 &         -0.96 &                                                                      yes &                                                                                                 5.30 &                                                                                                  5.14 &                                                                         no &                                                                        no &                                                           OB + stripped star \\
s013 &       O7.5 III  & 21.532 & -0.295 &                                                                    no info &      - &             - &                                                                        - &                                                                                                 4.87 &                                                                                                  5.42 &                                                                         no &                                                                       yes &                                                                          CHE \\
s015 &     O7.5 V + B? & 22.259 & -0.410 &                                                                        yes &  18.64 &         -0.18 &                                                                       no &                                                                                                 4.50 &                                                                                                  5.05 &                                                                        yes &                                                                        no &                                                                Stripped star \\
s018 &         O8 III  & 21.092 & -0.321 &                                                                    no info &      - &             - &                                                                        - &                                                                                                 4.97 &                                                                                                  5.35 &                                                                         no &                                                                       yes &                                                                          CHE \\
s023 &           O8 V  & 20.288 & -0.299 &                                                                    no info &      - &             - &                                                                        - &                                                                                                 5.36 &                                                                                                  4.96 &                                                                         no &                                                                       yes &                                                           OB + stripped star \\
s030 &       O8.5 III  & 21.008 & -0.322 &                                                                    no info &      - &             - &                                                                        - &                                                                                                 4.94 &                                                                                                  5.28 &                                                                         no &                                                                       yes &                                                                          CHE \\
s032 &         O8.5 V  & 21.216 & -0.326 &                                                                    no info &      - &             - &                                                                        - &                                                                                                 4.86 &                                                                                                  4.86 &                                                                         no &                                                                       yes &                                                                          CHE \\
s033 &         O8.5 V  & 21.654 & -0.271 &                                                                        yes &  18.92 &         -0.41 &                                                                       no &                                                                                                 4.83 &                                                                                                  4.86 &                                                                         no &                                                                        no &                                                                            ? \\
s040 & O9 III + mid-B? & 21.285 & -0.311 &                                                                        yes &  18.22 &         -0.35 &                                                                       no &                                                                                                 4.83 &                                                                                                  5.21 &                                                                        yes &                                                                         ? &                                                                            ? \\
s063 &         B0 III  & 20.727 & -0.295 &                                                                        yes &      - &             - &                                                                        - &                                                                                                 4.95 &                                                                                                  4.98 &                                                                         no &                                                                        no &                                                                            ? \\
s080 &         B1 III  & 21.471 & -0.292 &                                                                        yes &  18.38 &         -0.69 &                                                                      yes &                                                                                                 4.50 &                                                                                                  4.72 &                                                                         no &                                                                        no &                                                                            ? \\
s101 &         B3 III  & 21.254 & -0.328 &                                                                        yes &  18.28 &         -0.69 &                                                                      yes &                                                                                                 4.18 &                                                                                                  3.90 &                                                                         no &                                                                        no &                                                                            ? \\
s129 &       O V + neb & 21.237 & -0.310 &                                                                        yes &  17.81 &         -0.75 &                                                                      yes &                                                                                                 3.85 &                                                                                                  4.68 &                                                                         no &                                                                         ? &                                                                Stripped star \\
s164 &           OBA   & 22.253 & -0.436 &                                                                        yes &  18.99 &         -0.24 &                                                                       no &                                                                                                 3.13 &                                                                                                     - &                                                                         no &                                                                         ? &                                                                            ? \\
\bottomrule
\end{tabular}
}
\end{table*}

\section{Summary and conclusions}
\label{sec:conclusions}
Very low metallicity massive stars hold the key to interpreting processes from the earliest times of the Universe, such as the re-ionization epoch and early cosmic chemical enrichment. Their low metal content may lead to significant changes in their evolution, some of them with an extreme impact on their ionizing fluxes like the chemically homogeneous evolution channel. 
In contrast with the extensive studies in the Magellanic Clouds \citep[e.g. ][]{Ramirez-Agudelo2017, Sabin-Sanjulian2017, Ramachandran2019}, current references for the metal-poor Universe, massive stars at sub-SMC metallicities were very poorly studied. Building an extensive census was a fundamental first step to understanding the physics and evolution of these objects, and to ultimately extrapolating the properties of the first, metal-free stars.

In this work, we present the results of a spectroscopic survey of massive stars in the 1/10 \Zsun~galaxy Sextans~A built with observations from the 10.4-m telescope GTC. Our \OBStars~OB star catalogue is the largest census of massive stars at metallicities lower than the SMC to date.
We also provide the first list of candidate stars undergoing chemically homogeneous evolution or systems hosting a stripped star. Confirming even one single case of these scenarios will have critical implications for the stellar evolution theories and the interpretation of high-redshift galaxies. 
In addition, we identify \SBStars~sources within the sample displaying signs of multiplicity in their spectra.

Most of the sample OB stars are located in rich H~{\sc i} density regions producing large or small bubbles of ionized gas.
However, we also find massive stars in inconspicuous areas of the galaxy where the column density of the neutral gas is low. This was already pointed out by \citet{Garcia2019} in the south of Sextans~A, but we also detect them in the gas-void galactic centre and other external areas of the galaxy. Our extensive spectroscopic sample and future detailed maps of gas will enable disentangling the roles of neutral and molecular gas in star formation at very low metallicity and identify the mechanisms that have triggered star formation in unexpected regions.

We find high scatter in the colour-magnitude diagram, hinting that the internal extinction of Sextans~A is significant and non-uniform. Mapping the local reddening and estimating the stellar parameters and abundances of the massive stars of our catalogue by quantitative spectroscopic analyses 
is left for future work.

We have also assembled a sample of \BAI~BA supergiants that, with further quantitative analyses, will yield the abundances of $\alpha$- and Fe-group elements. Studying the chemical composition of blue stars 
will help us investigate the possible existence of chemical abundance variations across the galaxy and will shed light on the chemodynamical evolution of Sextans~A.

In addition, we observed \LateStarsSexA~late-type stars in the galaxy. Enhancing the statistics of evolved massive stars in LG galaxies allows us to quantify 
the role of metallicity in evolution.

This work presents the first extensive sample of sub-SMC metallicity massive stars. This catalogue is a fundamental first step for high-quality, high-resolution spectroscopic observations that will enable a thorough characterisation of stellar and wind properties of very metal-poor massive stars. These, in turn, will constrain stellar evolutionary models leading to more realistic predictions of stellar feedback for these stars, which is fundamental to interpreting observations of high-redshift galaxies.

\section*{Acknowledgements}
M. Lorenzo, M. Garcia and F. Najarro gratefully acknowledge support by grants PID2019-105552RB-C41 and MDM-2017-0737 Unidad de Excelencia "María de Maeztu"-Centro de Astrobiología (CSIC-INTA), funded by MCIN/AEI/10.13039/501100011033. M. Lorenzo ackowledges funding from grant PRE2019-087988 under project MDM-2017-0737-19-3, and by “ESF Investing in your future".
A. Herrero acknowledges support by the Spanish MCI through grant PGC-2018-0913741-B-C22 and the Severo Ochoa Program through \mbox{CEX2019-000920-S}.
M. Cerviño acknowledges financial support from the Spanish MCI through grant PID2019-107408GB-C41.
N. Castro gratefully acknowledges funding from the Deutsche Forschungsgemeinschaft \mbox{(DFG) – CA 2551/1-1}.

The work has made use of the \textsc{gtcmos} pipeline for the reduction of the GTC/OSIRIS spectroscopic data for which we thank its author Divakara Mayya. It also has made use of "Aladin sky atlas" developed at CDS, Strasbourg Observatory, France (\citealp{Bonnarel2000, Boch2014}); SIMBAD database, operated at CDS, Strasbourg, France
\citep{Wenger2000}; and NASA's Astrophysics Data System Bibliographic Services.

Finally, we want to thank our colleagues at the \textit{IAU Symposium 361: Massive Stars Near and Far} for their insightful questions and comments.

\section*{Data Availability}

The data underlying this article will be shared on reasonable request to the corresponding author.


\bibliographystyle{mnras}
\bibliography{BB_bibliography}


\newpage
\newpage
\newpage
\pagebreak[3]

\appendix
\section*{Appendix}


\section{Observation logs}
\label{sec:appx_logs}
This section collects the logs of our four observation campaigns, including the date and weather conditions and total exposure times (Tables \ref{table:mos1_log}-\ref{table:LSS3_log}). In addition, we provide a list of the observing blocks and super-observing blocks coadded to generate the final spectrum of the stars observed in the MOS1 run (Table \ref{tab:OBosOB}).

\begin{table*}
\caption{MOS1 observing log.}
\centering
\begin{tabular}{ccccccc}
\hline
\hline
\begin{tabular}[c]{@{}c@{}} Observing \\ block \end{tabular} &
Date & Seeing ('') & Transparency & Moon & Airmass & Exposure time (s)\\
\hline
OB01 (pre-image) & 2 Apr 2014 & 0.8 - 1.0 & Clear & Grey & 1.29 & - \\
OB02 & 16 Jan 2015 & 1.2 & Clear & Dark & 1.20 & 2950 \\
OB03 & 3 May 2014 & 1.1 & Photometric & Dark & 1.24 & 2950  \\
OB04 & 3 May 2014 & 1.2 & Photometric & Dark & 1.37 & 2950  \\
OB05 & 4 May 2014 & 0.9 & Clear & Grey & 1.57 & 2950  \\
OB06 & 5 May 2014 & 1.0 & Clear & Grey & 1.27 & 2950  \\
OB07 & 5 May 2014 & 1.0 & Clear & Grey & 1.42 & 2950  \\
OB08 & 5 May 2014 & 0.9 & Clouds & Grey & 1.72 & 2950  \\
OB09 & 6 May 2014 & 1.0 & Clear & Grey & 1.29 & 2950  \\
OB10 & 16 Jan 2015 & 1.0 & Clear & Dark & 1.26 & 2950  \\
OB11 & 16 Jan 2015 & 0.9 & Clear & Dark & 1.39 & 2950  \\
OB12 & 18 Jan 2015 & 1.1 & Clear & Dark & 1.20 & 2950  \\
OB13 & 18 Jan 2015 & 0.8 & Clear & Dark & 1.25 & 2950  \\
OB14 & 18 Jan 2015 & 0.8 & Clear & Dark & 1.39 & 2950  \\ 
\hline
\end{tabular}
\label{table:mos1_log}
\end{table*}

\begin{table*}
\centering
\caption{MOS2 observing log.}
\label{table:mos2_log}
\begin{tabular}{ccccccc}
\hline
\hline
\begin{tabular}[c]{@{}c@{}} Observing \\ block \end{tabular}  &
Date & Seeing ('') & Transparency & Moon & Airmass & Exposure time (s)\\
\hline
Night-1 & 16 Feb 2018 & 0.9 & Clear & Dark & 1.20 & 12600 \\
Night-2 & 17 Feb 2018 & 1.2 & Clear & Dark & 1.20 & 9000 \\
\hline
\end{tabular}

\end{table*}

\begin{table*}
\centering
\caption{LSS2 observing log.}
\label{table:lss2_log}
\begin{tabular}{ccccccccc}
\hline
\hline
\begin{tabular}[c]{@{}c@{}} Observing \\ block \end{tabular} & Date & Seeing ('') & Slit-width ('') & Transparency & Moon & Airmass & Exposure time (s)\\
\hline
OB01 & 18 Feb 2020 & 1.0 & 1.2 & Clear & Dark & 1.23 & 2806\\
OB02 & 18 Feb 2020 & 1.0 & 1.2 & Clear & Dark & 1.20 & 2806\\
OB03 & 18 Feb 2020 & 1.3 & 1.2 & Clear & Dark & 1.28 & 2806\\
OB04 & 18 Mar 2020 & 0.9 & 1.2 & Phot. & Dark & 1.68 & 2806\\
\hline
\end{tabular}

\end{table*}
\begin{table*}
\caption{LSS3 observing log.}
\centering
\begin{tabular}{cccccccc}
\hline
\hline
\begin{tabular}[c]{@{}c@{}} Observing \\ block \end{tabular} &
Date & Seeing ('') & Slit-width ('') & Transparency & Moon & Airmass & Exposure time (s) \\
\hline
OB01 & 16 Jan 2021 & 1.0 - 1.2 & 1.2 & Photometric & Dark & 1.25 & 3600 \\
OB06 & 16 Jan 2021 & 1.0 - 1.2 & 1.2 & Photometric & Dark & 1.20 & 3600 \\
OB04 & 16 Jan 2021 & 1.3 & 1.2 & Photometric & Dark & 1.25  & 3600 \\
OB05 & 16 Jan 2021 & 1.4 & 1.5 & Photometric & Dark & 1.41 & 3600 \\

OB03 & 17 Jan 2021 & 0.8 & 0.8 & Photometric & Dark & 1.30 & 3788 \\
OB02 & 17 Jan 2021 & 0.75 & 0.8 & Photometric & Dark & 1.20 & 3788 \\
OB13 & 17 Jan 2021 & 0.8 & 1.0 & Photometric & Dark & 1.25 & 6846 \\
OB17 & 17 Jan 2021 & 0.8 & 0.8 & Photometric & Dark & 1.75 & 5055 \\

OB18 & 18 Jan 2021 & 0.8 & 0.8 & Photometric & Dark & 1.48 & 3804 \\
OB15 & 18 Jan 2021 & 0.8 & 0.8 & Photometric & Dark & 1.21 & 6846 \\
OB11 & 18 Jan 2021 & 0.8-1.0 & 1.0 & Photometric & Dark & 1.29 & 6846 \\

OB44 & 19 Jan 2021 & 1.2-1.5 & 1.5 & Photometric & Dark & 1.25 & 4388 \\
OB31 & 19 Jan 2021 & 1.5 & 1.2 & Photometric & Dark & 1.20 & 6225 \\

OB45$^*$ & 9 Mar 2021 & 1.3-1.4 & 1.2 & Photometric & Dark & 1.22 & 7125 \\
OB46$^*$ & 10 Mar 2021 & 1.3-1.5 & 1.2 & Photometric & Dark & 1.50 & 7125 \\
OB47$^*$ & 9 Mar 2021 & 1.3-1.5 & 1.2 & Photometric & Dark & 1.30 & 2883 \\
OB49$^*$ & 10 Mar 2021 & 1.3-1.5 & 1.2 & Photometric & Dark & 1.21 & 7125\\
OB50$^*$ & 10 Mar 2021 & 1.3-1.5 & 1.2 & Photometric & Dark & 1.35 & 3704\\

\hline
\end{tabular}
\label{table:LSS3_log}

\begin{tablenotes}[flushleft]
    \item \small \textbf{Notes.} $^{(*)}$ observing blocks performed in service mode.
\end{tablenotes}


\end{table*}

\begin{table*}
\centering
\caption{List of the observing blocks and super-observing blocks coadded to generate the final spectrum of each star observed in the MOS1 run.}
\begin{tabular}{ccl}
\hline
\hline
ID & Exposure time (s) & Exposures\\ \hline
\Remark{MOS1.s02} & 26550 & OB02, sOB03(=OB03+OB04), OB05, OB06, OB10, sOB13(= OB12+OB13+OB14) \\
\Remark{MOS1.s04} & 26550 & OB02, OB03, OB04, OB05, OB06, OB07, OB10, OB12, OB13 \\
\Remark{MOS1.s06} & 26550 & OB02, OB03, OB05, OB06, OB07, OB10, OB12, OB13, OB14 \\
\Remark{MOS1.s03} & 26550 & OB02, OB03, OB04, OB05, OB06, OB07, OB10, OB12, OB13 \\
\Remark{MOS1.s01} & 26550 & OB02, OB03, OB04, OB05, OB06, OB07, OB10, OB12, OB13\\
\Remark{MOS1.s07} & 23600 & OB02, OB03, OB04, OB06, OB10, OB12, OB13, OB14 \\
\Remark{MOS1.s11} & 32450 & OB02, OB03, OB04, OB05, OB06, OB07, OB10, OB11, OB12, OB13, OB14 \\
\Remark{MOS1.s13} & 20650 & OB02, OB03, OB04, OB10, OB11, OB12, OB13 \\
\Remark{MOS1.s08} & 29500 & OB02, OB03, OB04, OB05, OB06, OB07, OB10, OB12, OB13, OB14 \\
\Remark{MOS1.s14} & 29500 & OB02, OB03, OB04, OB06, OB07, OB10, OB11, OB12, OB13, OB14 \\
\Remark{MOS1.s10} & 32450 & OB02, OB03, OB04, OB05, OB06, OB07, OB09, OB10, OB12, OB13, OB14 \\
\Remark{MOS1.s12} & 26650 & OB02, sOB03(= OB03+OB04), OB05, OB06,OB10, sOB13(= OB12+OB13+OB14) \\
\hline
\end{tabular}
\label{tab:OBosOB}
\end{table*}


\section{Spectra}
\label{sec:appx_Spectra}

In this section, we provide the GTC-OSIRIS spectra of all our targets. We first show the spectra for the blue massive stars with a full classification (corresponding to Table \ref{tab:cat_OB_highQ}), then for the blue massive stars with no defined spectral subtype (Table \ref{tab:cat_OB_lowQ}) and finally for the late type stars (Table \ref{tab:cat_late}).

\cleardoublepage

\begin{figure*}\includegraphics[width=\textwidth]{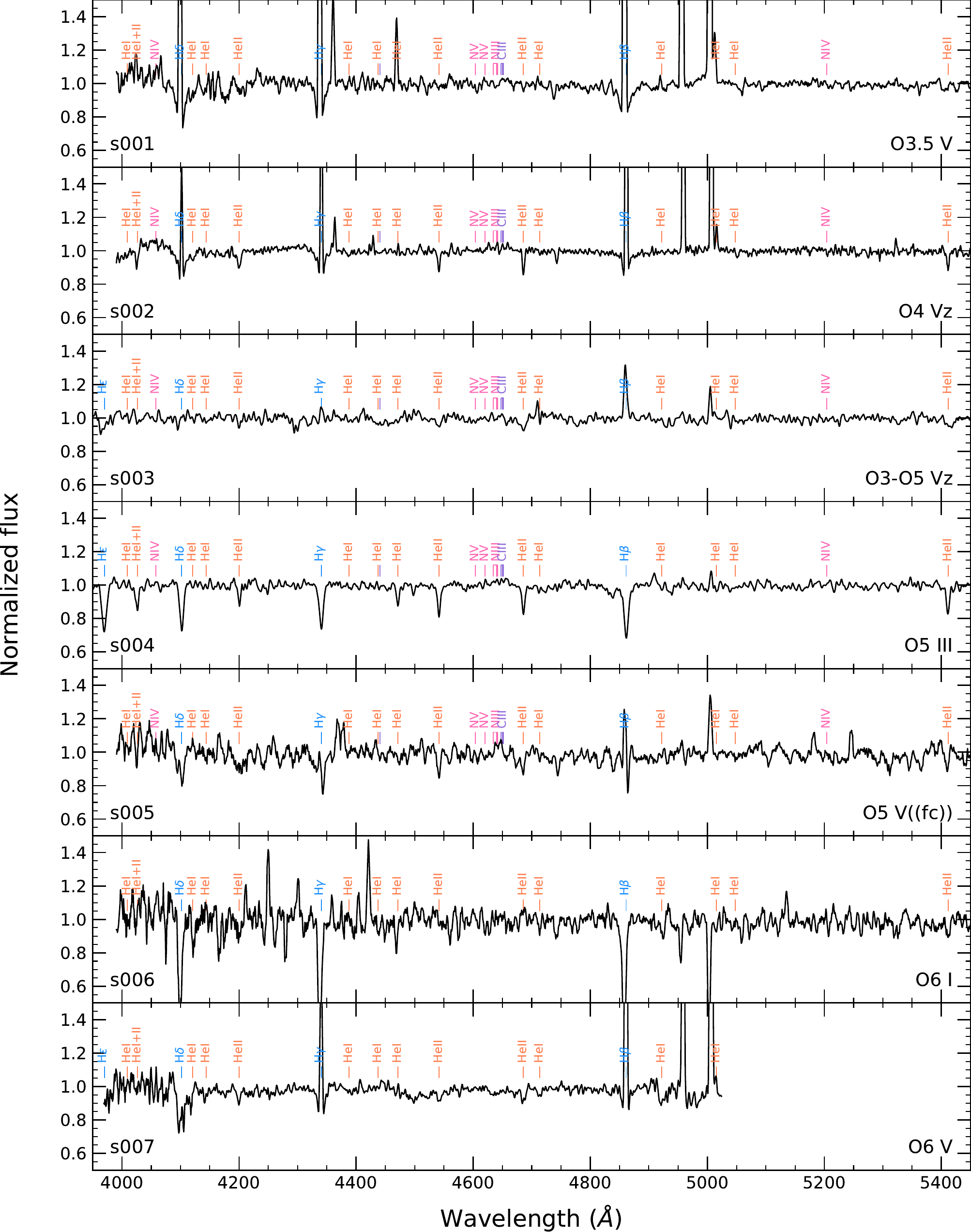}\caption{GTC–OSIRIS spectra of the blue massive stars of our catalogue with complete spectral classification. The data have been corrected by heliocentric and radial velocity, and smoothed for clarity based on the S/N of the spectrum.}\label{fig:OB_highQ_0}\end{figure*}

\begin{figure*}\includegraphics[width=\textwidth]{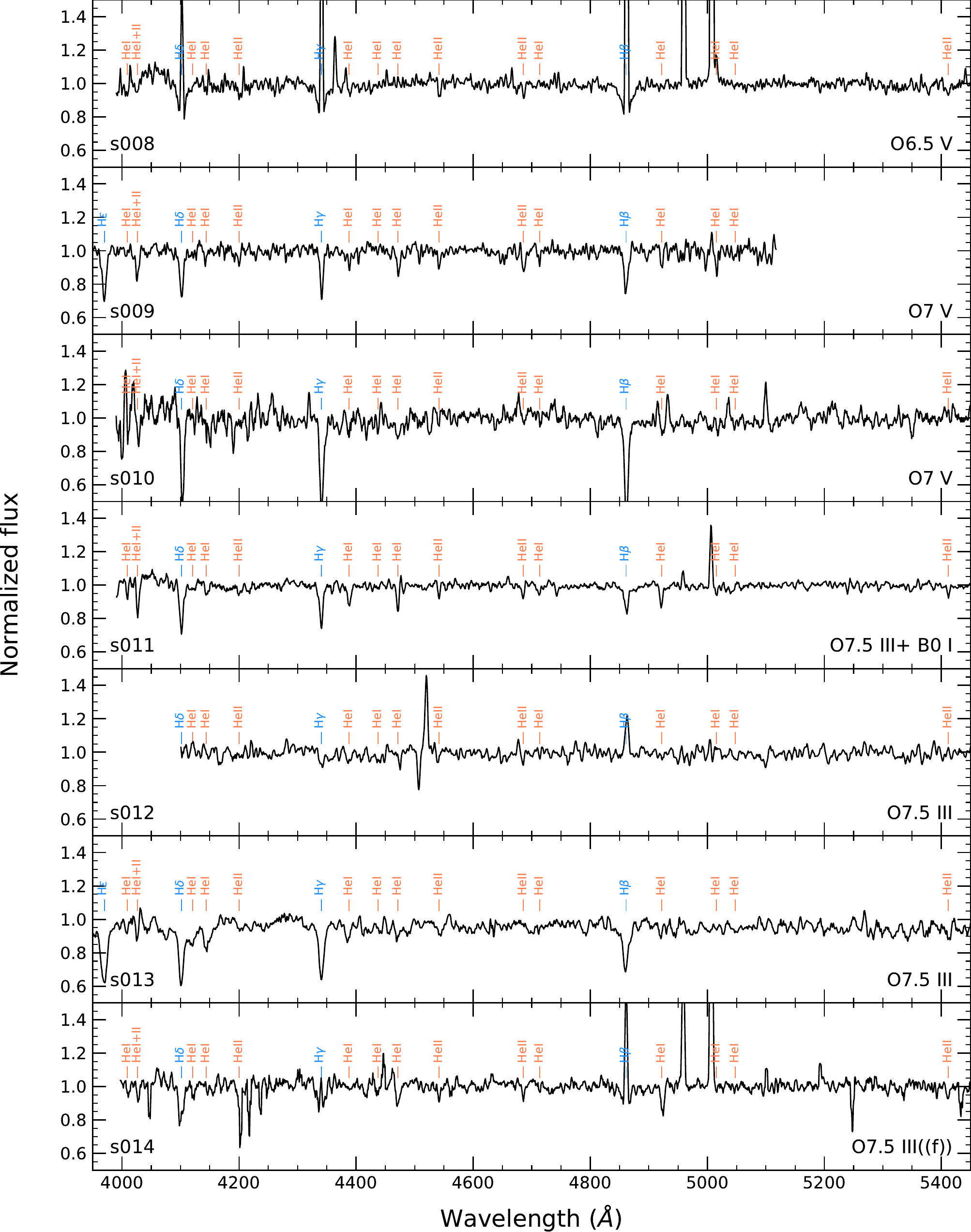}\caption{Same as Figure \ref{fig:OB_highQ_0}, continued.}\label{fig:OB_highQ_1}\end{figure*}

\begin{figure*}\includegraphics[width=\textwidth]{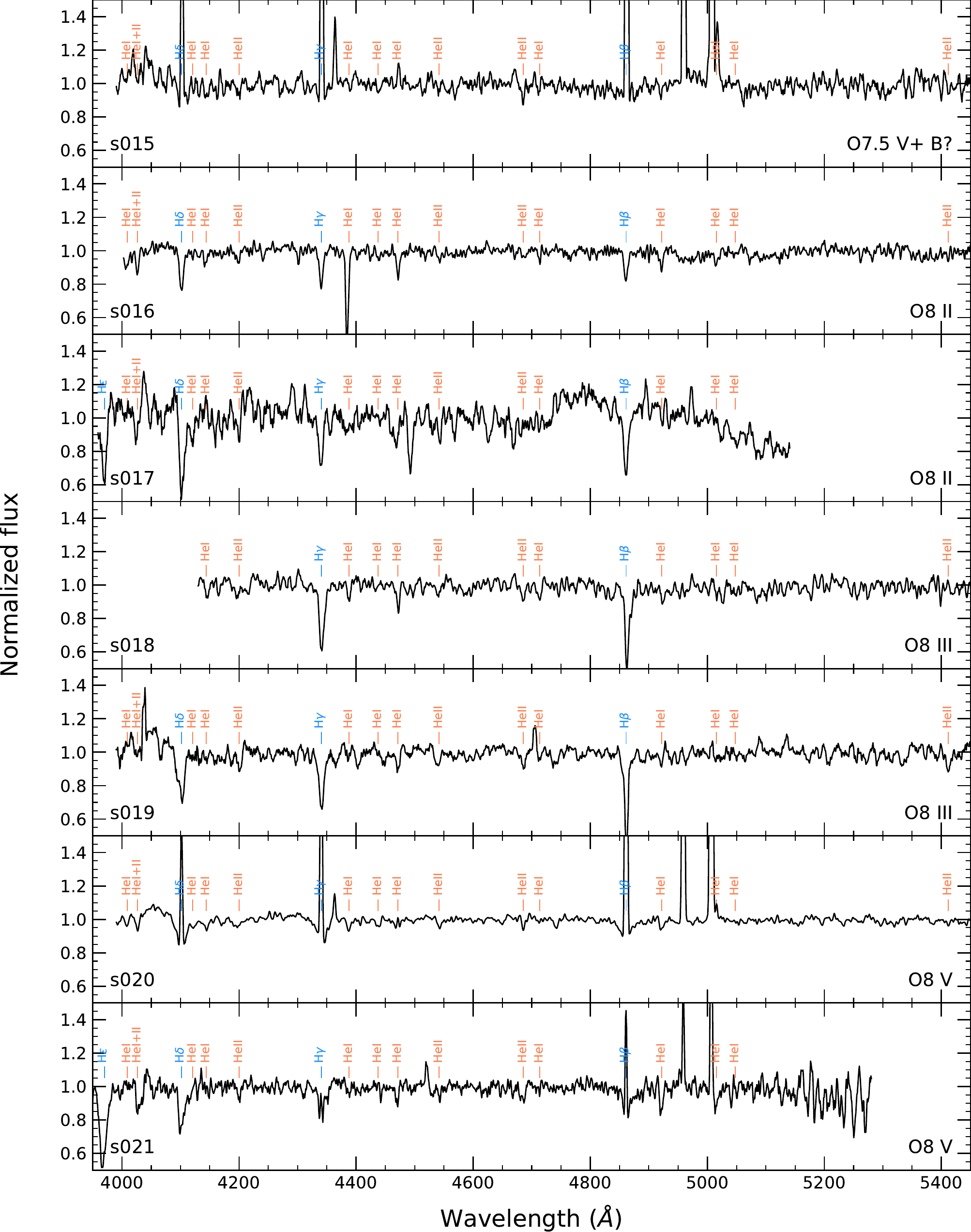}\caption{Same as Figure \ref{fig:OB_highQ_0}, continued.}\label{fig:OB_highQ_2}\end{figure*}

\begin{figure*}\includegraphics[width=\textwidth]{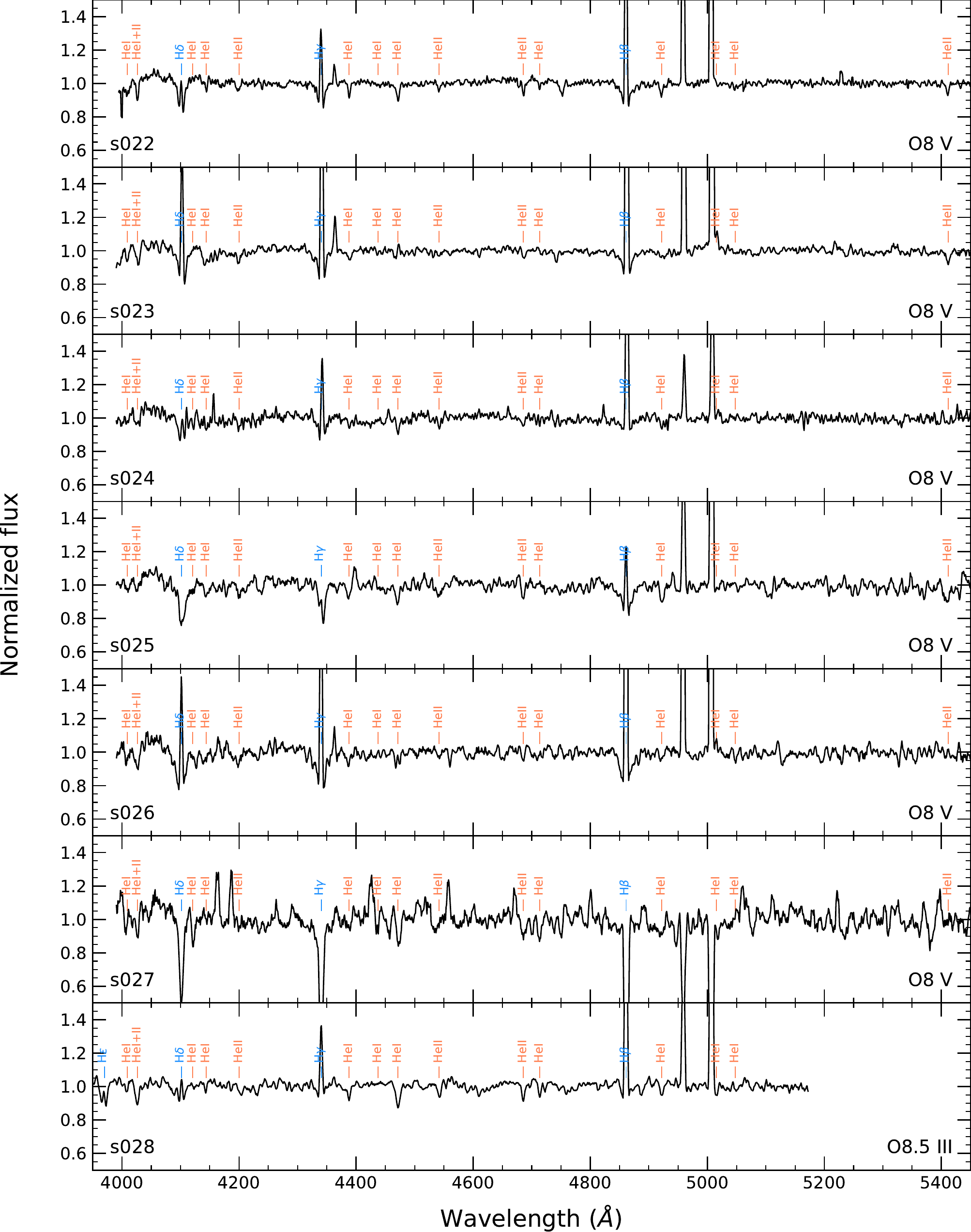}\caption{Same as Figure \ref{fig:OB_highQ_0}, continued.}\label{fig:OB_highQ_3}\end{figure*}

\begin{figure*}\includegraphics[width=\textwidth]{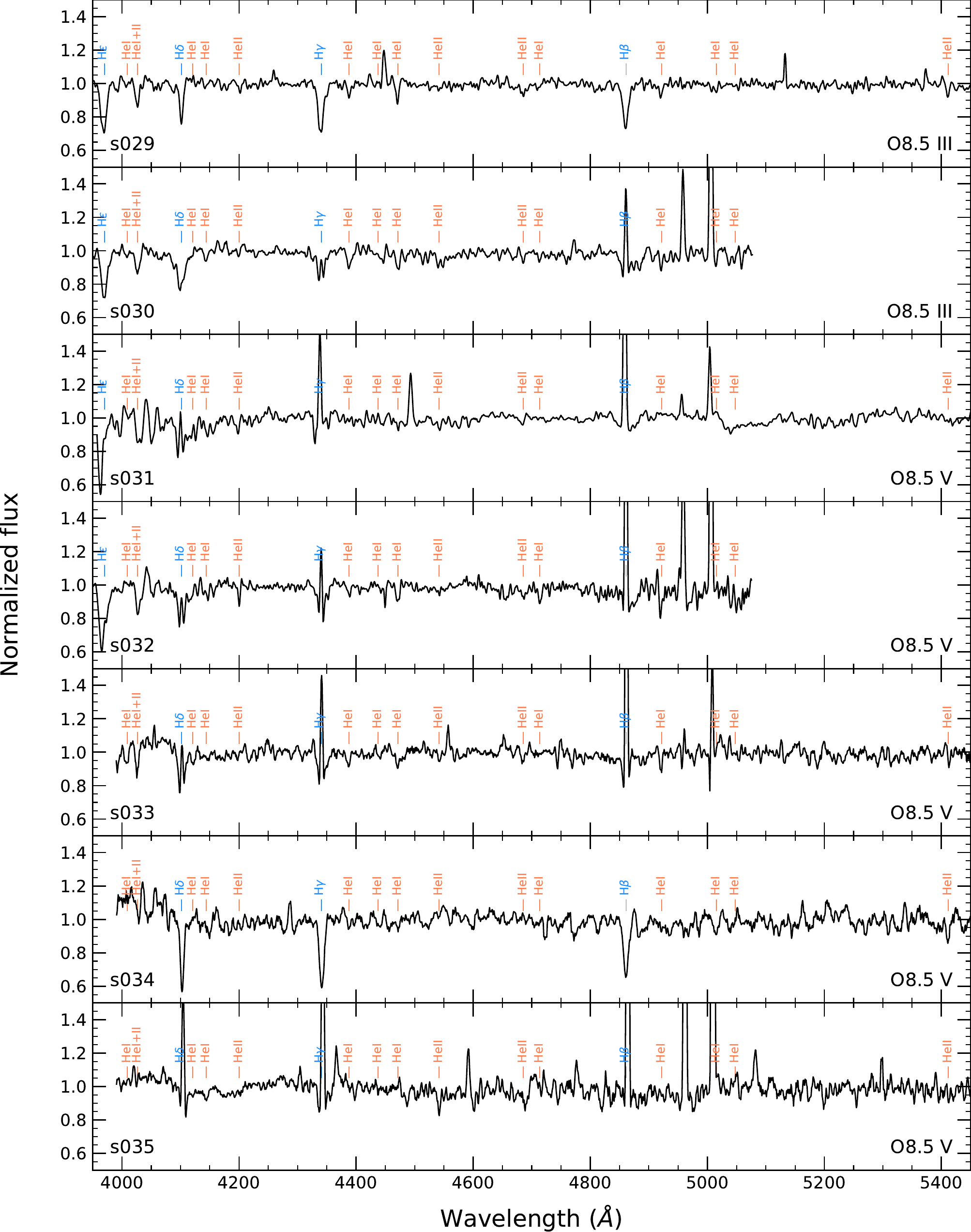}\caption{Same as Figure \ref{fig:OB_highQ_0}, continued.}\label{fig:OB_highQ_4}\end{figure*}

\begin{figure*}\includegraphics[width=\textwidth]{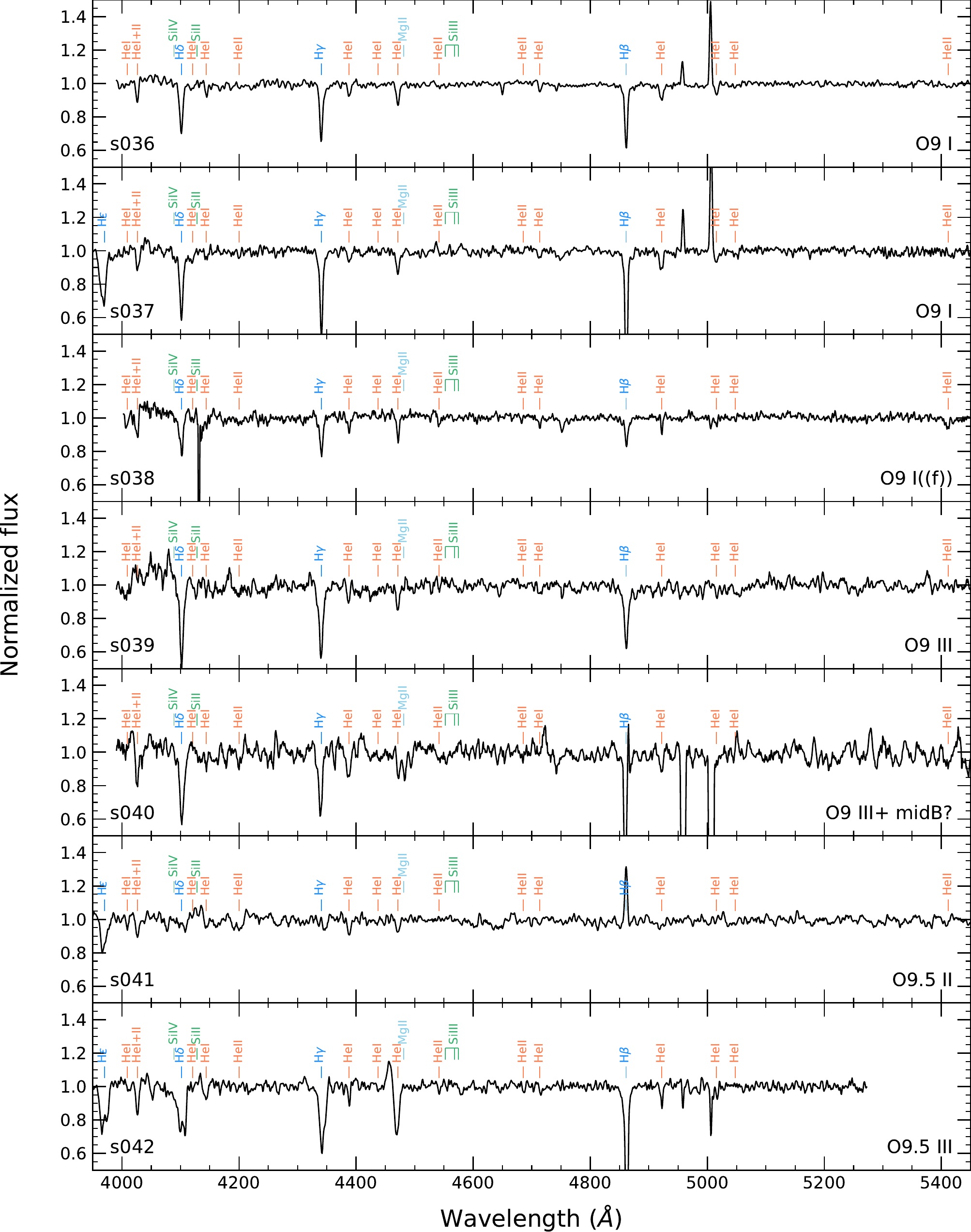}\caption{Same as Figure \ref{fig:OB_highQ_0}, continued.}\label{fig:OB_highQ_5}\end{figure*}

\begin{figure*}\includegraphics[width=\textwidth]{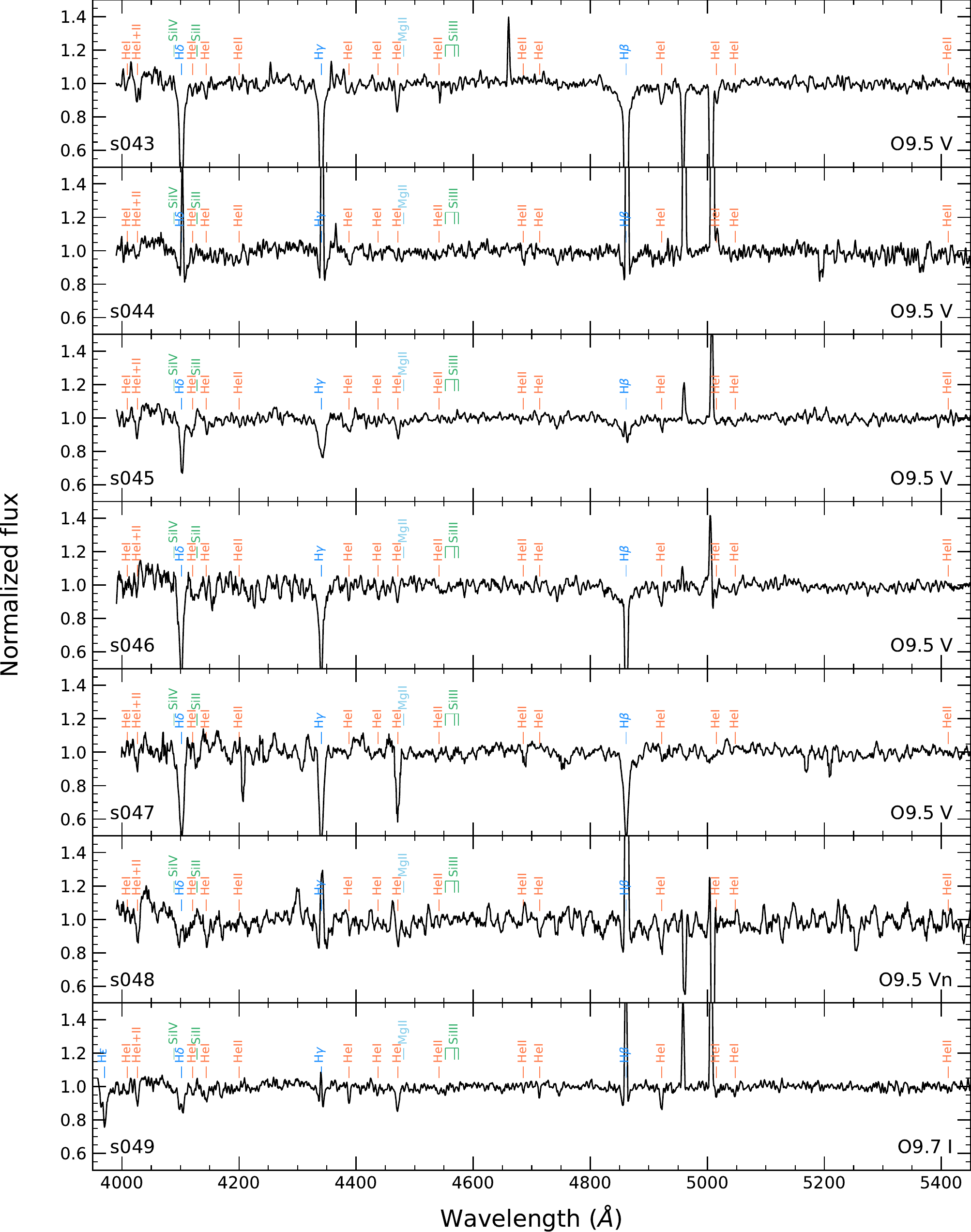}\caption{Same as Figure \ref{fig:OB_highQ_0}, continued.}\label{fig:OB_highQ_6}\end{figure*}

\begin{figure*}\includegraphics[width=\textwidth]{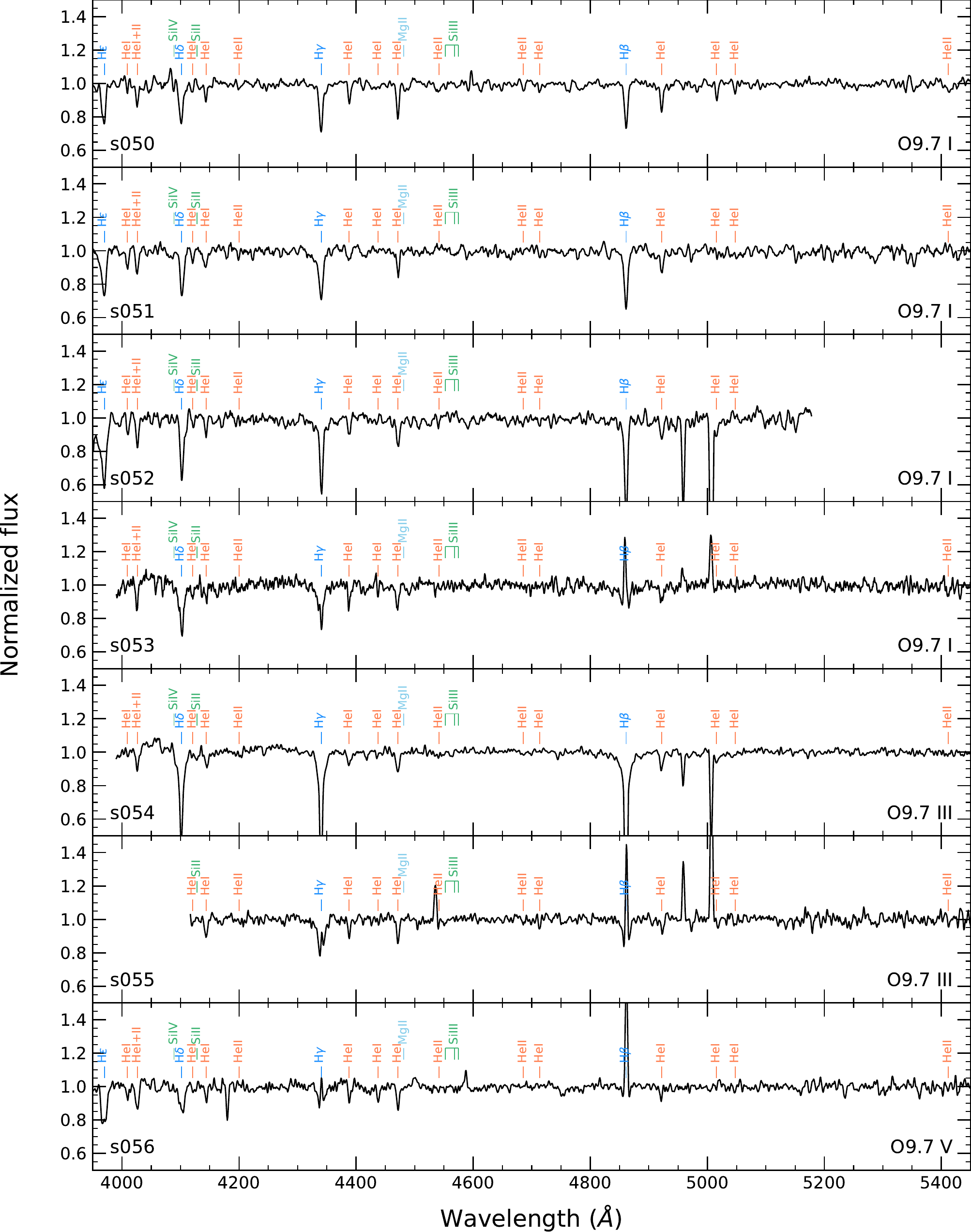}\caption{Same as Figure \ref{fig:OB_highQ_0}, continued.}\label{fig:OB_highQ_7}\end{figure*}

\begin{figure*}\includegraphics[width=\textwidth]{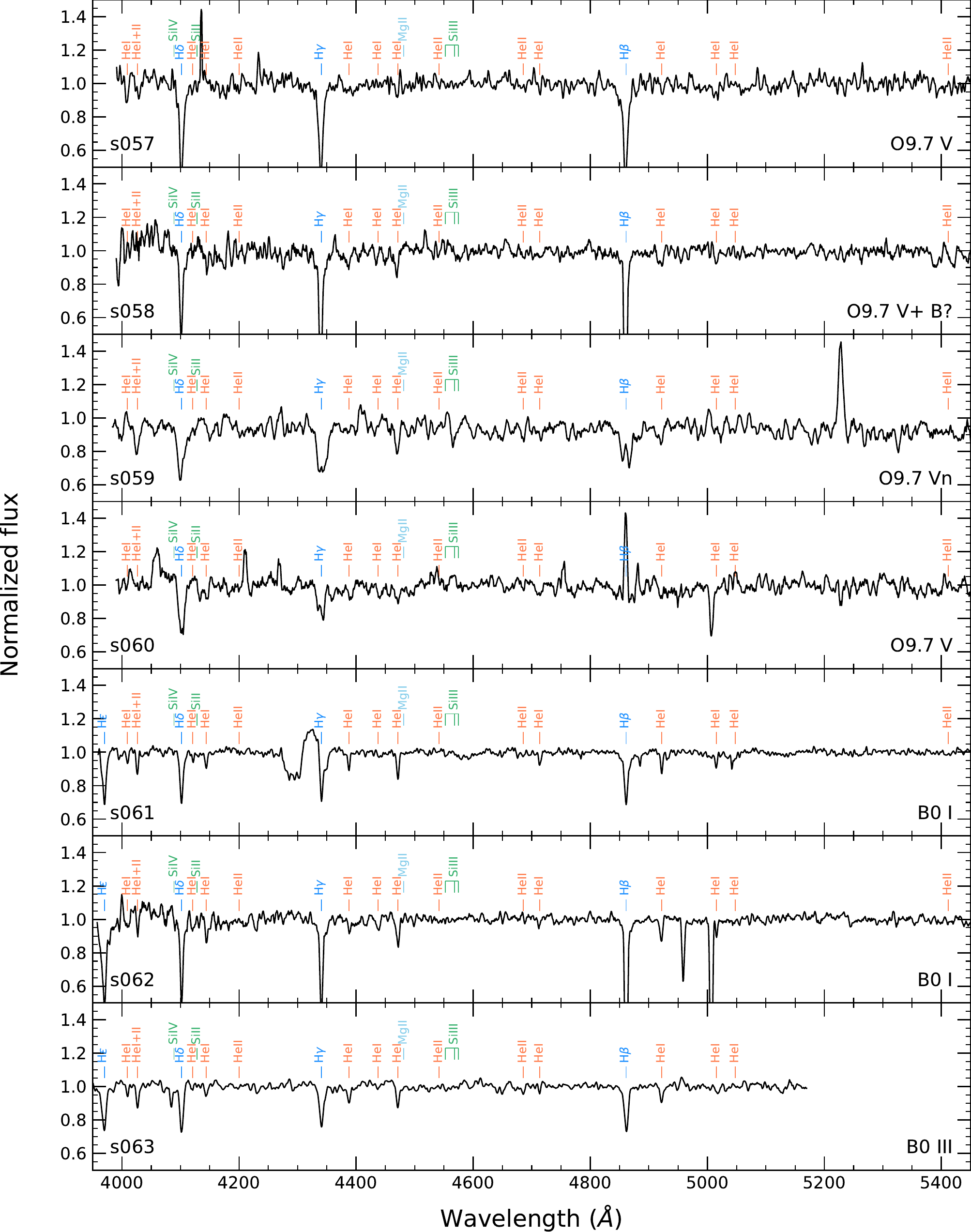}\caption{Same as Figure \ref{fig:OB_highQ_0}, continued.}\label{fig:OB_highQ_8}\end{figure*}

\begin{figure*}\includegraphics[width=\textwidth]{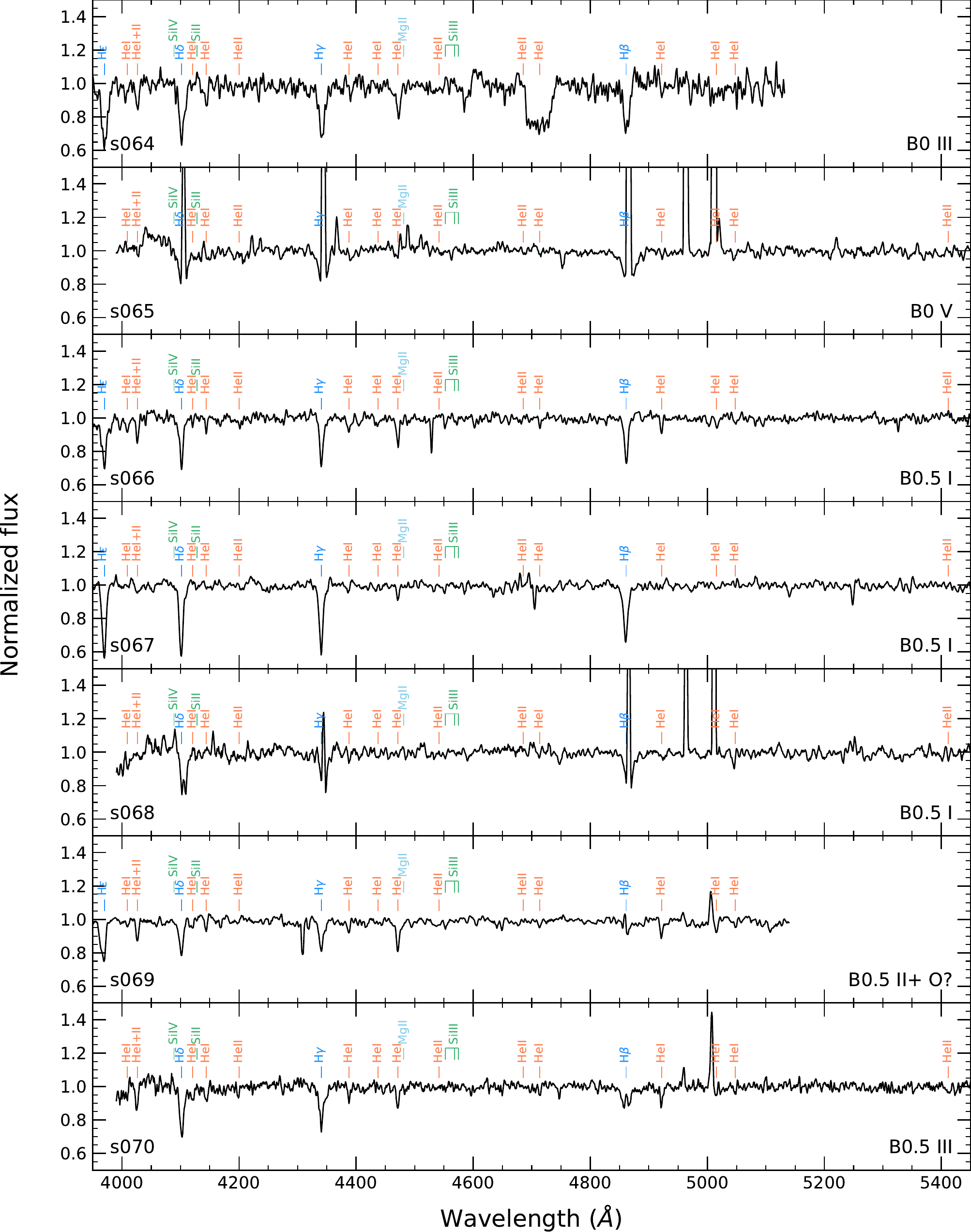}\caption{Same as Figure \ref{fig:OB_highQ_0}, continued.}\label{fig:OB_highQ_9}\end{figure*}

\begin{figure*}\includegraphics[width=\textwidth]{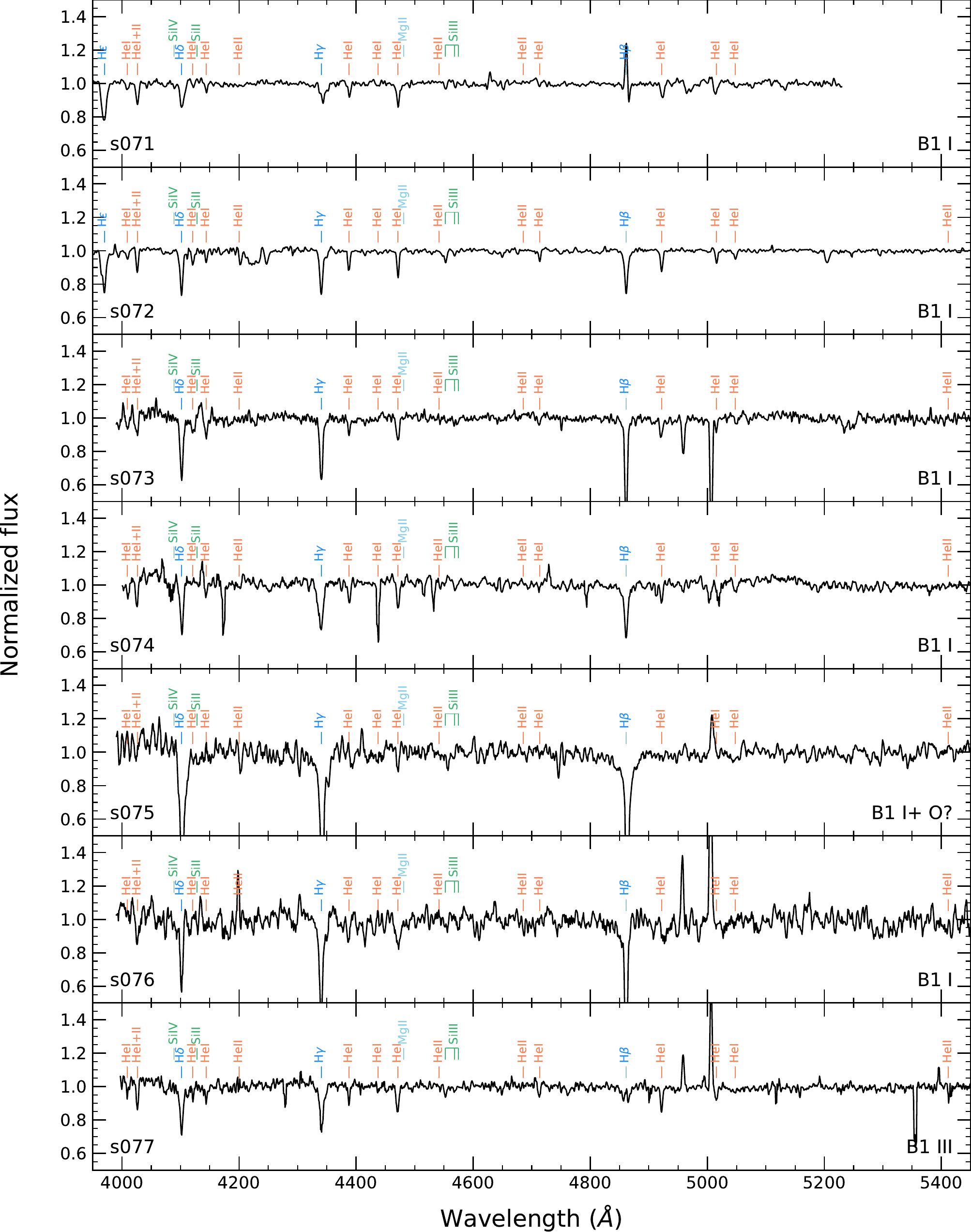}\caption{Same as Figure \ref{fig:OB_highQ_0}, continued.}\label{fig:OB_highQ_10}\end{figure*}

\begin{figure*}\includegraphics[width=\textwidth]{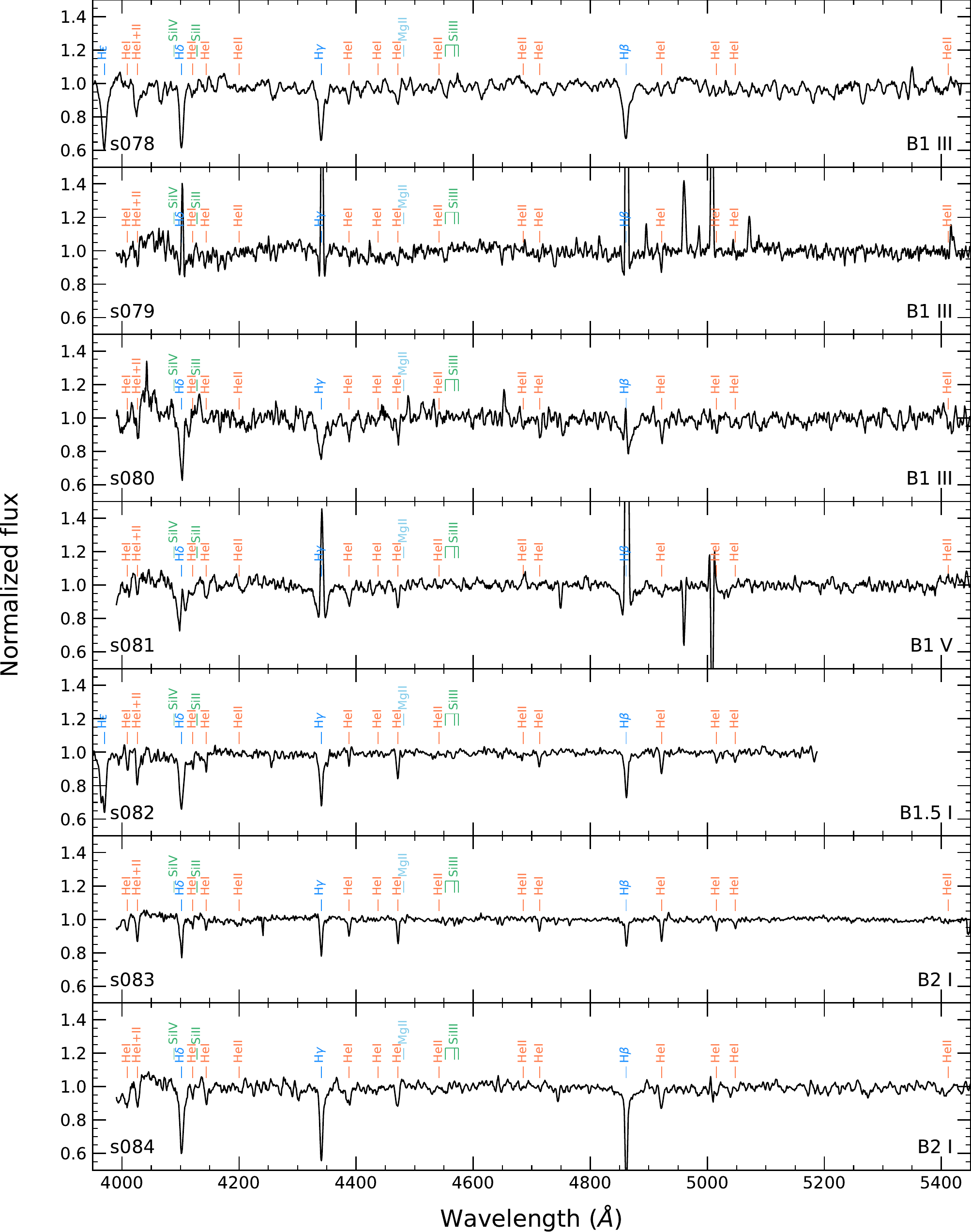}\caption{Same as Figure \ref{fig:OB_highQ_0}, continued.}\label{fig:OB_highQ_11}\end{figure*}

\begin{figure*}\includegraphics[width=\textwidth]{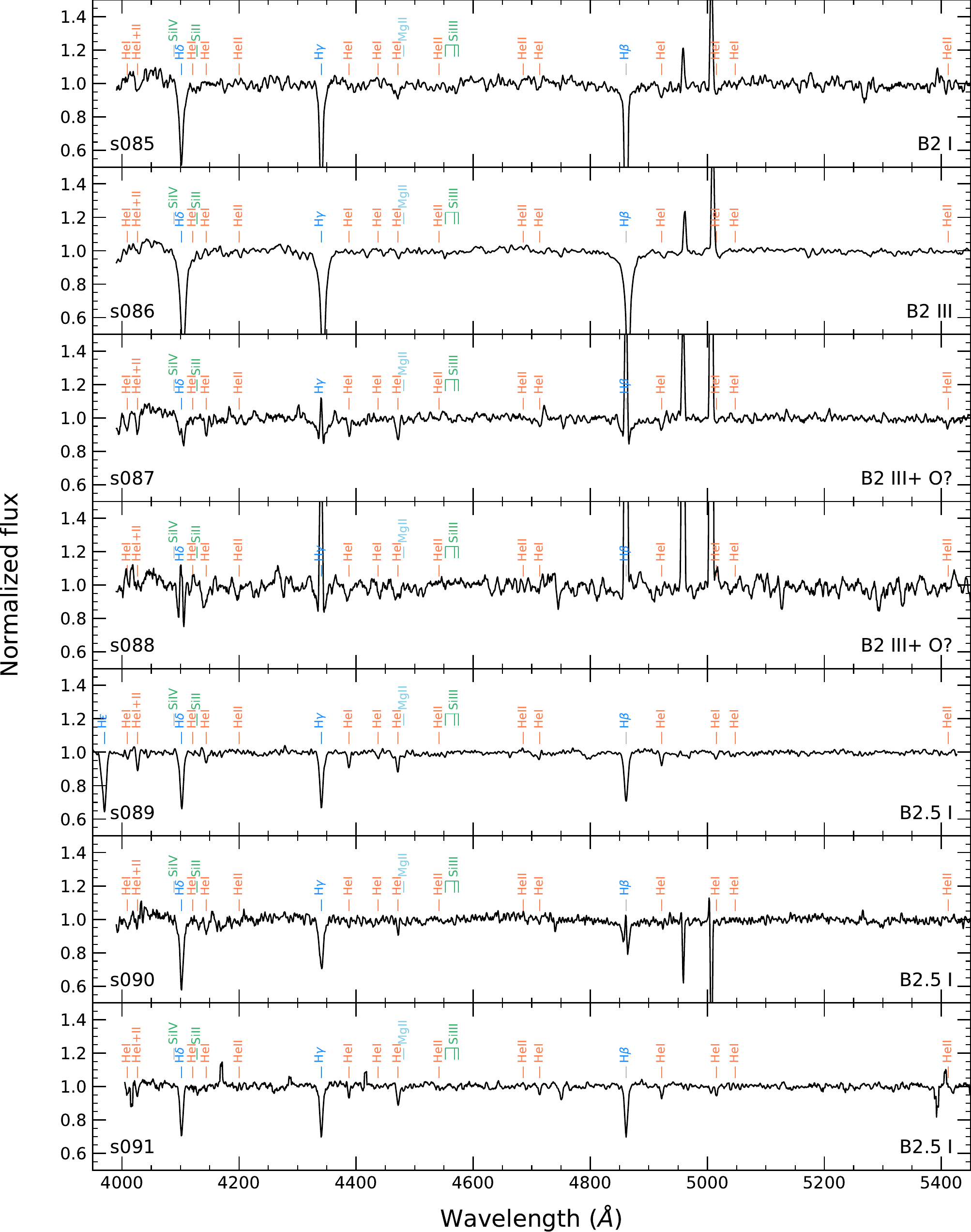}\caption{Same as Figure \ref{fig:OB_highQ_0}, continued.}\label{fig:OB_highQ_12}\end{figure*}

\begin{figure*}\includegraphics[width=\textwidth]{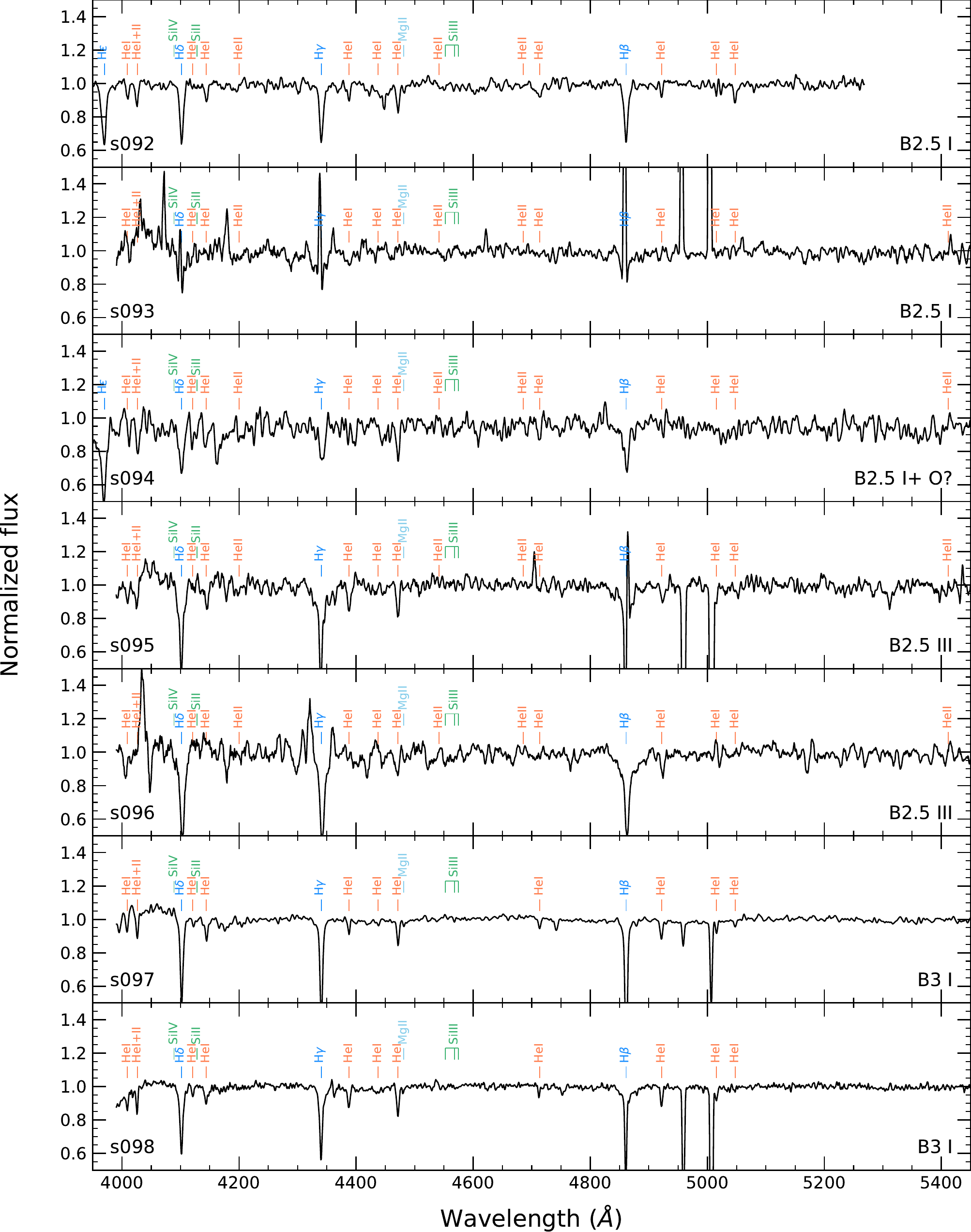}\caption{Same as Figure \ref{fig:OB_highQ_0}, continued.}\label{fig:OB_highQ_13}\end{figure*}

\begin{figure*}\includegraphics[width=\textwidth]{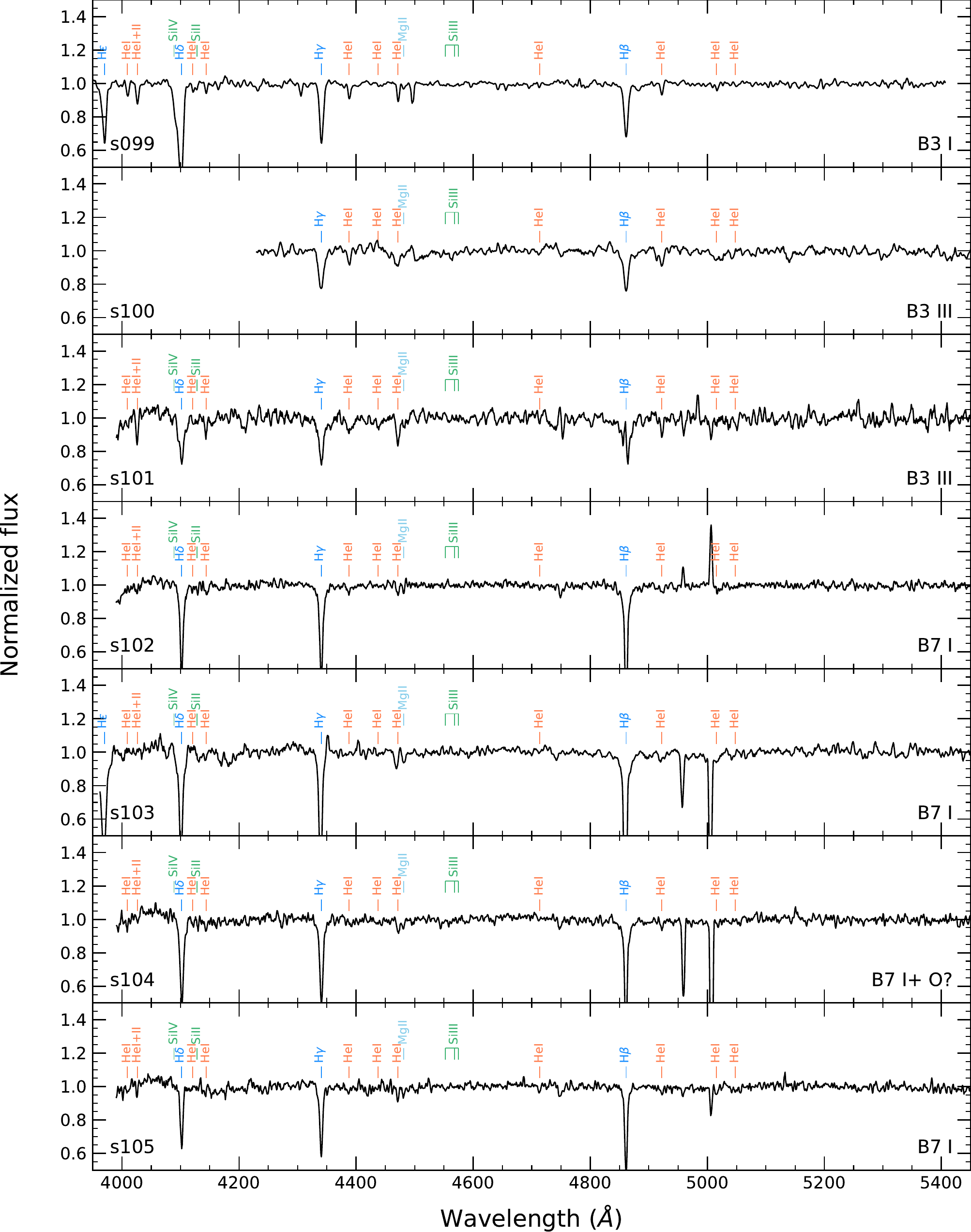}\caption{Same as Figure \ref{fig:OB_highQ_0}, continued.}\label{fig:OB_highQ_14}\end{figure*}

\begin{figure*}\includegraphics[width=\textwidth]{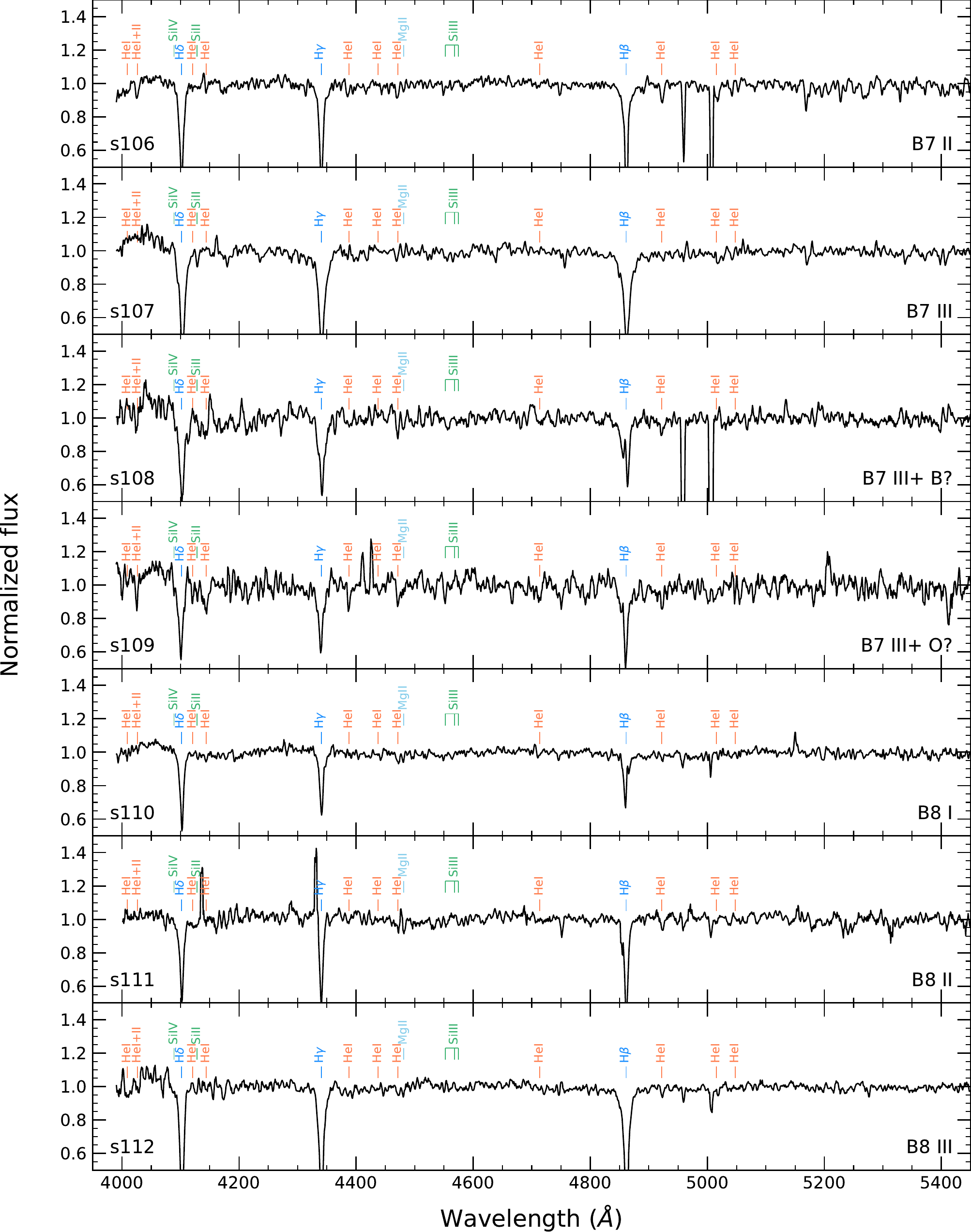}\caption{Same as Figure \ref{fig:OB_highQ_0}, continued.}\label{fig:OB_highQ_15}\end{figure*}

\begin{figure*}\includegraphics[width=\textwidth]{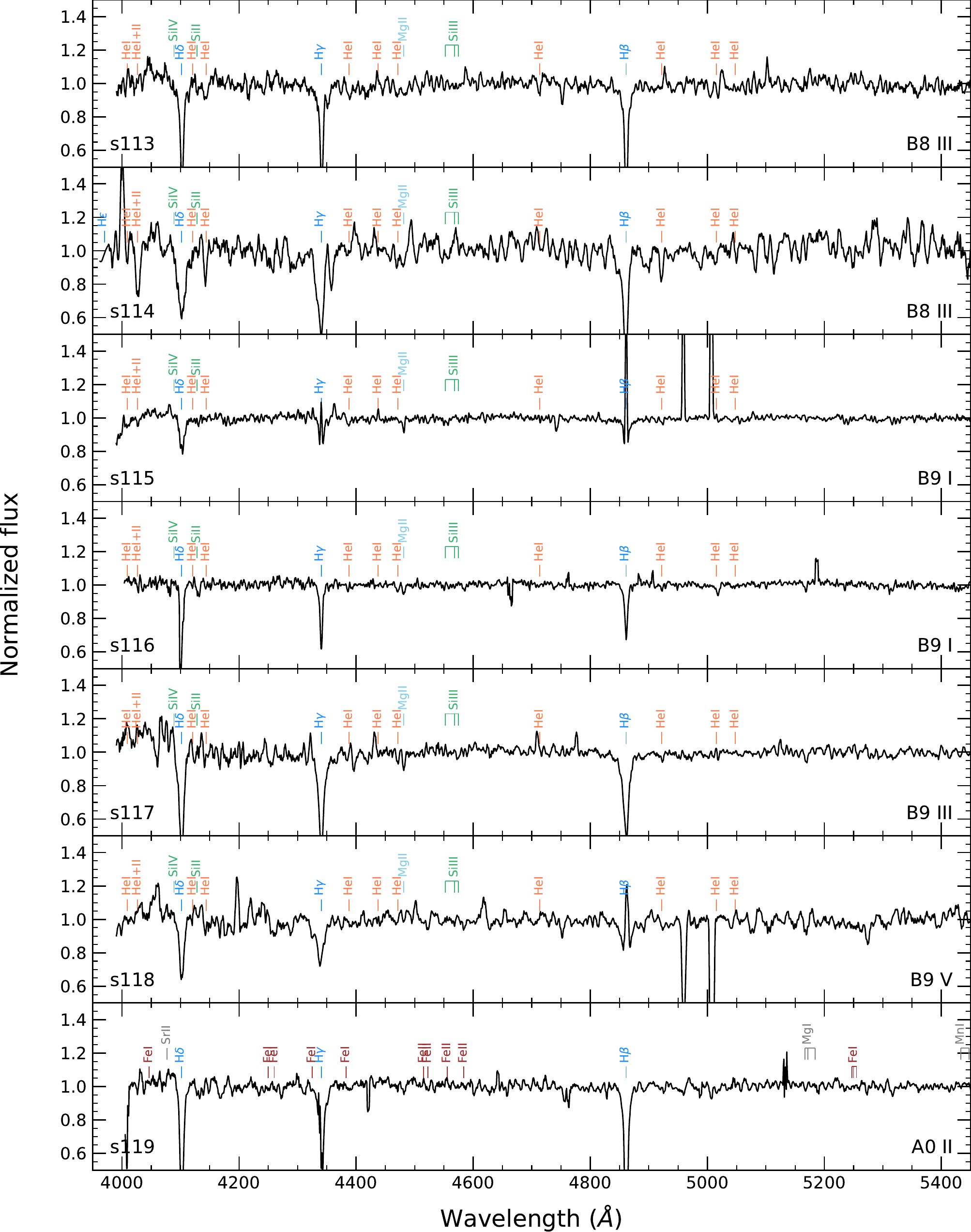}\caption{Same as Figure \ref{fig:OB_highQ_0}, continued.}\label{fig:OB_highQ_16}\end{figure*}

\begin{figure*}\includegraphics[width=\textwidth]{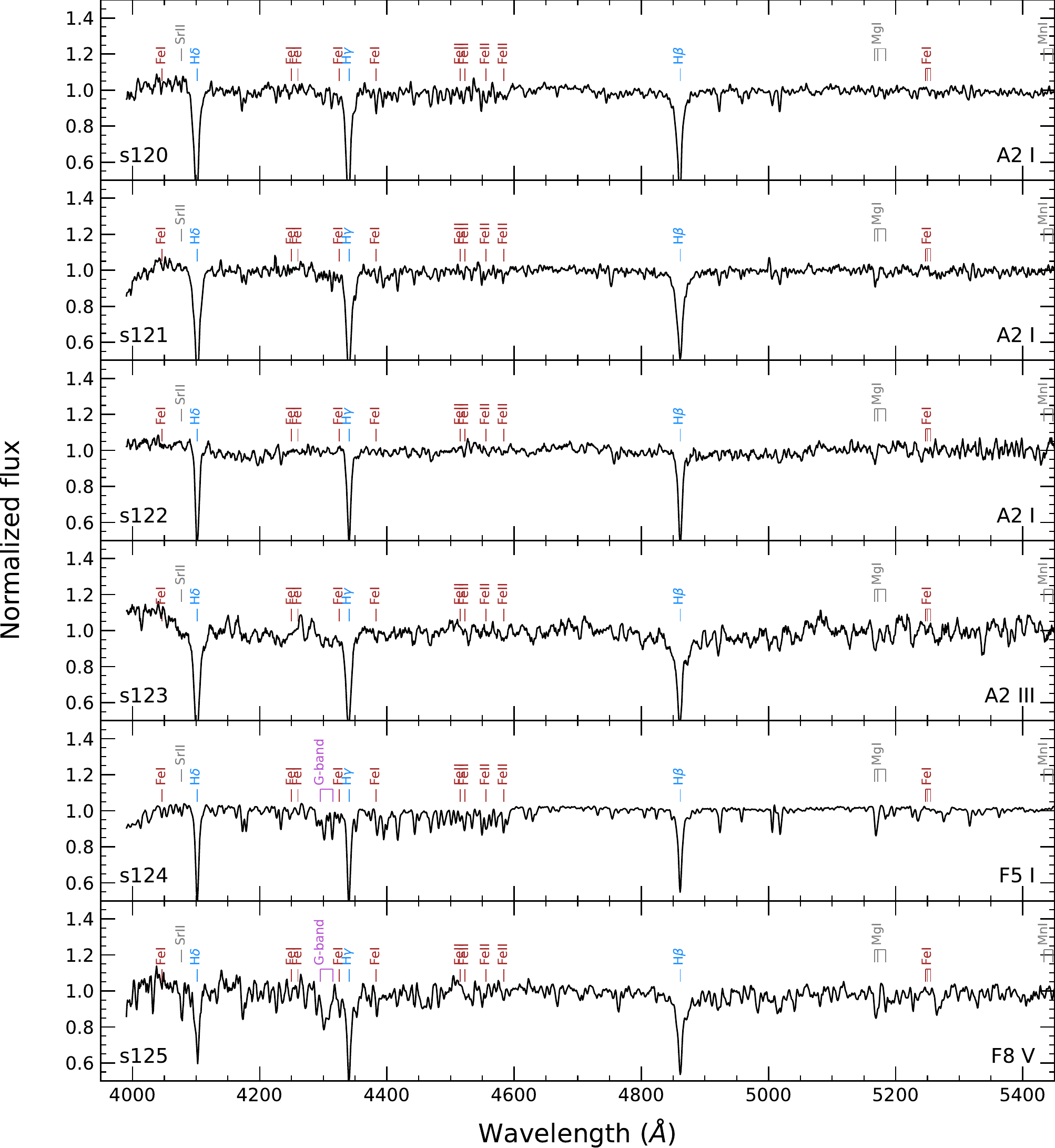}\caption{Same as Figure \ref{fig:OB_highQ_0}, continued.}\label{fig:OB_highQ_17}\end{figure*}

\begin{figure*}\includegraphics[width=\textwidth]{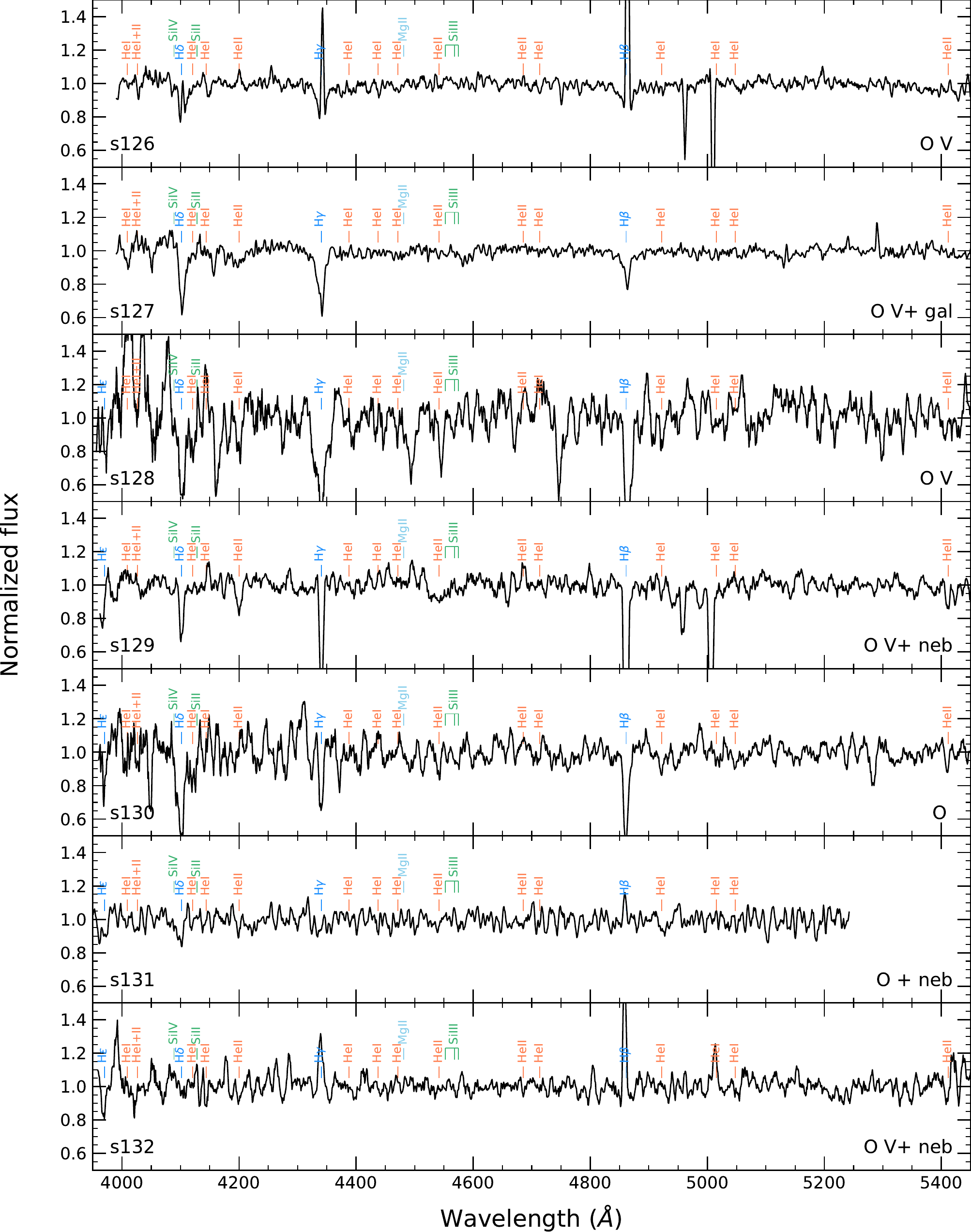}\caption{GTC–OSIRIS spectra of the sampled blue massive stars with no defined spectral subtype. The data have been corrected by heliocentric and radial velocity, and smoothed for clarity based on the S/N of the spectrum.}\label{fig:OB_lowQ_0}\end{figure*}

\begin{figure*}\includegraphics[width=\textwidth]{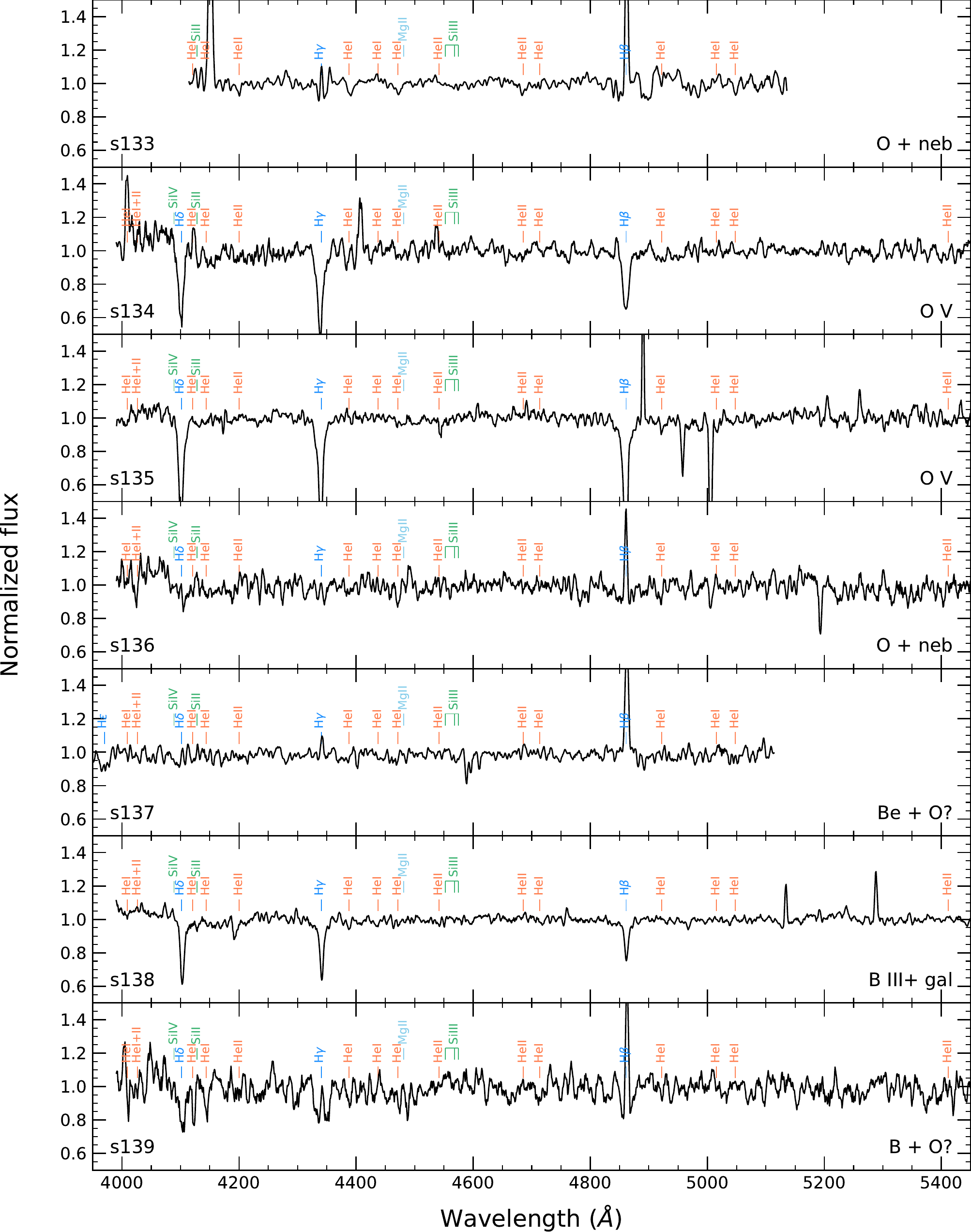}\caption{Same as Figure \ref{fig:OB_lowQ_0}, continued.}\label{fig:OB_lowQ_1}\end{figure*}

\begin{figure*}\includegraphics[width=\textwidth]{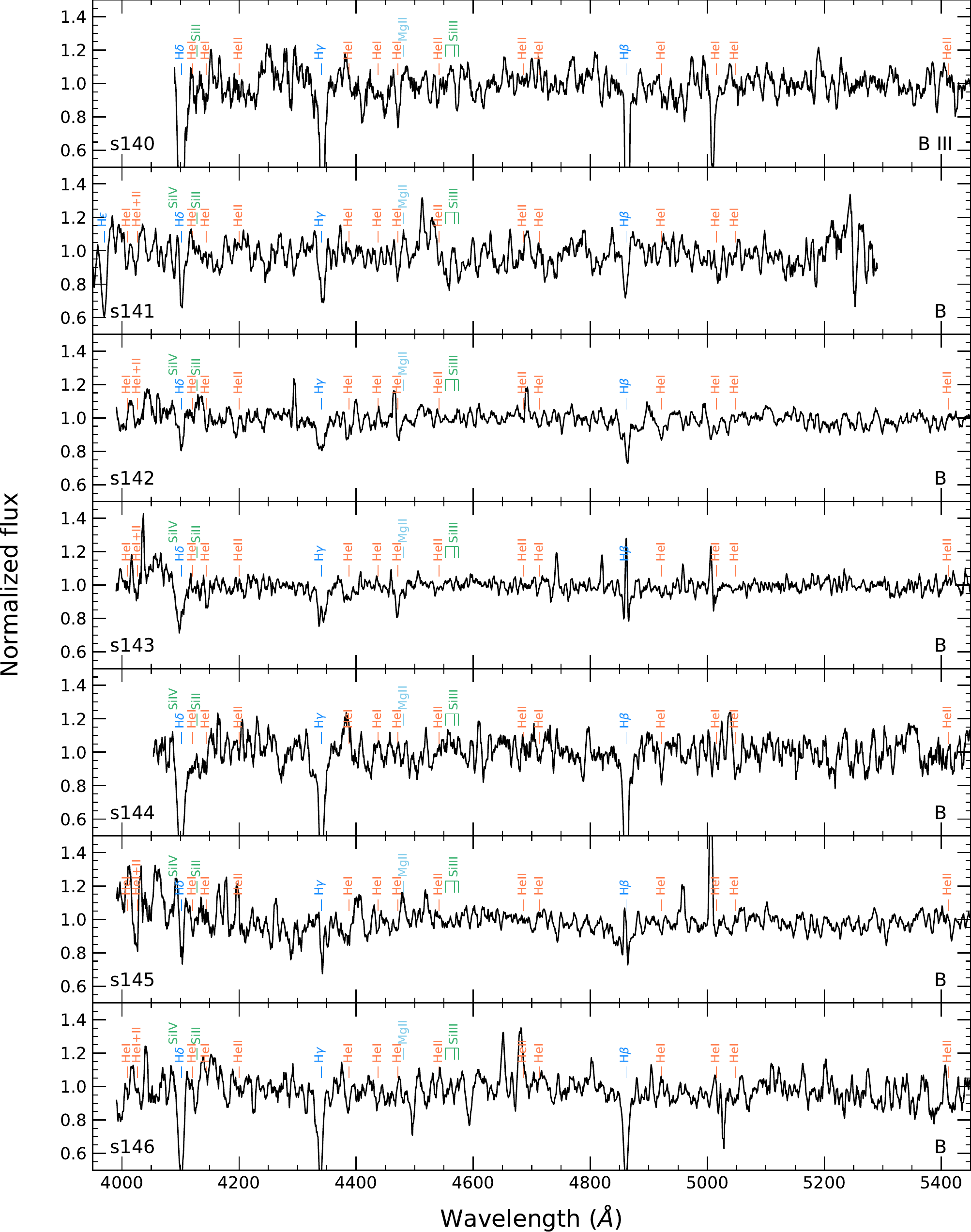}\caption{Same as Figure \ref{fig:OB_lowQ_0}, continued.}\label{fig:OB_lowQ_2}\end{figure*}

\begin{figure*}\includegraphics[width=\textwidth]{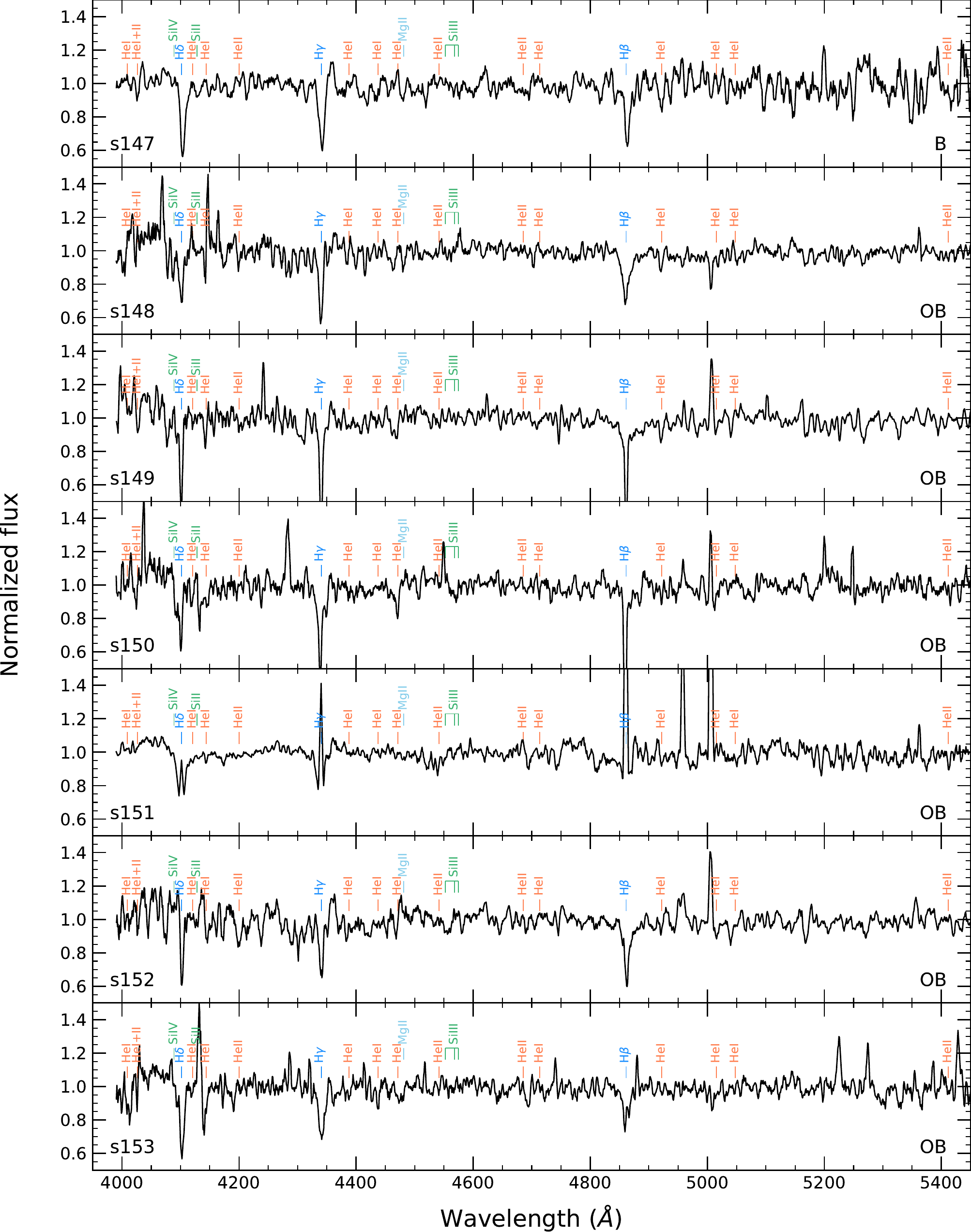}\caption{Same as Figure \ref{fig:OB_lowQ_0}, continued.}\label{fig:OB_lowQ_3}\end{figure*}

\begin{figure*}\includegraphics[width=\textwidth]{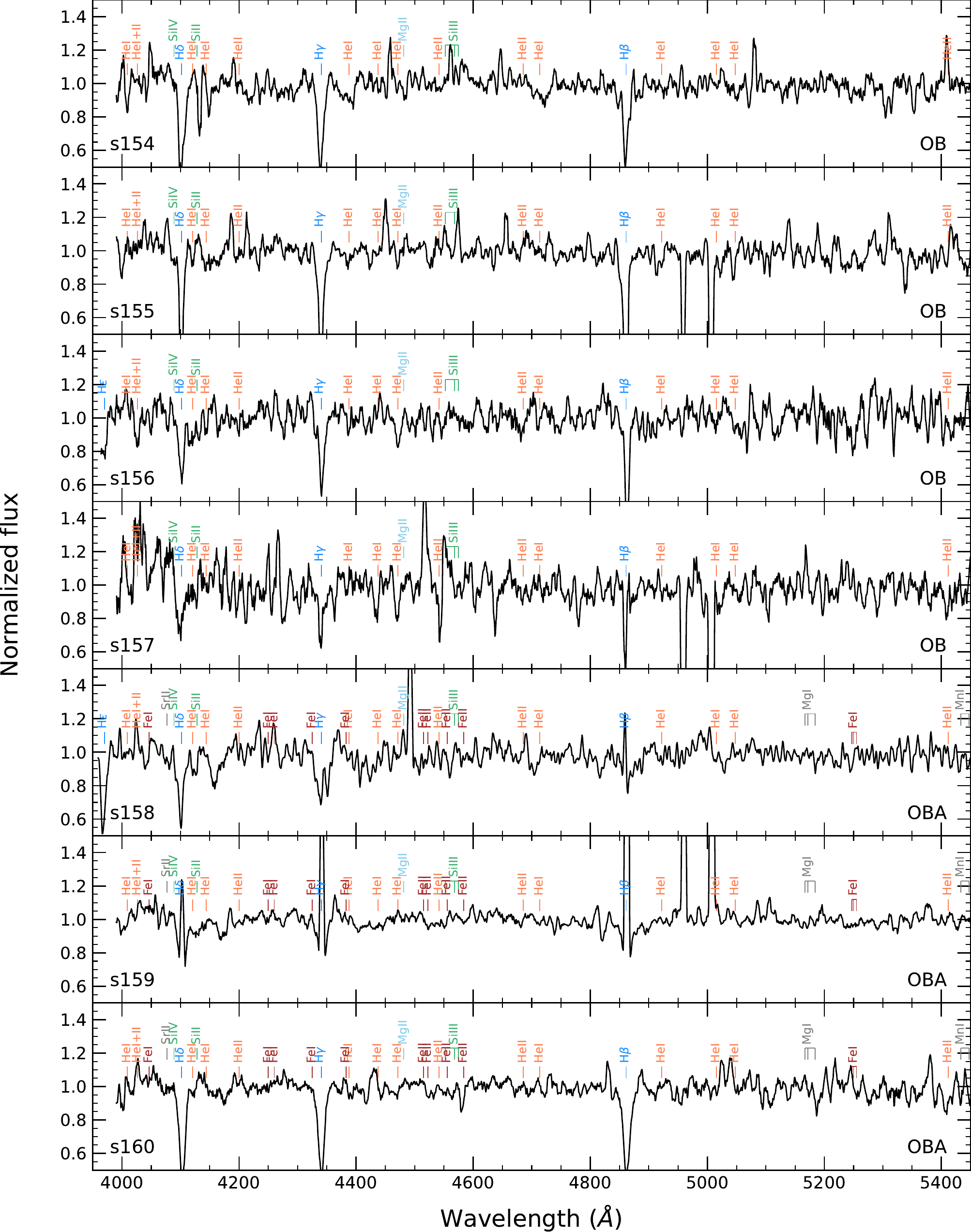}\caption{Same as Figure \ref{fig:OB_lowQ_0}, continued.}\label{fig:OB_lowQ_4}\end{figure*}

\begin{figure*}\includegraphics[width=\textwidth]{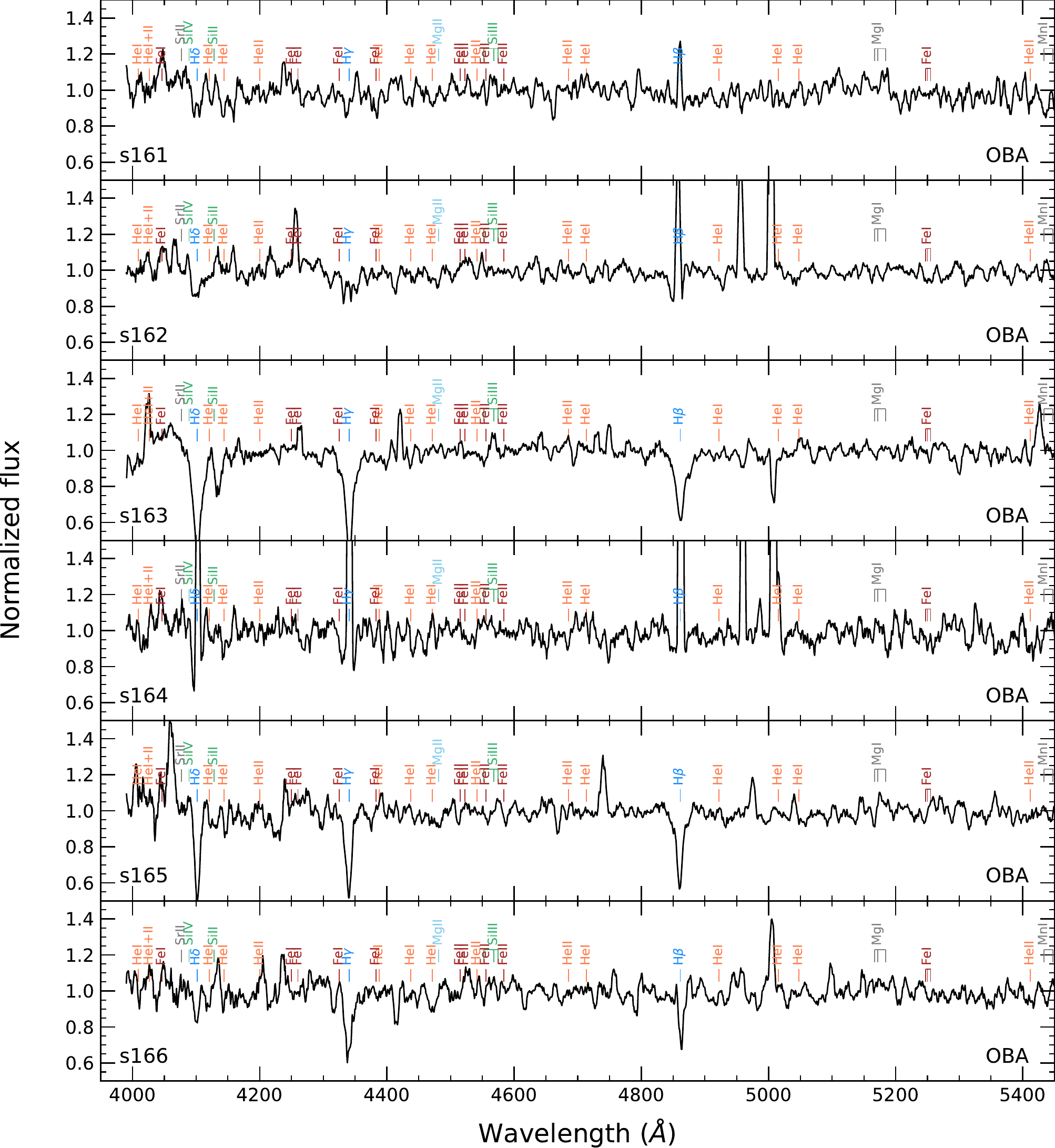}\caption{Same as Figure \ref{fig:OB_lowQ_0}, continued.}\label{fig:OB_lowQ_5}\end{figure*}

\begin{figure*}\includegraphics[width=\textwidth]{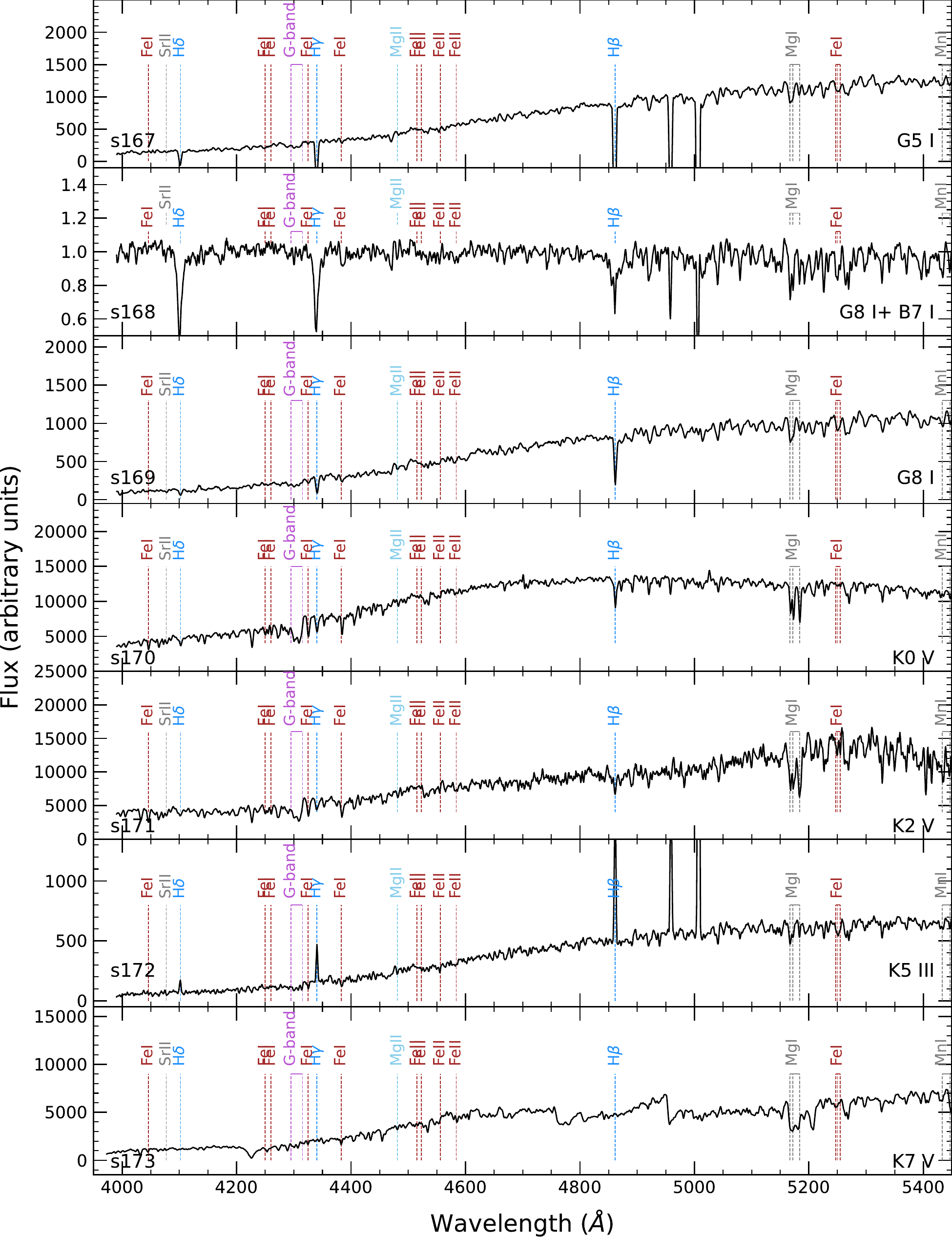}\caption{GTC–OSIRIS spectra of the late type stars in the sample. The data have been smoothed for clarity, and corrected by heliocentric and radial velocity.}\label{fig:late_0}\end{figure*}

\begin{figure*}\includegraphics[width=\textwidth]{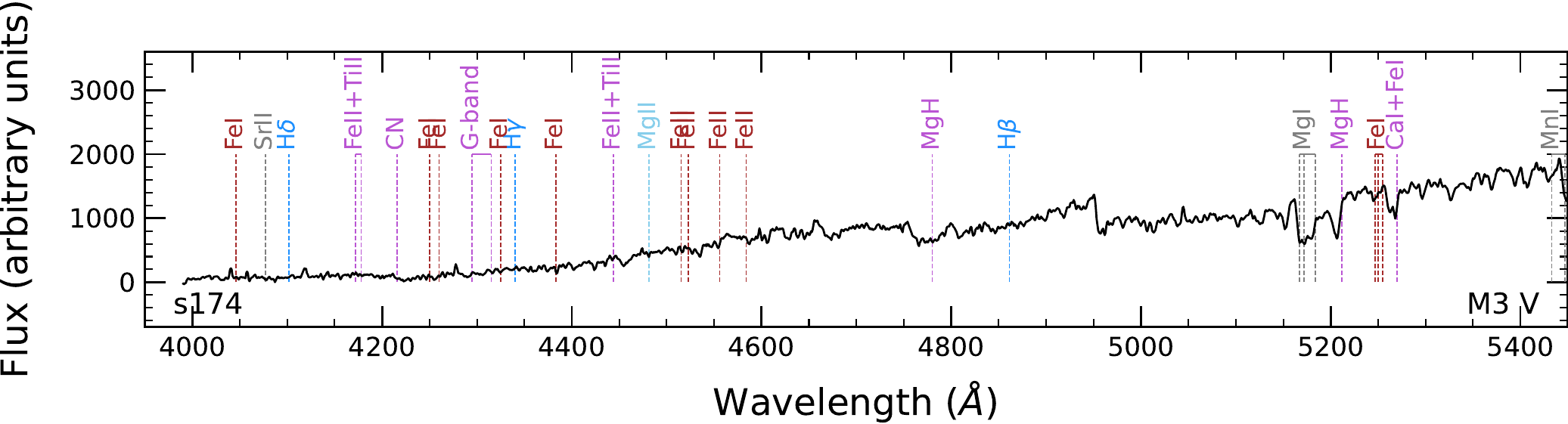}\caption{Same as Figure \ref{fig:late_0}, continued.}\label{fig:late_1}\end{figure*}

\cleardoublepage


\section{Comments on targets}
\label{sec:appx_IndividualNotes}

In this section, we provide information on individual stars regarding the observation and reduction procedure when it is needed to understand our spectral classification and the morphology of the given spectra.
We also detailed the process of spectral classification when it is problematic or when the source presents interesting features.

\subsection{Stars with a full classification}
\begin{figure} \centering \includegraphics[width=\hsize]{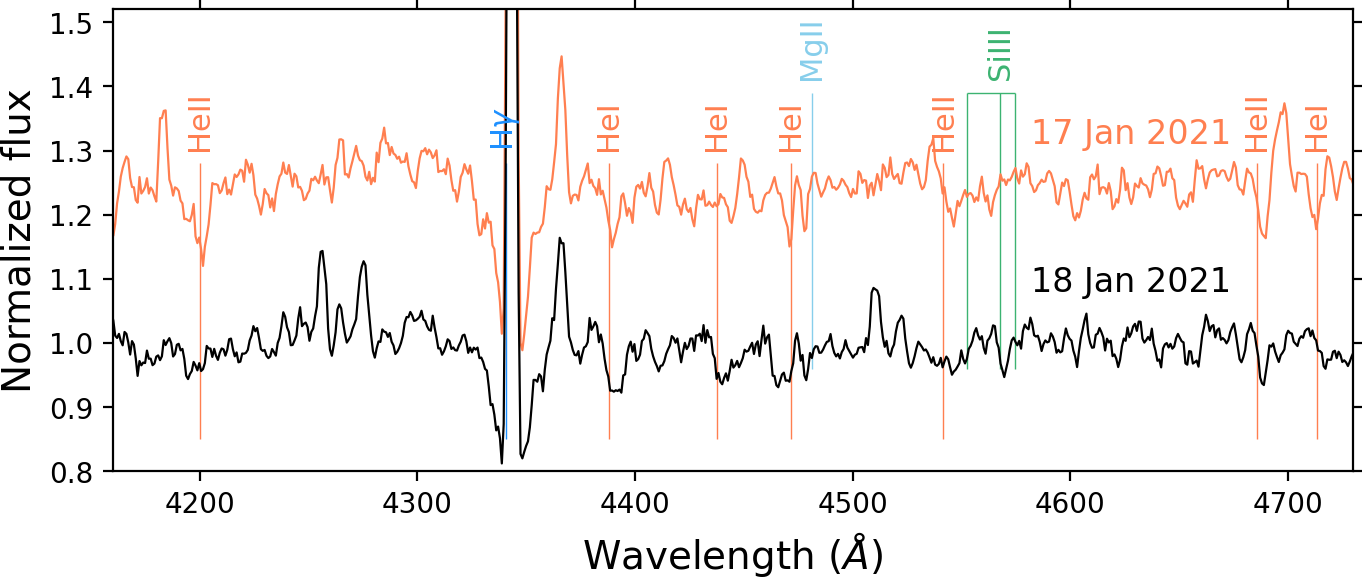}\caption{Observations of star \Remark{LSS3.OB17202} on two consecutive nights. We detect a change in the spectra features of the source. On the first night (in orange), \mbox{\HeII~4200} is detected in strong absorption, while on the second night (in black) it appears extinct, suggesting a possible binary system.}  \label{fig:plot_LSS3.OB17202} \end{figure}

\begin{figure} \centering \includegraphics[width=\hsize]{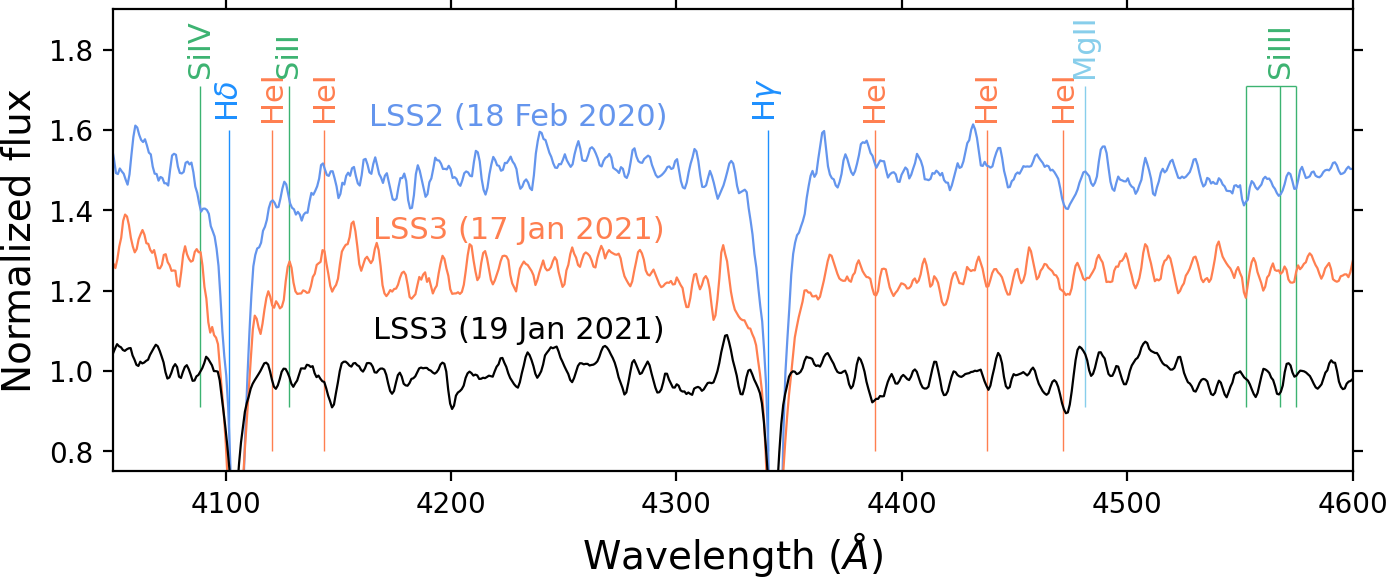}\caption{Observations of source \Remark{LSS3.OB17206} on three different dates. We measure radial velocity shifts of 100\kms~for the spectrum taken on the night of 17 January 2021 (in orange) and 50 \kms~for the one taken on 19 January 2021 (in black) with respect to the observation of 18 February 2020 (in blue), which points to an SB1.}  \label{fig:plot_LSS3.OB17206} \end{figure}

\textit{s001 (O3.5 V):} The spectrum is dominated by nebular emission lines, although it displays stellar wings for the Balmer series. \mbox{\HeII~4541} and \mbox{\HeII~4686} are present, but we do not detect any \HeI~lines.  The star is classified as O3.5 V based on the tentative detection of \mbox{\NV~4604} in absorption. Class V is assigned based on the width of the Balmer lines. We also note that \HeII~lines are weaker than expected, which may be due to the contamination from a nearby star that we detected at the reduction stage. The radial velocity was measured using \HeII~lines, and we note the resulting value is not consistent with the core of the Balmer lines, hence we mark it as a possible binary.
\newline \newline
\textit{s002 (O4 Vz):} The spectrum displays high nebular contamination, \mbox{\HeI~4471}, \mbox{\HeI~4920} and \mbox{\HeI~5015} lines in emission and strong emissions at [OIII]~4363~\AA, [OIII]~4959~\AA~and [OIII]~5007~\AA, fitting the typical locus of the sky lines. It also presents strong absorption of \HeII~lines. Since \mbox{\HeI~4471} is hidden by the nebular contamination, we classify this star by comparing the depths of \mbox{\HeII~4200}, \mbox{\HeI~4144} and \mbox{\HeI~4121}. As \mbox{\HeII~4686} in absorption is stronger than \mbox{\HeII~4541}, we assigned Vz as its luminosity class.
\newline \newline
\textit{s003 (O3-O5 Vz):} This is star s2 in \citet{Garcia2019}, who noted the broad \HeII~lines and the absence of \HeI~lines. The new reduction confirms the initial O3-O5 Vz spectral classification, further supported by the tentative detection of \mbox{\NIII~4634, 4640–4642} in emission. The updated spectrum also shows \hb~and \hg~in emission. However, a nebular or stellar origin cannot be discerned at the current resolution.
\newline \newline
\textit{s004 (O5 III):} This star was observed in two observing runs, MOS1 and MOS2, where no radial velocity variations were detected. The poorer S/N of the MOS2 observation prevented the co-addition of the two spectra. We reclassify the star with respect to \citet{Garcia2019} since the new reduction reveals stronger \mbox{\HeII~4541}. The luminosity class is reassigned to III based on the width of the Balmer lines, although \mbox{\HeII~4686} is in strong absorption. We note the new classification makes the star even more underluminous for its spectral type than reported in \citet{Garcia2019}.  \newline We observe a \vrad~mismatch between \mbox{\HeI~4471} and \mbox{\HeII~4541} in different observing blocks, and \mbox{\HeI~4471} is not detected in OB03-OB06, so we mark the star as a possible binary. 
\newline \newline
\textit{s005 (O5 V((fc))):} The spectrum presents an unidentified emission at 4375 \AA, a very broad \mbox{\HeII~4200} line and high nebular contamination shifted to the blue in the Balmer lines with respect to their cores. By comparing the strong absorption of \mbox{\HeII~4541} with the \mbox{\HeI~4471} line, we classify this star as O5. Nevertheless, this star might carry a later type since \mbox{\HeI~4471} may be affected by nebular contamination. The very broad Balmer lines and the relative strength between \mbox{\HeII~4686} and \mbox{\HeII~4541} indicate a luminosity class of V. This star presents a counterpart in the UV which additionally supports its O type. Lastly, due to the tentative presence of \mbox{\NIII~4634} and \mbox{\CIII~4647} in emission, we add the qualifier ((fc)).
\newline \newline
\textit{s006 (O6 I):} The star is classified as O6 due to the strong \mbox{\HeII~4200} and \mbox{\HeII~5411}~lines. The spectrum is severely affected by nebular oversubtraction, which leads to a different radial velocity (\mbox{\vrad~= 300 \kms}) for the Balmer lines and \mbox{\HeI~4471} absorption. Since stellar Balmer wings are dominated by nebular oversubtraction and \mbox{\HeII~4686} is not in absorption, luminosity class I is assigned. 
\newline \newline
\textit{s007 (O6 V):} This source is the central star of one of the southeast bubbles of region-B. It was observed in a very short slit of MOS2, which made unfeasible the selection of sky boxes for \textit{apall} background subtraction. The spectrum of one of the sky-slits was used instead, which results in a still incomplete nebular subtraction (see Balmer lines and \mbox{\HeI~4471}) and the loss of the bluest spectral range (\mbox{$\lambda$ $\lesssim$ 4050 \AA{}}). \newline We detected a \vrad~mismatch of $\sim$ 70 \kms~between different OBs, corrected prior to co-adding the spectra. The main \HeII~lines and some \HeI~lines seem double in the raw spectra of Night-1, suggesting it could be a binary- However, we cannot discard that this may be due to central nebular contamination by incomplete nebular subtraction at the current resolution. \newline This star was also observed in run LSS2, which was used to recover the broader \hd~profile, but it does not shed further information on the binary hypothesis. 
\newline \newline
\textit{s008 (O6.5 V):} This source is a blend of two unresolved stars identified in \citet{Bianchi2012}. The spectrum displays strong nebular contamination affecting the Balmer series and \mbox{\HeI~4471}. The \HeII~lines are in strong absorption and their relative strengths with the \HeI~lines (\mbox{\HeI~4387} and \mbox{\HeI~4144}) suggest an O6.5 type. Since the Balmer series are broad, we assigned a dwarf luminosity class. The spectrum may be contaminated by nearby stars.
\newline \newline
\textit{s009 (O7 V):} This star was observed in two observing runs, MOS2 and LSS3. We did not detect any radial velocity variations, and we discarded the co-addition of the spectra since there was no S/N improvement. The spectrum is plotted with smooth equals to 3 for clarity purposes. Without it, the component of nebular oversubtraction can be cleaned from the \mbox{\HeI~4471} core. This line is similar in strength to \mbox{\HeII~4541}, rendering type O7. Likewise, the feature around \mbox{\MgII~4481} is revealed as an extremely wide absorption, unlikely to originate in a stellar spectrum, so it is not considered for classification. Luminosity class V is assigned based on the strong absorption of \mbox{\HeII~4686}, although the width of the Balmer lines suggests class III.
\newline \newline
\textit{s010 (O7 V):} The spectrum is very noisy and shows normalization issues. Spectral type of O7 is assigned based on the relative strength of \mbox{\HeI~4471} and \mbox{\HeII~4541}. The adopted radial velocity is inconsistent with the Balmer series. \mbox{\HeII~4686} in absorption suggest class V.
\newline \newline
\textit{s011 (O7.5 III + B0 I):} The spectrum shows strong \HeII~lines, but also clear \mbox{\SiIV~4089} and perhaps \mbox{\SiIII~4552} and \mbox{\MgII~4481} lines. The target was observed in two different nights, but we have not detected changes in the spectral morphology or a radial velocity variation, hence we conclude that it is likely a blend of two stars. We assigned an O7.5 type to one of the stars based on the strength of its \HeII~lines compared with FASTWIND models (Lorenzo et al. in prep.) and considering that the strong \HeI~lines have contamination from the other star. Since \mbox{\HeII~4686} is present in absorption and its strength is similar to \mbox{\HeII~4541}, we assigned a giant luminosity class. Spectral type B0 I is assigned to the second star due to the presence of \mbox{\SiIV~4089} and its relative strength with respect to \mbox{\HeI~4121}.
\newline \newline
\textit{s012 (O7.5 III):} This star was observed in a short slit with little room for the selection of sky boxes for sky subtraction. This problem is further exacerbated by the black column artefact from the adjacent slit. As a consequence, sky subtraction is very poor. We found that the best solution was to use only the spectrum observed during the second night. \HeI~lines are wide, while \HeII~lines are narrow, and their \vrad~is slightly different. While it cannot be discarded that this is due to nebular oversubtraction, we mark the star as a possible binary. The spectral type was assigned on the premise that the \HeII~lines are stellar.
\newline \newline
\textit{s014 (O7.5 III((f))):} This star shows indications of increased He abundance and strong winds \citep{Camacho2016}.
\newline \newline
\textit{s015 (O7.5 V + B?):} The spectrum displays strong \HeII~lines consistent with a radial velocity of \mbox{\vrad~$\sim$ 225 \kms} and an O7.5 type. However, it also shows strong \mbox{\SiIII~4552}. In addition, the \HeI~lines suggest a different radial velocity. Therefore, we mark the star as a possible SB2 with a possible B star companion.
\newline \newline
\textit{s018 (O8 III):} The core of \hb~and \hg~is dominated by oversubtracted nebular lines, which also affect the \mbox{\HeI~4471} line. 
\newline \newline
\textit{s019 (O8 III):} This source is a blend of three stars included in the \citet{Bianchi2012}'s catalogue. We assign the photometry and ID of the brightest one. We clearly detect the Balmer series, some \HeI~lines and strong \HeII~lines. However, we note \mbox{\HeI~4471} is affected by nebular contamination. We classify the star as an O8 type based on the ratio of \mbox{\HeII~4541} and \mbox{\HeI~4387}, the strength of \mbox{\HeII~4200} and the presence of \mbox{\HeII~5411}. We assign luminosity class III because of the similar strength of \mbox{\HeII~4541} and \mbox{\HeII~4686}.
\newline \newline
\textit{s020 (O8 V):} Strong nebular contamination affecting the Balmer series, \mbox{\HeI~4471} and some other \HeI~lines. The spectrum displays \HeII~lines. The relative strength of \mbox{\HeII~4541} with \mbox{\HeI~4388} and \mbox{\HeII~4200} with \mbox{\HeI~4144} points to an O8 type. Since \mbox{\HeII~4686} is present in absorption and the Balmer lines are wide, we assign a dwarf luminosity class. \newline Two spectra were taken for this star in two consecutive nights. The spectrum of the second night seems to lack the absorption of \mbox{\HeII~4200} and \mbox{\HeII~4541} (see Figure \ref{fig:plot_LSS3.OB17202}). Since the star is located in a crowded region we simply marked it as an SB2 candidate or blend. However, this could be a short period eclipsing binary. We do not detect radial velocity variations between observations. 
\newline \newline
\textit{s021 (O8 V):} This star was observed in a narrow 0.63'' slit tilted from the N-S orientation by 3\textdegree{} in the MOS2 run. The slit was too short, compromising the selection of sky subtraction with \textit{apall}. We used the sky spectrum of one of the sky-slits as it yielded better results. The star was also observed in the observing run LSS3, no radial velocity variations were detected between the two runs. We discarded the combination of the spectra since no improvement in the S/N on the resulting spectrum was detected. \newline This star is a UV bright source in \citet{Bianchi2012}'s catalogue with a nearby star undetected in the UV. Hubble Space Telescope resolves them fully at F555W, but Massey's ground-based imaging could not separate them and provides one single detection for both. We included only the blue object in the slitlet, but some contamination is to be expected. In fact, the whole blend was included in the LSS3 run, and no significant difference is found between the spectrum of the MOS2 run and the blend, supporting the hypothesis that the source itself is also blended. There seems to be a mismatch between the radial velocities of \HeI~and \HeII~lines of $\sim$ 80 \kms, which could be caused by the blend or by nebular contamination at the core of the stellar lines.
\newline \newline
\textit{s022 (O8 V):} This source was classified as O9.5~III-V by \citet{Camacho2016}, but we reclassify it as an O8~V based on the strength of \mbox{\HeII~4200} and \mbox{\HeII~5411}, consistently with the classification criteria of this paper. We note that the observed magnitude would instead suggest class I. HST photometry reveals this source is a blend of two stars identified in \citet{Bianchi2012}.
\newline \newline
\textit{s023 (O8 V):} Strong \HeII~lines are detected. Since \mbox{\HeI~4471} is in emission due to high nebular contamination, we based our classification on the strength of \mbox{\HeII~4200}, the presence of \mbox{\HeII~5411} and the ratio of \mbox{\HeII~4541} with \mbox{\HeI~4388}. The strong detection of \mbox{\HeII~4686} and the lack of \SiIV~suggests a dwarf luminosity class. The spectrum could be contaminated by the nearby star \Remark{LSS3.OB11201}. 
\newline \newline
\textit{s024 (O8 V):} Strong nebular contamination. The spectrum displays strong lines of \HeII. The relative strength of \mbox{\HeII~4541} and \mbox{\HeII~4200} with respect to \mbox{\HeI~4471} and \mbox{\HeI~4388} suggests an O8 type. We assign V as its luminosity class due to the broadening of the Balmer series and the presence of \mbox{\HeII~4686} in absorption.
\newline \newline
\textit{s025 (O8 V):} We detect \HeII~lines. The relative strength of \mbox{\HeII~4541} with \mbox{\HeI~4471} and \mbox{\HeI~4388}, and the detection of \mbox{\HeII~5411} points to an O8 type. Since \mbox{\HeII~4686} is in strong absorption, we classify this star as dwarf.
\newline \newline
\textit{s026 (O8 V):} This source is a blend of three stars identified in \citet{Bianchi2012}'s catalogue. We assign the photometry and ID of the brightest one. The spectrum displays strong, broad \HeII~lines and extremely wide Balmer lines. The \mbox{\HeI~4471} is strongly affected by nebular contamination. Based on the strength ratio of \mbox{\HeII~4200} with respect to \mbox{\HeIandII~4026}, we classify the star as O8. We assign class V due to the broadening of the Balmer series and the strong \mbox{\HeII~4686} absorption.
\newline \newline
\textit{s027 (O8 V):} Noisy spectrum. We classify the star as O8 dwarf based on the strength of \mbox{\HeII~4200} and \mbox{\HeII~4686}. However, \mbox{\HeII~4541} is absent. We also note that all He lines are very broad. 
\newline \newline
\textit{s029 (O8.5 III):} This source is s03 in \citet{Garcia2019}, but an updated reduction is presented in this work. The new spectrum shows an emission at 4450 \AA{} due to a slit-ghost. \mbox{\HeI~4471} seems double due to nebular contamination in some observing blocks but was not discarded in the classification, and \mbox{\HeII~4541} is not seen in all the selected OBs. The target is marked as a possible SB1 since we detected radial velocity variations of more than 200 \kms~in \mbox{\HeII~4541} between different OBs. These variations were corrected prior to co-adding the spectra of individual OBs. The updated spectrum exhibits \mbox{HeII~4200} and \mbox{HeII~5411}, and narrower Balmer lines, hence the source is reclassified as O8.5 III. 
\newline \newline
\textit{s030 (O8.5 III):} The slit was tilted by 7.1\textdegree{}, which results in moderate degradation of the spectral resolution. This star was observed in the same slit as \Remark{MOS2.s29}. Targets are very close to each other, and some contamination is expected. The stars are located within an H~{\sc ii} bubble and nebular subtraction was incomplete, as evinced by the core of the Balmer lines. 
\newline \newline
\textit{s031 (O8.5 V):} This star was observed within a slit tilted by 8\textdegree{} to include also star \Remark{MOS2.s33}. The inclination results in degraded spectral resolution. This star was close to the slit edge, and sky subtraction using a sky-slit provided better results than the other stars from the slit. Yet, nebular subtraction is not fully satisfactory, as demonstrated by the strong emission that overlaps the wide Balmer lines and the possibly nebular oversubtraction at the core of He lines. Spectral classification is further hindered by poor S/N as expected from the faint magnitude. The derived radial velocity is high, suggesting it may be a binary.
\newline \newline
\textit{s032 (O8.5 V):} This source encompasses two stars unresolved in \citet{Massey2007}'s catalogue, and it is possibly contaminated by \Remark{MOS2.s40}, which is also in the same slit. The slit was tilted by 7.1\textdegree{}, which results in moderate degradation of the spectral resolution. For the second night, we found that the sky-slit provided better background subtraction, although that implies that \mbox{$\lambda$ < 4100 \AA{}} can only be recovered from Night-1, and errors are larger. Despite this action, nebular subtraction is incomplete in the Balmer lines, and the \HeI~lines display nebular oversubtraction at their cores. 
\newline \newline
\textit{s033 (O8.5 V):} The spectrum shows strong nebular contamination. We assign an O8.5 type based on the relative strength of \mbox{\HeII~4541} with \mbox{\HeI~4471} and the presence of \mbox{\HeII~5411}. The detection of \mbox{\HeII~4686} in absorption and the broad Balmer lines suggest a dwarf luminosity class.
\newline \newline
\textit{s034 (O8.5 V):} The spectrum displays \mbox{\HeII~4541} that, compared with the \mbox{\HeI~4471} line suggest an O8.5 type. \mbox{\HeII~5411} is also detected. Based on the broadening of the Balmer series, we assign a luminosity class V, but we note that this is in contradiction with the absence of \mbox{\HeII~4686} in absorption.
\newline \newline
\textit{s035 (O8.5 V):} At the assigned \vrad~of 170 \kms, we detect a strong absorption of \mbox{\HeII~4541}, partially affected by nebular oversubtraction. We classify the source as O8.5 due to the relative strength of \mbox{\HeI~4144} and \mbox{\HeII~4200}. The Balmer series shows nebular contamination and seems inconsistent with the adopted \vrad.
\newline \newline
\textit{s036 (O9 I):} This star was observed in three observing runs, MOS2, LSS2 and LSS3. No variations in spectral features or radial velocities were detected. In MOS2, it was included in a very short slit that hindered sky subtraction. In LSS2, the star was off-slit, and exposure times were shorter, yet the sky subtraction was slightly better. The \mbox{\HeII~4541} transition is detected completely in emission in MOS2, but the stellar component is partially recovered in LSS2 and LSS3. After checking against the spectral lines, we co-added the spectra to increase the spectral S/N. We note that the \mbox{\HeII~4541} line is lost in the co-added spectrum, but we used the LSS2 line strength for spectral classification. The strong absorption at 4650 \AA{} is due to an artefact detected in MOS2 run, but not in the LSS2 or LSS3 runs.
\newline \newline
\textit{s037 (O9 I):} This star was observed in runs LSS1, MOS2 and LSS2. \citet{Camacho2016} classified this star, its OB623, as an O8~Ib. We detect no significant morphological or radial-velocity variations in the subsequent MOS2 and LSS2 runs. In MOS2, the spectrum was taken in a too-short slit that compromised the sky subtraction with \textit{apall}. For the spectrum of the second night, with worse effective seeing, we instead used the sky spectrum from one of the sky-slits, although sky subtraction was incomplete. Sky subtraction in LSS2 is not fully satisfactory either. After checking against spectral variations, we co-added the LSS2 and MOS2 spectra to increase the S/N and discarded the LSS1 spectrum that only introduced noise in the co-addition. \newline The strong emission close to \mbox{\HeII~4541} is remnant from the sky-spectrum and precludes the detection of \mbox{\HeII~4541}. This line is detected in absorption both in LSS1 and LSS2. Alternatively, \mbox{\HeII~4200} is clearly observed in the co-added spectrum and \mbox{\HeII~4686} is still neither detected in emission nor absorption, thus pointing to class I. Considering the strength of the \mbox{\HeII~4200} and \mbox{\SiIV~4089} lines relative to \HeI~transitions, we reclassify the star as O9~I.
\newline \newline
\textit{s038 (O9 I((f))):} This star was classified as O9.7~I((f)) by \citet{Camacho2016}, who also noted indications of strong winds and increased He abundance. In this work, we reclassify the star as O9~I((f)) based on the strength of \mbox{\HeII~5411} and \mbox{\HeII~4200}, more consistently with the current criteria. 
\newline \newline
\textit{s039 (O9 III):} An unknown emission detected near \mbox{\HeII~4200}, at 4183 \AA. The presence of \mbox{\HeII~4541} and its relative strength with respect \mbox{\HeI~4471} suggest an O9 type. The lack of \mbox{\HeII~4686} points to a supergiant class, but the width of the Balmer series and the lack of \SiIV~suggest luminosity class III.
\newline \newline
\textit{s040 (O9 III + mid-B?):} The spectrum is noisy and presents significant nebular contamination. Because of the relative strength of \mbox{\HeII~4200} with \mbox{\HeI~4387}, we assign an O9 type. The source also shows weak \mbox{\HeII~4541} and \mbox{\HeII~5411}. The width of the Balmer lines and the relative strength of \mbox{\HeII~4541} with \mbox{\HeII~4686} points to the adopted class III, although the faint magnitude suggests class V. The \mbox{\MgII~4481} and \mbox{\SiIII~4552} absorptions would be consistent with a secondary component at \vrad~= 300 \kms~and mid-B spectral type. Space photometry from \citet{Bianchi2012} cannot resolve this source. 
\newline \newline
\textit{s041 (O9.5 II):} This star was observed in two observing runs, MOS1 and MOS2, where no radial velocity variation was detected. The poorer S/N of the MOS2 observation prevented the co-addition of the two spectra.
\newline \newline
\textit{s042 (O9.5 III):} An emission at 4450 \AA{} is only seen in the observing blocks taken in the first year (as in \Remark{MOS1.s03}), and it may be due to a slit-ghost. In addition, an anomalous sky line emission is detected at 900 \kms~of \hd~and \hg~in some observing blocks, which leads to an artificial broadening of the lines. \mbox{\HeI~4471} was not considered for classification since it seems to be an artefact. Apart from the MOS1, the source was also observed in the LSS3 observing run, where \mbox{\HeI~4471} has a much weaker absorption. No radial velocity variations were detected, but since the LSS3 spectrum has worse S/N and poorer spectral resolution, the spectra were not co-added. HST photometry reveals the source is a blend of two stars.
\newline \newline
\textit{s043 (O9.5 V):} The source is a blend of two unresolved stars identified in \citet{Bianchi2012}. It was observed in the LSS2 and LSS3 observing runs, and we detected no radial velocity variation. Since the S/N of the LSS2 spectrum was poor, the data were not co-added. The spectrum shows strong lines of \HeI, perhaps \mbox{\HeII~4200} and no metallic lines. The absorption found at $\sim$ 4542 \AA~is an artefact of reduction. Based on the presence of \mbox{\HeII~4200} and the width of the Balmer lines, we classify it as O9.5 V.
\newline \newline
\textit{s044 (O9.5 V):} \mbox{\HeI~4471} is absent due to nebular contamination. We clearly detect the Balmer series, some \HeI~lines, and \mbox{\HeII~4541} and \mbox{\HeII~4686} in absorption. We classify the star as an O9.5 dwarf due to the relative strength of \mbox{\HeII~4541} with respect to \mbox{\HeI~4387} and the width of the Balmer series. \hg, \hd~and \mbox{\HeI~4471} show hints of a second component at higher radial velocity (\mbox{$\sim$ 600 \kms}).
\newline \newline
\textit{s045 (O9.5 V):} The detection of \mbox{\HeII~4200} and \mbox{\HeII~4541} points to an O9.5 type. \mbox{\HeI~4388} is anomalously broad. The star is classified as a dwarf because of the broadening of the Balmer wings. However, we note that this is in contradiction with the absence of \mbox{\HeII~4686}. 
\newline \newline
\textit{s046 (O9.5 V):} This source is a blend of two unresolved stars identified in \citet{Bianchi2012}. We classify it as O9.5 based on the relative strength of \mbox{\HeII~4200}/\mbox{\HeII~4144} since \mbox{\HeII~4541} seems to be affected by noise. Luminosity class V is assigned due to the width of the Balmer lines.
\newline \newline
\textit{s047 (O9.5 V):} We reclassify the star from an A5 II in \citet{Camacho2016} to an O9.5 V as we note the presence of \mbox{\HeII~4686} and \mbox{\HeII~4541}.
\newline \newline
\textit{s048 (O9.5 Vn):} We classify the star as O9.5 based on the relative strength of \mbox{\HeI~4471} and \mbox{\HeII~4541}. Luminosity class Vn is assigned due to the width of the Balmer lines. 
\newline \newline
\textit{s049 (O9.7 I):} This source was observed in two observing runs, LSS2 and LSS3. We found a radial velocity variation of $\sim$30 \kms~between the two runs. However, we did not co-add the spectra since we did not detect any improvement in the S/N. The spectrum shows \mbox{\HeII~4541} and similarly strong \mbox{\SiIII~4552}, we assigned a compromise spectral type of O9.7 supergiant. HST photometry reveals this source is a blend of two unresolved stars identified in \citet{Bianchi2012}.
\newline \newline
\textit{s050 (O9.7 I):} This star is OB321 in \citet{Camacho2016}, where they assigned O9.7~I((f)). Since it is brighter than most of the MOS2 targets ($V$ = 19.609 mag), it was observed in a narrow slit to increase resolution and constrain better its stellar parameters. The S/N of this spectrum is better than the LSS1 program, so we replaced OB321 with this one in our database. \mbox{\NIII~4634} does not exhibit emission in the improved quality spectrum, so we removed the ((f)) qualifier. 
\newline \newline
\textit{s051 (O9.7 I):} This star was observed in LSS2 and MOS2 observing runs. We detect no radial velocity variations and discard the spectrum from the LSS2 run due to poor S/N.
\newline \newline
\textit{s052 (O9.7 I):} This star was observed together with \Remark{MOS2.s21} in a very short slit. For the first night, we found that the subtraction of one of the sky-slits provided better results. For the second night, \textit{apall} sky subtraction provided better results, but only one box was used (located in the middle of the slit). \HeII~lines are clearly seen only in the spectrum of the second night, but we detected no apparent radial velocity variation between both nights.
\newline \newline
\textit{s053 (O9.7 I):} We detect radial velocity inconsistencies between different Balmer lines and different \HeI~lines, which may be explained by nebular contamination. Adopting the systemic velocity of the galaxy, the presence of \mbox{\HeII~4200} and lack of \mbox{\SiIII~4552}, \mbox{\MgII~4481} and \mbox{\HeII~4686} suggest an O9.7 I. The spectrum may be contaminated by the near star \Remark{LSS3.OB31203}.
\newline \newline
\textit{s054 (O9.7 III):} The star was observed in both LSS2 and LSS3 runs. We did not detect radial velocity variations between the two observations, and we co-added the spectra to achieve higher S/N. Only \HeI~lines are clearly detected. There might be hints of \mbox{\HeII~4541}, \mbox{\HeII~4686} and \mbox{\SiIV~4089}, hence we classify the star as O9.7. Luminosity class III is assigned due to the width of the Balmer series. 
\newline \newline
\textit{s055 (O9.7 III):} The main star (J101053.90-044111.0) is contaminated by a nearby one with very blue $Q$-colour (J101053.94-044110.1) and only 0.2 mag fainter, slightly outside the slit. The selection of sky-boxes with \textit{apall} was hampered by this source and, as a consequence, the nebular subtraction is incomplete, as seen clearly in the core of the Balmer lines, but also \mbox{\HeII~4686}. The emission at $\sim$ 4540 \AA{} that affects \mbox{\HeII~4541} has an unknown origin and does not seem to be a cosmic ray.
\newline \newline
\textit{s056 (O9.7 V):} In the MOS2 run, this star was observed in a very short slitlet that severely handicapped sky subtraction (e.g. see the central nebular emission lines in Balmer and possibly \HeI~oversubstraction). It was also observed in the MOS1 run, but nebular subtraction was similarly poor. Considering that the star is far from the large ionized structures of the galaxy, the strong nebular contamination is quite puzzling. Alternatively, the star could be an Oe star. After checking against spectral line or flux continuum level variations, we co-added the spectra for increased S/N. There is an absorption feature at $\sim$ 4200 \AA{} of unknown origin that is detected in the spectrum of both observing runs, making it unlikely for it to be an artefact. The subtracted sky spectrum was checked, and there are no features at those wavelengths. \mbox{\HeII~4541} is present but weak, so the star is classified as O9.7. Luminosity class V is assigned based on the width of the Balmer lines. We note that \mbox{\HeII~4686} should be strong in absorption in an O9.7 V but is missing in the spectrum of MOS2.
\newline \newline
\textit{s057 (O9.7 V):} This star was observed in LSS2 and LSS3 observing runs, where no radial velocity variations were detected. The spectra were co-added to improve S/N. We note the absence of \mbox{\HeII~4686}, in contradiction with the luminosity class Vn, indicated by the Balmer wings.
\newline \newline
\textit{s058 (O9.7 V + B?):} Very noisy spectrum. The O9.7 type is assigned because of the presence of \HeII~lines and \mbox{\SiIII~4552}. The strong \HeI~lines at \vrad~= 150 \kms suggest a secondary component.
\newline \newline
\textit{s059 (O9.7 Vn):} This star is located in the galactic outskirts, and it is the farthest of the southern sources from the galactic centre. Contrastingly, it shows nebular lines at the core of the Balmer series and GALEX UV emission. We assign spectral type O9.7 based on the detection of \mbox{\HeII~4686} in absorption. The modifier "n" is used because of the abnormally broad Balmer lines.
\newline \newline
\textit{s060 (O9.7 V):} Possible blend, which hampers the determination of its radial velocity. The target does not show \mbox{\HeII~4541}, but it presents \mbox{\HeII~4200} and \mbox{\HeII~4686} in absorption. The relative strength of \mbox{\HeII~4200} with respect to \mbox{\HeI~4144} indicates an O9.7 type. Due to the broadening of the Balmer series and \mbox{\HeII~4686} in absorption, we assign class V. The broad \HeI~lines would be consistent with the source being a blend or an unresolved binary. 
\newline \newline
\textit{s061 (B0 I):} The spectrum presents a deep depression at 4320 \AA{} due to a defect in the CCD. The star seems to have N and O.
\newline \newline
\textit{s063 (B0 III):} \mbox{\SiIV~4089} is affected by an artefact and therefore cannot be used for classification. Since there are no \HeII~lines except for \mbox{\HeII~4686}, and \mbox{\SiIII~4552} is very weak, type B0 is assigned. The relative \mbox{\HeII~4686} to \mbox{\HeI~4713} line ratio indicates luminosity class III.
\newline \newline
\textit{s064 (B0 III):} This star was observed in two runs, MOS2 and LSS3, where no radial velocity variations were detected. We discarded the combination of the observations and used only the MOS2 observation since we did not find S/N improvement in the co-addition. In this observing run, the star was observed in a very short slit, and sky subtraction was compromised. The wavelength of the core of the Balmer lines and the strongest \HeI~lines is consistent with the systemic velocity of the galaxy (\mbox{\vrad~= 325 \kms}). An instrumental defect at \mbox{$\sim$ 4700 \AA} prevents any conclusion on \mbox{\HeII~4686} and \mbox{\HeI~4713}. There could be a second component at \mbox{\vrad~$\sim$ 625 \kms}~showing \mbox{\SiIII~4552} and \mbox{\HeII~4541}.
\newline \newline
\textit{s065 (B0 V):} The wings of the Balmer series are abnormally broad, hinting it may be a binary star. The component with the quoted \mbox{\vrad~$\simeq$ 100 \kms} would be consistent with the blue wing of the Balmer lines and would show \HeI~lines, but no metallic lines.  We note the detection of \mbox{\HeI~4471} and \mbox{\MgII~4481} is hampered by artefacts. We tentatively classify this star as a B0~V. A second component with \mbox{\vrad~$\simeq$ 700 \kms} would be consistent with the red wing of the Balmer lines and \mbox{\SiIII~4552}. The emissions lines at $\sim$ 4230 \AA~and  $\sim$ 4480~\AA~are reminiscents of the Be star VFTS 822 from \citet{Evans2015}. 
\newline \newline
\textit{s068 (B0.5 I):} This source is a blend of two stars, as revealed by HST photometry \citep{Bianchi2012}.  It was observed in two consecutive nights with observations separated 19 hours. The spectra are very different, so we treated them as different stars (\Remark{LSS3.OB17203} and \Remark{LSS3.OB18203}). This component is consistent with \mbox{\vrad $\simeq$ 0 \kms} and spectral type B0.5 I because of the absence of \mbox{\MgII~4481} and the relative strength of \mbox{\SiIII~4552} with respect to \mbox{\HeI~4387}. We note the abnormally broad Balmer wings that can be caused by the blend.
\newline \newline
\textit{s069 (B0.5 II + O?):} This star was observed in the LSS3 and MOS2 runs, and no radial velocity variations were detected. The co-addition of the spectra do not improve the S/N, so we only use the MOS2 observation. The detection of \mbox{\SiIII~4552} and the strength of the \HeI~lines suggest B0.5 spectral type, although the possible presence of \mbox{\HeII~4200}, \mbox{\HeII~4541} and \mbox{\NV~4603, 4620} hints an early-O component at roughly the same \vrad.
\newline \newline
\textit{s070 (B0.5 III):} This star is OB524 from \citet{Camacho2016} and it was also observed within a tilted slit of MOS1 and in the LSS3 run. We obtained a deeper, higher-resolution spectrum in the LSS3 observing run, which will be used from now on. No significant radial velocity variations have been detected, although the MOS1 spectrum could show \mbox{\MgII~4481}. We reclassified the star from B0 to B0.5. Lines \mbox{\HeII~4200} and \mbox{\HeII~5411} exhibit abnormal profiles and were discarded for the classification.
\newline \newline
\textit{s071 (B1 I):} This source was observed in MOS1 and LSS3 observing runs. We detected a radial velocity variation of \mbox{$\sim$ 100 \kms}, indicative of an SB1 system. The poor S/N of the LSS3 data prevented us from co-adding the spectra, and we only use the MOS1 observation. In this observing run, the star was observed in a slit tilted 49\textdegree~from the North-South axis, which translates into decreased spectral resolution. If all the spectra of this source are co-added, the He and Si lines show an additional absorption at the blue part of the line core. Therefore, we applied radial velocity shifts to the individual OBs used, aligning them with respect to \mbox{\HeI~4471}.
\newline \newline
\textit{s072 (B1 I):} This source was observed in three observing runs, MOS1, MOS2 and LSS1 (here as OB421 in \citet{Camacho2016}). We detect radial velocity variations of \mbox{$\leq$ 60 \kms}~between runs and between the observing blocks of the MOS1 run, executed more spread over time than the MOS2 and LSS1. We do not find spectral type variations between observations, hence we mark the source as an SB1 candidate. \newline Since this source was brighter, we used a 0.63'' slit in MOS2. However, we could not benefit from the increased spectral resolution due to the comparatively poorer S/N, and we convolved the spectrum to a lower resolution. The poorer S/N could be due to the very short slitlet that, together with a bad column artefact on the right, may have led to problematic nebular subtraction. We co-added the data shifting the spectra to the restframe velocity (adopting the \vrad~of the MOS2 spectra). \newline The deep depression at \mbox{$\sim$ 4200 \AA{}} is due to a defect in the CCD. It is detected in both MOS1 and MOS2 observing runs since the star was observed in the same position of the CCD (MOS1 and MOS2 use the same pre-imaging). It is remarkably rich in oxygen based on the strength of \mbox{\OII~4651}, considering the star is in the outskirts and far from all the regions of active star formation.
\newline \newline
\textit{s073 (B1 I):} The lack of \HeII~lines and the presence of only weak \mbox{\SiIII~4552} points to a B1 type. The width of the Balmer series suggests a supergiant luminosity class.
\newline \newline
\textit{s075 (B1 I + O?):}  The spectrum shows lines at different radial velocities and very broad Balmer wings. At \mbox{\vrad~$\sim$ 200 \kms}, \mbox{\HeI~4471}, \mbox{\SiIV~4089} and \mbox{\MgII~4481} are detected and indicate type B1. Luminosity class I is assigned based on the \mbox{\SiIV~4089} /\mbox{\HeI~4121} ratio. The broad Balmer wings suggest a second component at \mbox{\vrad~$\sim$ 500 \kms}, which would be consistent with \mbox{\HeII~4200} and \mbox{\HeII~4686} lines. However, we cannot discard the star has luminosity class III at this S/N.
\newline \newline
\textit{s076 (B1 I):} We note the anomalous profile of \mbox{\HeI~4471}, which may be due to unknown nebular oversubstraction or the low S/N of the spectrum. The lack of \mbox{\MgII~4481} and \mbox{\SiIV~4089} points to a B1 type. Since the \mbox{\SiIII~4552} is weaker than \mbox{\HeI~4388}, we classify this star as a supergiant. In addition, we expect some contamination from the nearby star, \Remark{LSS3.OB06203}.
\newline \newline
\textit{s077 (B1 III):} This source was observed in two observing runs, LSS1 and LSS3, and we detected a possible radial velocity variation of 50 \kms. The poorer S/N of the LSS3 observation prevented us from co-adding both spectra. HST-WFPC2-F555W images reveal the source is a blend. 
\newline \newline
\textit{s078 (B1 III):} This star was observed in two observing runs, MOS2 and LSS3, where no radial velocity variations were detected. We do not combine the data since we do not find S/N improvements and use only the observation of the MOS2 run. In this observing run, there is a nearby red star that should be blocked by the mask. We discard that the absorption at 4486 \AA{} is caused by \mbox{\MgII~4481}.
\newline \newline
\textit{s079 (B1 III):} Spectrum of two unresolved stars that may be causing inconsistencies in the measurement of the radial velocity. At the systemic radial velocity of the galaxy, we observe \HeI~lines and strong \mbox{\SiIII~4552}, but no \mbox{\SiIV~4089}, \mbox{\SiII~4128} or \mbox{\MgII~4481}, hence we assign B1 type. The relative strength of \mbox{\HeI~4387} and \mbox{\SiIII~4552} points to luminosity class III.
\newline \newline
\textit{s080 (B1 III):} The lack of \mbox{\SiII~4129} and \mbox{\MgII~4481} suggests it is a B1. We assign class III, although this is inconsistent with the abnormally broad wings of \hb. We expect some contamination from the nearby stars \Remark{LSS3.OB01202} and \Remark{LSS3.OB01204}. 
\newline \newline
\textit{s081 (B1 V):} This source is a blend of two unresolved stars, which may explain the broad Balmer series profiles. Although the extraction failed to resolve the blend, we classify the dominant component as B1 due to the presence of \mbox{\SiIII~4552} and the strong \HeI~lines. Luminosity class V is assigned based on the relative strength of \mbox{\SiIII~4552} with \mbox{\HeI~4388}.
\newline \newline
\textit{s082 (B1.5 I):} The star was observed in a very short slitlet, and the selection of the sky-box for \textit{apall} subtraction was further hindered by the bad-column artefact from the neighbouring slit hosting star \Remark{MOS2.s32}. For the second night, the subtraction of a sky-slit was preferred. There is an artificial emission at \mbox{$\sim$ 4350 \AA}. The star could show \OII~ absorption lines (see e.g. red wing of \hg).
\newline \newline
\textit{s083 (B2 I):} This source was observed in the LSS1 observing run, corresponding to OB222 in \citet{Camacho2016}'s catalogue, and in the LSS3 run on the night of the 17th of January of 2021. Between the two observations, we detected a significant radial velocity variation of \mbox{115 \kms}, but no difference in its spectral features, which indicates an SB1 system. We co-added the spectra after correcting the radial velocity variation to achieve higher S/N. We reclassify the target from B1 to B2 because of the strength of  \mbox{\MgII~4481} and \mbox{\SiII~4128}.
\newline \newline
\textit{s084 (B2 I):} The weak \mbox{\MgII~4481} and the relative strength of \mbox{\SiII~4128} with respect to \mbox{\SiIII~4552} and \mbox{\HeI~4121} suggests a B2 type.
\newline \newline
\textit{s085 (B2 I):} This source was observed in the LSS1 observing run, corresponding to OB222 in \citet{Camacho2016}'s catalogue, and in the LSS3 run on the night of the 17th of January of 2021. Between the two observations, we detected a significant radial velocity variation of \mbox{115 \kms}, but no difference in its spectral features, which indicates an SB1 system. We co-added the spectra after correcting the radial velocity variation to achieve higher S/N. We reclassify the target from B1 to B2 because of the strength of  \mbox{\MgII~4481} and \mbox{\SiII~4128}.
\newline \newline
\textit{s086 (B2 III):} This source was observed in the LSS2 observing run and in two nights of the LSS3 run. We find consistent radial velocities variations of \mbox{$\sim$ 100 \kms}~between the LSS2 spectrum (18 February 2020) and the LSS3 observation of the night of 17 January 2021. However, the LSS3 spectrum of 19 January 2021 shows a radial velocity different by \mbox{$\sim$ 50 \kms}, suggestive of an SB1 system. We shifted all of them to the restframe velocity and co-added the spectra, shown in Figure \ref{fig:plot_LSS3.OB17206}. \newline \mbox{\SiIII~4552}, \mbox{\HeI~4471} and other \HeI~lines are present. There could be some \mbox{\MgII~4481} which, together with its relative strength compared with \SiIII, indicates B2. We assigned class III based on the width of the Balmer lines, although we note that the existence of a second component could also contribute to a broader profile. 
\newline \newline
\textit{s087 (B2 III + O?):} This source is a blend, as revealed by HST imaging.  The relative strength of \mbox{\HeI~4471} and \mbox{\MgII~4481} points to a B2 type. We tentatively assign class III because of the broad wings of the Balmer lines. However, this broadening could also be explained by a nearby O dwarf, which would also contribute the \mbox{\HeII~4200} and \mbox{\HeII~5411} lines to the blended spectrum. 
\newline \newline
\textit{s088 (B2 III + O?):}  The star is classified as B2 because of the presence of \mbox{\SiIII~4552} and the ratio of \mbox{\HeI~4471} and \mbox{\HeI~4387}. We note that the \HeI~lines are broad and consistent with a second component at \mbox{\vrad $\sim$  100 \kms} which could also show \mbox{\HeII~4200}. Because of the broad Balmer wings and faint absolute magnitude (\mbox{$M_V$ = -3.987 mag}), we assign class III. 
\newline \newline
\textit{s089 (B2.5 I):} This star was observed with a slit tilted by -25\textdegree{} from the North-South axis, which translates into poorer spectral resolution. We reclassify the star with respect to \citet{Camacho2016} from a B0 type to a B2.5 type due to the ratios of \mbox{\MgII~4481}/\mbox{\SiIII~4552} and \mbox{\SiIII~4552}/\mbox{\SiII~4128}, and marked it as a binary candidate due to this change in spectral features. We assign class I based on the wings of the Balmer lines, and its absolute magnitude (\mbox{$M_V$ = -6.430 mag}) is consistent. 
\newline \newline
\textit{s090 (B2.5 I):} This source is a blend of two unresolved stars. B2.5 spectral type is assigned based on the relative strength of \mbox{\HeI~4471} with \mbox{\MgII~4481}. We note the \HeI~lines are abnormally weak.
\newline \newline
\textit{s091 (B2.5 I):} The star was observed in two observing runs, LSS1 and LSS3, where no radial velocity variations were detected. The poorer S/N of the LSS3 observation prevented us from co-adding both spectra.
\newline \newline
\textit{s092 (B2.5 I):} Its spectrum only displays clear lines of \HeI~and the Balmer series. Although \mbox{\SiIII~4552} and \mbox{\MgII~4481} are very weak, their relative line strength implies a B2.5 type. Lacking indicators for luminosity class, since the Balmer lines seem heavily contaminated by nebular oversubtraciton, we assign class I based on its absolute magnitude (\mbox{$M_V$ = -5.714 mag}).
\newline \newline
\textit{s093 (B2.5 I):}   This source is a blend of two stars, as revealed by HST photometry.  It was observed in two consecutive nights with observations separated 19 hours. The spectra are very different, so we treated them as different stars (\Remark{LSS3.OB17203} and \Remark{LSS3.OB18203}). This observation is consistent with a B2.5 supergiant and a radial velocity of 450 \kms. We note the abnormally broad Balmer wings that can be caused by the blend. 
\newline \newline
\textit{s094 (B2.5 I + O?):} The Balmer series and \HeI~lines are consistent with the restframe velocity of the galaxy and we classify the main component as B2.5 I. However, \HeII~lines are tentatively detected at \mbox{\vrad~= 650 \kms}. This component could also show \NV~and \NIII~lines. Therefore, we mark the source as a possible binary. 
\newline \newline
\textit{s095 (B2.5 III):} We classify this star as a B2.5 type due to the presence of \mbox{\MgII~4481} and its relative strength with respect \mbox{\SiIII~4552}. We assign luminosity class III due to the width of the Balmer lines.
\newline \newline
\textit{s096 (B2.5 III):} Noisy spectrum showing the Balmer series. The relative strength of \mbox{\MgII~4481} and \mbox{\SiIII~4552} suggests B2.5 type. Due to the width of the Balmer lines, we assign luminosity class III.
\newline \newline
\textit{s097 (B3 I):} We classify the star as B3 supergiant based on the relative strength of \mbox{\HeI~4471} and \mbox{\MgII~4481}, and the width of the wings of the Balmer lines. We note that no Si lines are detected in the spectrum.
\newline \newline
\textit{s098 (B3 I):} This star presents \mbox{\MgII~4481} and \mbox{\SiIII~4552}. The relative strength of \mbox{\SiIII~4552} with \mbox{\MgII~4481} is consistent with a B3 type.
\newline \newline
\textit{s099 (B3 I):} The left wing of \hd~is affected by an artefact that prevents using \mbox{\SiIV~4089} for spectral classification. No lines of \SiIII, or \HeII~are seen. We assign B3 because of the absence of \mbox{\SiIII~4552} and the relative strength of \mbox{\MgII~4481} and \mbox{\HeI~4471}. Lacking indicators for luminosity class, since the Balmer lines seem affected by nebular oversubtraciton, we assign class I based on the absolute magnitude.
\newline \newline
\textit{s100 (B3 III):} The spectrum shows the Balmer series, \HeI~lines, and no clear Si or \HeII~lines. Type B3 was assigned based on the \mbox{\MgII~4481}/\mbox{\HeI~4471} ratio. The luminosity class is assigned based on the width of Balmer lines. 
\newline \newline
\textit{s101 (B3 III):} Its spectrum only displays clear lines of \HeI, \MgII~and the  Balmer series. We classify this star as a B3 giant based on the relative strength of \mbox{\HeI~4471} with respect  \mbox{\MgII~4481} and the width of its Balmer lines.
\newline \newline
\textit{s102 (B7 I):} We classify the star as B7 because of the relative strength of \mbox{\MgII~4481} with respect to \mbox{\HeI~4471} and \mbox{\SiII~4128} with \mbox{\HeI~4121}. Due to the width of The Balmer lines, we assign a supergiant luminosity class, which is consistent with its apparent magnitude.
\newline \newline
\textit{s103 (B7 I):}   This source is a blend of two identified stars in \citet{Bianchi2012}.
\newline \newline
\textit{s104 (B7 I + O?):} The \textit{apall} extraction indicated that the star could be a blend. The relative strengths of \mbox{\SiII~4128} with \mbox{\HeI~4121} and of \mbox{\MgII~4481} with \mbox{\HeI~4471} indicate a B7 type. The spectrum also shows \mbox{\HeII~4541} consistent with a secondary component.
\newline \newline
\textit{s105 (B7 I):} This object was observed in LSS3 and off-slit LSS2. We did not registered radial velocity variations and decided to use only the LSS3 observation since no S/N improvement was achieved in the co-addition. The most prominent spectral features are the Balmer series, \mbox{\HeI~4026}, and \mbox{\HeI~4471}, although their profiles are dominated by the oversubtracted nebular components. On the other hand, \mbox{\HeI~4144} and \mbox{\HeI~4920} seem to be mostly stellar.  We classify the star as B7 supergiant based on the relative strength of \mbox{\MgII~4481} and \mbox{\HeI~4471}. 
\newline \newline
\textit{s106 (B7 II):} The source was observed in the LSS1 run with a radial velocity of \mbox{$\sim$ 335 \kms}. It was also included in the LSS3 run where we register a different radial velocity of \mbox{$\sim$ 250 \kms}~thus we marked it as SB1. We keep the latter spectrum, because of the better S/N and reclassify the target as B7 II. In addition, HST photometry reveals the source is a blend of two stars.
\newline \newline
\textit{s107 (B7 III):} The strong absorption of the \mbox{\SiII~4128} and the relative strength of \mbox{\HeI~4471} and \mbox{\MgII~4481} suggest a B7 type. We note, however, there are some radial velocity inconsistencies. We assigned class III based on the broadening of the Balmer lines.
\newline \newline
\textit{s108 (B7 III + B?):} This star was observed in the LSS2 run and in 2 observing blocks of LSS3, no significant velocity variation was detected. Only the spectrum with the highest S/N was used (LSS3). Spectral type B7 was assigned based on the relative strength of \mbox{\MgII~4481} with respect to \mbox{\HeI~4471}. The width of the Balmer lines suggests class III. There could be a second component at \mbox{\vrad $\sim$ 450 \kms}~showing \mbox{\SiIII~4552}.
\newline \newline
\textit{s109 (B7 III + O?):} The source is classified as B7 based on the relative strength of \mbox{\HeI~4121} with respect to \mbox{\SiII~4128} and the ratio of \mbox{\MgII~4481} to \mbox{\HeI~4471}. The broad Balmer lines suggest a giant luminosity class. The spectrum shows strong \mbox{\HeII~5411} and possibly \mbox{\HeII~4686} in absorption, hence we mark it as an SB2 candidate.
\newline \newline
\textit{s110 (B8 I):} The relative strengths of \mbox{\SiII~4128} with \mbox{\HeI~4121} and of \mbox{\MgII~4481} with \mbox{\HeI~4471} indicate a B8 type.
\newline \newline
\textit{s111 (B8 II):} \citet{Camacho2016} proposed that the absolute magnitude of this star suggests it may be a central star of a planetary nebula (CSPN) still in a low-excitation phase.
\newline \newline
\textit{s112 (B8 III):} The equal depth of \mbox{\MgII~4481} and \mbox{\HeI~4471} points to a B8 type. The broad Balmer lines signal a giant luminosity class.
\newline \newline
\textit{s113 (B8 III):} Noisy spectrum. The strong presence of \mbox{\SiII~4128} and the similar depths of \mbox{\MgII~4481} and \mbox{\HeI~4471}, along with a broad Balmer series, indicate a B8 giant.
\newline \newline
\textit{s114 (B8 III):} The source is classified as B8 based on the relative strength of \mbox{\HeI~4471} to \mbox{\MgII~4481}. 
\newline \newline
\textit{s115 (B9 I):} Since the \mbox{\MgII~4481} line is strong, and \HeI~lines are weak but present, we classify this star as a B9. The luminosity class is assigned based on the width of its Balmer lines.
\newline \newline
\textit{s117 (B9 III):} Due to the relative strength of the \mbox{\MgII~4481} with \mbox{\HeI~4471} lines and the broad Balmer series, we classify this star as a B9 giant.
\newline \newline
\textit{s118 (B9 V):} The spectrum is very noisy. We classify it as B9 due to the detection of \mbox{\MgII~4481} and \mbox{\SiII~4128}. The broad Balmer wings suggest class V, although we note this is inconsistent with the observed \mbox{$V$ = 21.98} magnitude.
\newline \newline
\textit{s120 (A2 I):} We detect \FeII~and \SiII~lines slightly stronger than the A0 II star \Remark{LSS1.OB525} (see also \citet{Hosek2014}) and we assign an A2 type. The star is classified as a supergiant because of its absolute magnitude (\mbox{$M_V$ $\sim$ -6.5 mag}). The star was observed in the first two observing nights of LSS3 and no radial velocity variations are detected. 
\newline \newline
\textit{s121 (A2 I):} We detect \FeII~lines and \SiII~lines consistent with \Remark{LSS3.OB01101}, hence we classify it as A2 I type. 
\newline \newline
\textit{s122 (A2 I):} Similar spectrum to \Remark{LSS3.OB01204}.
\newline \newline
\textit{s123 (A2 III):} Noisy spectrum. Classified as A2 based on resemblance to star \Remark{LSS3.OB01101}.
\newline \newline
\textit{s124 (F5 I):} This star was observed in two observing runs, LSS1 (as OB621) and LSS3. We did not detect radial velocity variations, and we co-added the spectra to increase the S/N. The star is also included in \citet{Kaufer2004}'s catalogue where they classify the star as an F hypergiant.
\newline \newline
\textit{s167 (G5 I):} This star is identified as Sextans~A~6 in \citet{Britavskiy2019}, where it was classified as late G -- early K~I. It is located in region--B, where ionized hydrogen is abundant, and the Balmer lines show sky over subtraction in our spectrum. The radial velocity and absence of parallaxes and proper motions of the star further support that the star belongs to Sextans~A. We reclassified the star as G5 I based on the relative strength of the $G$--band compared to \mbox{\FeI~4325}.
\newline \newline
\textit{s168 (G8 I + B7 I):} At the time of spectral extraction, we noticed that the FWHM of the source was wider than the PSF of the instrument, which suggests the source may be a blend. This is actually seen in the observed spectrum. At \mbox{$\lambda$ > 4800 \AA}, the spectrum is consistent with a G8~I. At bluer wavelengths, the spectrum displays strong lines of \mbox{\SiII~4128} and \mbox{\MgII~4481} at \mbox{\vrad~= 250 \kms}, and their relative strength with \mbox{\HeI~4121} and \mbox{\HeI~4471}, respectively, points a B7 type.
\newline \newline
\textit{s169 (G8 I):} This star is identified as Sextans~A~4 in \citet{Britavskiy2019}, where it was classified as K1-3 I. It is located in region--B, where ionized hydrogen is abundant, and Balmer lines show sky over subtraction in our spectrum. Its radial velocity and lack of parallaxes and proper motions further support the star belongs to Sextans~A. We reclassified the star as G8~I based on the relative strength of the $G$--band compared to \mbox{\FeI~4325}.
\newline \newline
\textit{s172 (K5 III):} This star is located in region--B, where ionized hydrogen is abundant. The presence of nebular emission lines and the radial velocity of the star supports that the star belongs to Sextans~A. The star is classified as K5~III based on the strength of the $G$--band, although we note that \hg~can not be used for comparison because it is filled with nebular emission.
\newline \newline
\textit{s173 (K7 V):} The extreme flux loss at \mbox{$\lambda$ > 5000 \AA} is likely due to the star being slightly off-slit.
\subsection{O stars}
\begin{figure} \centering \includegraphics[width=\hsize]{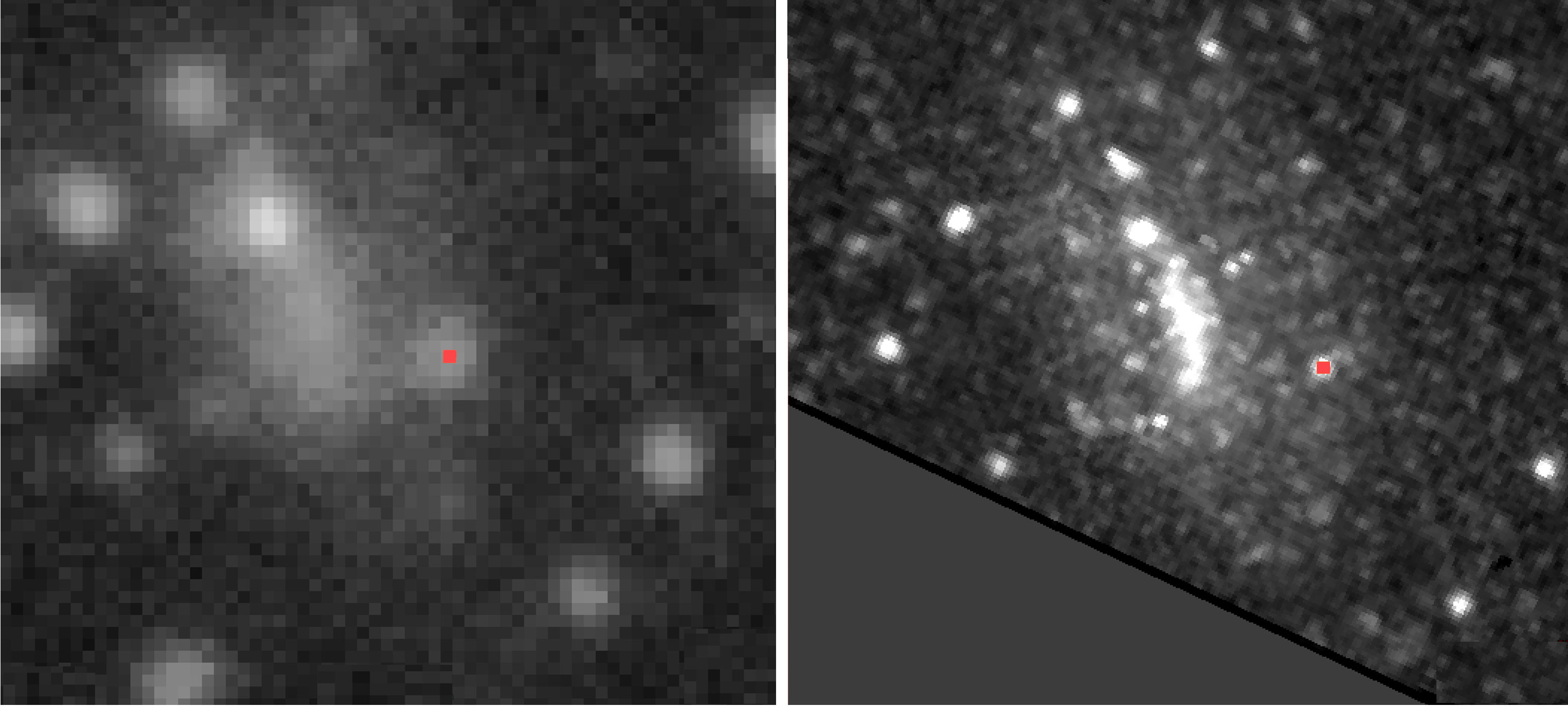}\caption{Images in the $V$--band by \citet{Massey2007} (left panel) and HST-WFPC2-F555W--band from Proposal ID 5915, Cycle 5, PI E. Skillman (right panel). The red square marks the location of \Remark{LSS3.OB13208}.}  \label{fig:images_LSS3.OB13208} \end{figure}

Here we report stars for which \HeII~lines are detected.
\newline \newline
\textit{s126 (O V):} HST photometry reveals the source is a blend of three unresolved stars identified in \citet{Bianchi2012}. The tentative presence of \mbox{\NV~4603}, \mbox{\NV~4620} and \mbox{\CIV~4441} would correspond to an O2 type (see e.g. star 30 Dor/P871 in \citet{Walborn2002}). However, we note the weak emission of \mbox{\HeII~4686} and the absence of \mbox{\HeII~4541}, which may be due to nebular contamination. On the other hand, \mbox{\HeII~4200} displays strong emission, unseen in typical spectra of early O stars. We assign luminosity class V because of the broad Balmer lines, although this is in contradiction with \mbox{\HeII~4686} in emission or filled.
\newline \newline
\textit{s127 (O V + gal):} This star overlaps with the projection of a background galaxy, which may explain why its colours are not compatible with the adopted spectral type. HST imaging is consistent with this source being a star from Sextans~A (see Fig. \ref{fig:images_LSS3.OB13208}).  \newline \newline The spectrum shows high S/N, although \HeII~lines are weak (which may be caused by the continuum dilution by the background galaxy) and \HeI~lines are absent due to nebular contamination. Class V is assigned based on the width of the Balmer lines, although we note the absence of \mbox{\HeII~4686}, which should exhibit strong absorption at this class. 
\newline \newline
\textit{s128 (O V):} This star is very faint and was not registered by \citet{Massey2007}'s catalogue, although \citet{Bianchi2012}'s HST catalogue provides a source in that location with \mbox{$V$ = 22.67 mag}. \hb~and \hg~are strong in absorption, and there are absorption features at the expected location of \mbox{\HeII~4200} and \mbox{\HeII~4541}. The latter would be stronger than \mbox{\HeI~4471}. Therefore we classify the star as a potential O dwarf.
\newline \newline
\textit{s129 (O V + neb):}   Noisy spectrum dominated by nebular features. The only He lines unambiguously detected are \mbox{\HeII~4200}, \mbox{\HeII~5411} and \mbox{\HeII~4922}. The relative strength of these features suggests a mid- to early-O type star. Considering the visual magnitude of the star, we assign a conservative O~V classification, although we note the narrow, nebular over-subtracted Balmer lines. 
\newline \newline
\textit{s132 (O V + neb):}   The spectrum shows very poor S/N resulting from the off-slit allocation of this \mbox{$V$ = 21.35 mag} star. It is dominated by nebular emission lines. Yet, it shows some absorption features. The radial velocity was estimated from \mbox{\HeII~4686}. Other absorption features, most notably \mbox{\HeII~4200}, concur with this radial velocity. 
\newline \newline
\textit{s133 (O  + neb):} This star was observed in a very short slit that prevented the use of local sky boxes for sky subtraction with \textit{apall}. We used the sky spectrum from the sky slits, which resulted in the loss of the extremes of the spectral range. \mbox{\HeII~4200} and \mbox{\HeII~4686} are strong. \mbox{\HeII~4541} is not detected but could be filled by nebular emission. We note that the \HeII~lines, and also \mbox{\HeI~4471} are broad as well. This star is far from the optical centre of the galaxy and the most prominent diagnostics of star formation. 
\newline \newline
\textit{s134 (O V):} Very noisy spectrum. We detected the Balmer series and tentatively the \HeI~lines and \mbox{\HeII~4686}. There may be contamination from a nearby star. There is an unknown emission at 4403 \AA.
\newline \newline
\textit{s135 (O V):} The spectrum might present possible contamination from the adjacent stars in the slit, \Remark{LSS3.OB44205} and \Remark{LSS3.OB44207}. There is a strong \mbox{\HeII~4541} line, although it shows inconsistent radial velocity. 
\newline \newline
\textit{s136 (O  + neb):} We classify this star as O based on the \HeI~lines and \mbox{\HeII~4686} in absorption. 
\subsection{B stars}
\begin{figure} \centering \includegraphics[width=\hsize]{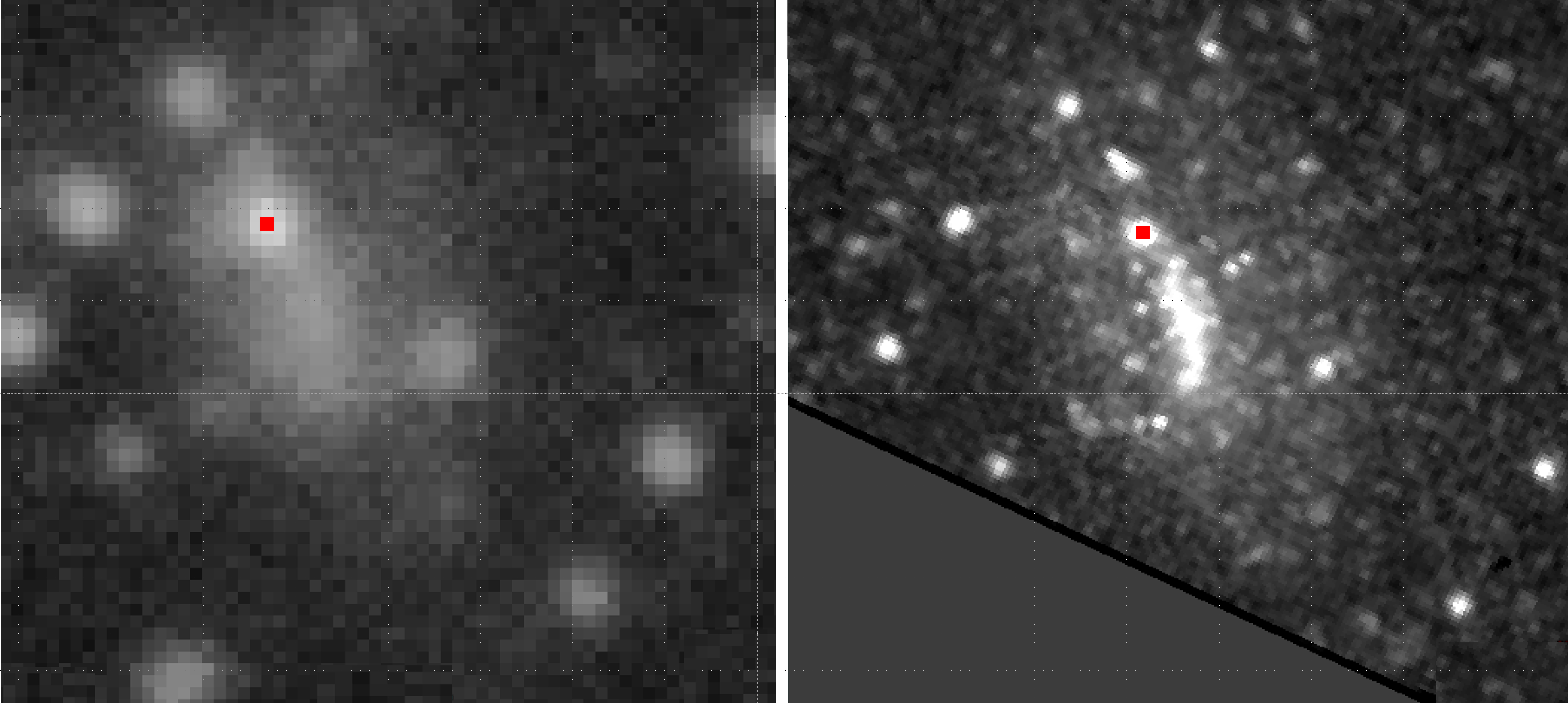}\caption{Images in the $V$--band by \citet{Massey2007} (left panel) and HST-WFPC2-F555W--band from Proposal ID 5915, Cycle 5, PI E. Skillman (right panel). The red square marks the location of \Remark{LSS3.OB13207}.}  \label{fig:images_LSS3.OB13207} \end{figure}

Here we report stars for which \HeI~lines are detected.
\newline \newline
\textit{s137 (Be  + O?):}    This spectrum displays strong emission at \hb, \hg~and \hd. The star is located far from the H~{\sc ii}-rich regions, and \hg~and \hb~are wider than the expected width of nebular lines, suggesting a stellar origin. The Balmer series and \HeI~lines are consistent with the systemic radial velocity of the galaxy. There may be a secondary component showing \HeII~and Si lines at 0 \kms.
\newline \newline
\textit{s138 (B III + gal):} This source spatially overlaps a background galaxy (see Fig. \ref{fig:images_LSS3.OB13207}), which contributes to its total flux (see e. g. redshifted lines of \hb~and [O~{\sc iii}]). The shown spectrum exhibits stellar Balmer lines and very weak \HeI~lines, probably due to the dilution of the stellar flux. We classify the star as a B giant based on the width of the Balmer lines. \newline The target was observed in two different nights within the LSS3 run. The spectra of the second night were discarded since they were taken at worse seeing conditions and suffer from a larger contribution of the background galaxy. We note, however, that we detect a radial velocity variation at the Balmer lines of $\sim$150 \kms~between nights.
\newline \newline
\textit{s139 (B  + O?):} The extremely broad Balmer lines, together with a tentative detection of \HeII~lines at \mbox{\vrad $\sim$ 400\kms} suggest a possible O-type component.
\newline \newline
\textit{s140 (B III):} Off-slit star. \mbox{\HeI~4471} and other \HeI~lines are clearly detected. \mbox{\SiIII~4552} could be present. Luminosity class III is assigned as a trade-off between the width of the Balmer lines and its visual $V$ magnitude.
\newline \newline
\textit{s141 (B):} Faint star, not intended in the survey, and off-slit. Its trace is lost at red wavelengths ($\lambda$ > 5300 \AA{}) during the extraction. 
\newline \newline
\textit{s144 (B):} Very faint ($V$ = 21.535 mag), off-slit star. Balmer lines show broader wings with cores affected by nebular oversubtraction. Some \HeI~lines and possibly \mbox{\SiIII~4552} are detected. 
\newline \newline
\textit{s145 (B):} Very noisy spectrum. Balmer lines and some \HeI~lines are detected. There may be contamination from a nearby star.
\newline \newline
\textit{s146 (B):} Very noisy spectrum, only the Balmer series, \SiII~and some lines of \HeI~are detected.
\newline \newline
\textit{s147 (B):} Very noisy spectrum. The Balmer series is detected, together with some \HeI~lines, and perhaps \mbox{\SiII~4128} and \mbox{\MgII~4481}.
\subsection{OB stars}
Here we report stars for which \HeI~lines and tentatively \HeII~lines are detected.
\newline \newline
\textit{s148 (OB):} The spectrum could show \mbox{\HeII~4200} and broad lines of \mbox{\HeI~4387} and \mbox{\HeI~4920}. 
\newline \newline
\textit{s150 (OB):} The spectrum may show \mbox{\HeII~4541}. 
\newline \newline
\textit{s151 (OB):} This spectrum is a blend of three unresolved stars. 
\newline \newline
\textit{s152 (OB):} Balmer lines, some \HeI~lines and perhaps \mbox{\HeII~4200} are detected. There may be contamination from a nearby star.
\newline \newline
\textit{s154 (OB):} We clearly detect the Balmer series, some \HeI~lines and perhaps \mbox{\HeII~4541}.
\newline \newline
\textit{s156 (OB):} This star is very faint and it was not registered by Massey's catalogue, although \citet{Bianchi2012}'s HST catalogue provide \mbox{$V$ = 22.18 mag}. Its spectral S/N is poor as expected for such a faint star. \newline The Balmer lines seem to be relatively broad with the core affected by nebular oversubtraction, which would be offset in the \vrad~space from the broader stellar component. The radial velocity differs by \mbox{$\sim$ 200 \kms} from the galaxy's systemic velocity, but after this correction we detect \HeI~lines at their restframe wavelengths. \mbox{\HeII~4541} is not detected but \mbox{\HeII~4200} is. On the other hand, there is no clear \mbox{\SiIII~4552} but there seem to be Si absorption by the wings of \hd. 
\subsection{OBA stars}
Here we report stars for which only the Balmer series is clearly observed.
\newline \newline
\textit{s158 (OBA):} This star was observed within a slit tilted 8\textdegree{} to include also star \Remark{MOS2.s34}. The inclination results in degraded spectral resolution. The signal to noise ratio is extremely poor. 
\newline \newline
\textit{s159 (OBA):} This source is a blend of two unresolved stars based on HST photometry. 
\newline \newline
\textit{s161 (OBA):} Noisy spectrum with strong nebular contamination. Only \hd~is clearly detected.
\newline \newline
\textit{s162 (OBA):} There may be contamination from a nearby star. 
\newline \newline
\textit{s163 (OBA):} There may be contamination from a nearby star. 
\newline \newline
\textit{s165 (OBA):} We assigned the systemic velocity of the galaxy. 


\bsp	
\label{lastpage}
\end{document}